\documentclass[a4paper,11pt]{article}

\usepackage{fullpage}
\usepackage{color}
\usepackage{latexsym}
\usepackage{mathrsfs}
\usepackage{amsmath,amssymb,amsthm}
\usepackage{bbm} 

\usepackage{tikz-cd}                   
\usepackage{xspace,bbold}              
\usepackage[margin=1in]{geometry}      
\usepackage[utf8]{inputenc} 

\numberwithin{equation}{section}       

\usepackage{jheppub}                   

\makeatletter
\gdef\@fpheader{\ }                    
\makeatother

\newtheorem{thm}{Theorem}
\newtheorem*{cor}{Corollary}

\newcommand{\dd}{\mathrm{d}}
\newcommand{\me}{\mathrm{e}}
\newcommand{\ii}{\mathrm{i}}

\newcommand{\der}{\partial}

\newcommand{\bbZ}{\mathbb{Z}}
\newcommand{\bbR}{\mathbb{R}}
\newcommand{\bbC}{\mathbb{C}}

\newcommand{\gperp}{\mathfrak{g}^\perp}

\DeclareMathOperator{\U}{\mathit{U}}
\DeclareMathOperator{\SU}{\mathit{SU}}
\DeclareMathOperator{\OO}{\mathit{O}}
\DeclareMathOperator{\SO}{\mathit{SO}}
\DeclareMathOperator{\USp}{\mathit{USp}}
\DeclareMathOperator{\SL}{\mathit{SL}}
\DeclareMathOperator{\GL}{\mathit{GL}}

\DeclareMathOperator{\Spin}{\mathit{Spin}}
\DeclareMathOperator{\so}{\mathfrak{so}}

\newcommand{\np}{n}
\newcommand{\nm}{r}

\newcommand{\rep}[1]{\boldsymbol{#1}}
\newcommand{\repp}[2]{(\rep{#1}, \rep{#2})}
\newcommand{\id}{\mathbb{1}}

\DeclareMathOperator{\cc}{c}

\DeclareMathOperator{\im}{Im}

\DeclareMathOperator{\vol}{vol}

\DeclareMathOperator{\Lie}{Lie}

\newcommand{\Gs}[1]{\Gamma(#1)}

\def\cV{\mathcal{V}}
\newcommand{\ul}{\underline}

\newcommand{\comm}[2]{\left[#1,#2\right]}

\DeclareMathOperator{\Comm}{C}
\DeclareMathOperator{\Heis}{\mathrm{Heis}}

\newcommand{\Com}[2]{\Comm_{#2}(#1)}

\newcommand{\BLie}[2]{\big[#1,#2\big]}

\newcommand{\Dgen}{{D}}
\newcommand{\Lgen}{{L}}

\DeclareMathOperator{\adj}{ad}

\newcommand{\GM}[2]{\big<#1,#2\big>}

\newcommand{\LC}{\nabla}

\newcommand{\Edd}{\mathit{E_{d(d)}}}
\newcommand{\Hd}{\mathit{H_d}}
\newcommand{\dHd}{\mathit{\tilde{H}_d}}

\newcommand{\Ex}[1]{\mathit{E}_{#1(#1)}}

\newcommand{\Gst}{G_S}
\newcommand{\tGst}{\tilde{G}_S}
\newcommand{\GN}{G_\mathcal{N}}
\newcommand{\Ggauge}{G_{\text{gauge}}}

\newcommand{\Tint}{{T_{\text{int}}}}

\newcommand{\stimes}{\!\times\!}
\newcommand{\splus}{\!\oplus\!}

\newcommand{\Mscal}{\mathcal{M}_{\textrm{scal}}}

\newcommand{\be}{\begin{equation}}
\newcommand{\ee}{\end{equation}}
\newcommand{\bea}{\begin{eqnarray}}
\newcommand{\eea}{\end{eqnarray}}
\newcommand{\nn}{\nonumber}

\newcommand{\bpm}{\begin{pmatrix}}
\newcommand{\epm}{\end{pmatrix}}

\newcommand{\diff}{\mathrm{d}}

\newcommand{\rme}{\mathrm{e}} 

\def\ct{c_\theta}
\def\st{s_\theta}
\def\cc{c_\chi}
\def\sc{s_\chi}
\def\cp{c_\phi}
\def\sp{s_\phi}

\def\Cfl{C^{\rm f\,\!l}}

\usepackage{booktabs}

\title{Systematics of consistent truncations from generalised geometry} 

\author[a]{Davide Cassani,}
\emailAdd{davide.cassani@pd.infn.it}
\author[b]{Gr\'egoire Josse,}
\emailAdd{josse@lpthe.jussieu.fr}
\author[b]{Michela Petrini,}
\emailAdd{petrini@lpthe.jussieu.fr}
\author[c]{and Daniel Waldram}
\emailAdd{d.waldram@imperial.ac.uk}

\affiliation[a]{INFN, Sezione di Padova, Via Marzolo 8, 35131 Padova, Italy}
\affiliation[b]{Sorbonne Universit\'e, UPMC Paris 05, UMR 7589, LPTHE, 75005 Paris, France}
\affiliation[c]{Department of Physics, Imperial College London,
Prince Consort Road, London, SW7 2AZ, UK}

\abstract{
We present a generalised geometry framework for systematically constructing consistent truncations of ten- and eleven-dimensional supergravity preserving varying fractions of supersymmetry. Truncations arise when there is a reduced structure group $\Gst$ of the exceptional generalised geometry, such that the intrinsic torsion is a $\Gst$-singlet. The matter content of the truncated theory follows from group-theoretical arguments, while the gauging is determined by the sub-algebra of generalised diffeomorphisms generated by the $\Gst$-singlet vectors. After discussing the general ideas across different spacetime dimensions and amounts of supersymmetry, we provide detailed formulae for truncations to gauged half-maximal supergravity in five dimensions. In particular, we establish an expression for the generalised metric on the exceptional tangent bundle, which determines the scalar truncation ansatz. As applications, we show that this formalism gives a simple derivation of a new consistent truncation of type IIB supergravity on $\beta$-deformed Lunin--Maldacena geometries, yielding half-maximal supergravity coupled to two vector multiplets, and of the truncation of eleven-dimensional supergravity on Maldacena--N\'u\~nez geometries, given by $S^4$ twisted over a Riemann surface, which leads to half-maximal supergravity coupled to three vector multiplets.
}

\begin{document}

\maketitle

\section{Introduction}

A common problem in string theory and supergravity is how to derive lower-dimensional effective theories.  
Given a Kaluza--Klein reduction on a compact manifold,  a consistent truncation is a procedure to truncate  the infinite tower of  Kaluza--Klein states to a finite set in a consistent way, such that solutions of equations of motion of the truncated system are also always solutions of the original theory. In other words, the dependence of the higher-dimensional fields on the internal manifold factorises out once the truncation ansatz is plugged in the equations of motion. The classic example, known as a Scherk--Schwarz reduction, is when the internal space is a group manifold $\mathcal{G}$ (or a quotient $\mathcal{G}/\Gamma$ thereof by a freely-acting discrete group $\Gamma$)~\cite{Scherk:1979zr}. Consistency is a consequence of keeping only modes invariant under the group action. Aside from these cases, consistent truncations are relatively rare and hard to construct, see for instance~\cite{Duff:1984hn,Cvetic:2000dm}. Classic examples of consistent truncations on spaces that are not group manifolds are the truncations of eleven-dimensional supergravity on $S^7$~\cite{deWit:1986oxb} and on  $S^4$~\cite{Nastase:1999kf} both leading to a maximally supersymmetric truncated theory. 

Recently, the reformulation of supergravity using Generalised Geometry and Exceptional Field Theory has provided a new framework for giving a systematic geometrical description of maximally supersymmetric consistent truncations, both of conventional Scherk--Schwarz type and the exotic sphere truncations~\cite{Lee:2014mla,Baron:2014yua,Hohm:2014qga,Baron:2014bya,Baguet:2015sma,Inverso:2017lrz}. In particular, the notion of a generalised parallelisation allows one to show that all known such truncations are a form of generalised Scherk--Schwarz reductions and to prove the long-standing conjecture of the  consistency of type IIB supergravity on $S^5$~\cite{Lee:2014mla,Ciceri:2014wya,Baguet:2015sma}. Extensions of these ideas have also recently been considered in the case of half-maximal truncations in~\cite{Malek:2016bpu,Ciceri:2016hup,Malek:2017njj,Malek:2018zcz,Malek:2019ucd}, mostly focused on reductions to seven- and six-dimensional supergravities, although~\cite{Malek:2017njj} also discusses more general cases. An appealing feature of the maximal generalised Scherk--Schwarz reductions is that one can determine the lower-dimensional supergravity directly from the generalised geometry, a priori of any explicit substitution into the equations of motion.
It is therefore natural to ask whether generalised geometry can give a similar characterisation of generic consistent truncations with any amount of supersymmetry. 

In this paper, we derive such a unified framework for constructing consistent truncations with different amounts of supersymmetry (including non-supersymmetric truncations), based on the $G$-structure of the generalised geometry. The key requirement is that the so-called ``intrinsic torsion''~\cite{Coimbra:2014uxa} of the $G$-structure contains only singlets. This formalism allows one to easily determine all the features of the lower-dimensional gauged supergravity, such as the amount of supersymmetry, the coset manifold of the scalars, the number of gauge and tensor fields, and the gauging, all directly from the geometry. It also provides a general proof of the conjecture of~\cite{Gauntlett:2007ma}, stating that to any supersymmetric solution to ten- or eleven-dimensional supergravity of the warped product form AdS$_D \times_{\rm w} M$, there is a consistent truncation to pure gauged supergravity in $D$ dimensions containing that solution and having the same supersymmetry. As we will see, this statement follows from observing that supersymmetric AdS$_D \times_{\rm w} M$ solutions always define a ``maximal'' supersymmetric generalised $G$-structure, and the $G$-invariant tensors then can be used to define a consistent truncation.  When the actual generalised $G$-structure is a subgroup of the maximal one, we show that one may go further and obtain a consistent truncation which includes matter multiplets and in some cases preserves more supersymmetry than the vacuum. 

The structure of the paper and the main results are as follows. In Section~\ref{sec:gen-formalism} we describe the general ideas and apply them to a number of simple cases, notably identifying the maximal $G$-structure for a given amount of supersymmetry and thus proving the conjecture of~\cite{Gauntlett:2007ma}, and also deriving the field content of the supersymmetric truncations that arise from reductions to $D=4$ and $D=5$ on conventional $G$-structure manifolds, reproducing a number of known results in the literature. In Section~\ref{sec:half_max_str_5d} we focus on truncations leading to half-maximal supergravity in five dimensions, which are based on $\Ex{6}$ generalised geometry. This case was first considered in the general analysis of~\cite{Malek:2017njj}, but here we give a number of new results. In particular, we show that the relevant $\SO(5-n)\subset\SO(5,5)\subset \Ex{6}$ generalised structure is fully specified by a set of $6+n$ generalised vectors on the internal manifold. We argue that if the algebra of generalised diffeomorphisms (that is, diffeomorphisms together with form-field gauge transformations) generated by these vectors closes with constant coefficients, then the generalised structure has singlet intrinsic torsion and the consistent truncation exists. The resulting five-dimensional half-maximal supergravity is coupled to $n$ vector multiplets, and its gauge algebra is the one generated by the $6+n$ generalised vectors. We give detailed formulae based on these vectors specifying the full bosonic truncation ansatz.  In particular, we provide an expression for the generalised metric on the internal manifold, which gives the complete scalar truncation ansatz. This is one of the main results of our work.

In Section~\ref{sec:IIBtrunc} we apply our formalism to consistent truncations of type IIB supergravity on five-dimensional manifolds preserving half-maximal supersymmetry (that is, 16 out of 32 supercharges). We first illustrate how the formalism works by reproducing the truncation of type IIB supergravity  on squashed Sasaki--Einstein manifolds derived in~\cite{Cassani:2010uw,Gauntlett:2010vu}. This is half-maximal supergravity coupled to two vector multiplets and with a $\U(1)\times {\rm Heis}_3$ gauging, where ${\rm Heis}_3$ denotes the Heisenberg group. Then we argue that when the Sasaki--Einstein manifold is toric, the exact same truncated theory is also obtained by deforming the internal geometry via the TsT transformation of~\cite{Lunin:2005jy} with parameter $\beta$. Another way to say this is that we TsT-transform the full truncation ansatz, rather than just the AdS solution. We thus obtain a continuous family of uplifts of the $\U(1)\times {\rm Heis}_3$ gauged five-dimensional supergravity, parameterised by~$\beta$. At the technical level, this is shown by exploiting the fact that the TsT transformation has a simple action in generalised geometry via a bi-vector field. It was recently shown in the $S^5$ case that such backgrounds admitted a truncation to minimal gauged supergravity (8 supercharges)~\cite{Liu:2019cea}. Our result shows that they in fact admit a much larger truncation to half-maximal supergravity with two vector multiplets.

 In Section \ref{sec:MNsection} we derive a consistent truncation of eleven-dimensional supergravity on Maldacena--N\'u\~nez geometries where $S^4$ fibers over a Riemann surface \cite{Maldacena:2000mw}, leading to half-maximal supergravity coupled to three vector multiplets and with a $\U(1)\times \mathit{ISO}(3)$ gauge algebra. We note that the existence of such a consistent truncation, as well as an analysis of its sub-truncations and vacua, was very recently proven using a different approach, considering the explicit truncation directly from seven-dimensional maximal gauged supergravity~\cite{Cheung:2019pge}. We conclude in Section~\ref{sec:Conclusions} outlining some directions of future research including some more consistent truncations that it would be interesting to explore using our approach.


\section{Consistent truncations from $G$-structures}
\label{sec:gen-formalism}

\subsection{Conventional $G$-structure constructions}
\label{sec:conv-gen-form}

Before turning to the generalised geometry picture, let us review the role of conventional $G$-structures in consistent truncations. Through the study of several cases such as~\cite{Gauntlett:2009zw,KashaniPoor:2007tr,Cassani:2009ck,Cassani:2010uw,Gauntlett:2010vu,Cassani:2011fu,Cassani:2012pj}, it is now understood that any $G$-structure with constant, singlet intrinsic torsion leads to a consistent truncation.

The idea is as follows. In conventional Scherk--Schwarz reductions on a group manifold $M=\mathcal{G}$ all the higher-dimensional fields in the theory can be decomposed into representations of $\mathcal{G}$. By keeping all the singlet representations and nothing else, one ensures that the truncation is consistent, since products of singlet representations can never source the non-singlet representations that were truncated away. However, this argument extends: the key point is not that the manifold has isometries but that the structure group $\Gst$ is reduced, since this allows one to decompose all tensor fields into $\Gst$ representations and then keep only those fields transforming as singlets. In the case of a group manifold the structure group is trivial since the manifold is parallelisable, but more generally one can consider cases with larger structure groups.\footnote{The same symmetry argument used for Scherk--Schwarz reductions implies that dimensional reductions on coset manifolds $M=\mathcal{G}/\mathcal{H}$ keeping all $\mathcal{G}$-invariant Kaluza--Klein modes and nothing else are consistent. In this case, there is a nice connection with the other argument given above, based on the $G$-structure of $M$. Indeed one can show that if $\mathcal{H}$ contains no nontrivial invariant subgroup of $\mathcal{G}$, then $\mathcal{G}/\mathcal{H}$ admits a $\mathcal{G}$-invariant $\mathcal{H}$-structure (see e.g.~\cite[App.$\:$A]{Koerber:2008rx}). The $\mathcal{G}$-invariant truncation and the truncation based on singlets of the $\mathcal{H}$-structure then coincide.} Explicitly, one has 
\begin{thm}
   \label{th:standard}
Let $M$ be a $d$-dimensional manifold with a $\Gst$-structure defining a set of invariant tensors $\{\Xi_i\}$ with $\Gst\subset \OO(n)$ and only constant, singlet intrinsic torsion. Any field theory has a consistent truncation on $M$ defined by expanding all fields in terms of the invariant tensors.
\end{thm}
\noindent
If the theory includes spinors, then the $\Gst$-structure lifts to a  $\tGst\subset\Spin(d)$ structure and we can include fermions in the truncation by expanding any spinor fields in terms of spinors invariant under $\tGst$. 

To explain this in a little more detail, first recall that a choice of $\Gst$-structure on a $d$-dimensional manifold $M$ is a reduction of the structure group. Formally, a $\Gst$-structure defines a $\Gst$-principal sub-bundle $P$ of the $\GL(d,\bbR)$ frame bundle. In most cases, the structure can equivalently be defined by a set of $\Gst$-invariant, nowhere vanishing tensors $\{\Xi_i\}$. The existence of a $\Gst$-structure means that all tensor fields can be decomposed into irreducible representations of $\Gst$. For example, a choice of $\Gst=\OO(d)$ structure defines a subset of orthonormal frames, or equivalently is defined by an invariant metric tensor $g$. A given $\Gst$-structure $P$ is characterised by its intrinsic torsion. If $\Gst\subset O(d)$, this is defined in the following simple way (see for example~\cite{Salamon}). Since $\Gst\subset O(d)$ the structure defines a metric $g$ and hence a corresponding Levi--Civita connection $\LC$. Acting on each invariant tensor $\Xi_i$ we have
\begin{equation}
\label{eq:nabla-Xi}
\begin{aligned}
   \LC_m {\Xi_i}^{n_1\dots n_r}{}_{p_1\dots p_s}
   &= K_m{}^{n_1}{}_q {\Xi_i}^{q\dots n_r}{}_{p_1\dots p_s} + \dots
   + K_m{}^{n_r}{}_q {\Xi_i}^{n_1\dots q}{}_{p_1\dots p_s} \\ & \qquad
   - K_m{}^q{}_{p_1} {\Xi_i}^{n_1\dots n_r}{}_{q\dots p_s} + \dots 
   - K_m{}^q{}_{p_s} {\Xi_i}^{n_1\dots n_r}{}_{p_1\dots q} ,
\end{aligned}
\end{equation}
which uniquely defines $K_m{}^n{}_p$ as a section of $T^*M\otimes \gperp$ with $m$ and $n,p$ denoting the $T^*M$ and $\gperp$ indices respectively. Here we have decomposed $\Lambda^2T^*M\simeq \so(d)=\mathfrak{g}\oplus\gperp$ with $\mathfrak{g}$ the Lie algebra of $\Gst$. Note that the $T^*M\otimes \mathfrak{g}$ part is missing in $K$ because, by definition $\Xi_i$ is $\Gst$-invariant. The tensor $K$ defines the intrinsic torsion $(\Tint)_{mn}{}^p=K_n{}^p{}_m-K_m{}^p{}_n$. Note that equivalently one can define a new torsionful connection $\tilde{\nabla}=\nabla-K$ that is compatible with the structure, that is $\tilde{\nabla}\Xi_i=0$ for all $\Xi_i$. The intrinsic torsion $\Tint$ is then the torsion of $\tilde{\nabla}$. In general $\Tint$ will decompose into $\Gst$ representations, known as the ``torsion classes'' of the $\Gst$-structure. Note that in many examples, the invariant tensors $\Xi_i$ are all differential forms and the intrinsic torsion is completely determined by the exterior derivatives $\dd\Xi_i$. 
 
As for reduction on group manifolds, the proof of theorem~\ref{th:standard} is very straightforward. By expanding in terms of invariant tensors, all the fields one keeps transform as singlets under the structure group, with the only dependence on the internal space coming from the $\{\Xi_i\}$. Furthermore since the intrinsic torsion has only singlet components (and is independent of the internal space) any derivative of a field is given by the right-hand side of~\eqref{eq:nabla-Xi} and is itself an expansion in terms of singlets. So long as we keep all possible singlets and nothing else, given the equations of motion can be written as generalised tensors, the truncation is then necessarily consistent, since products of singlet representations can never source the non-singlet representations that were truncated away.

Focusing on  the gravity sector, the scalars and vector fields in the consistent truncation appear in the following way. Recall that the choice of metric parameterises a $\GL(d,\bbR)/\OO(d)$ coset. To count the number of $\Gst$ singlets in the metric we can use the commutant of $\Gst$ in $\GL(d,\bbR)$ and $\OO(d)$. The scalars in the consistent truncation coming from the metric thus parameterise 
\begin{equation}
   \text{metric scalars}  \quad \Leftrightarrow \quad  H \in  \frac{\Com{\Gst}{\GL(d,\bbR)}}{\Com{\Gst}{\OO(d)}} \, ,
\end{equation}
where $\Com{A}{K}$ denotes the commutant of $A\subset K$ inside $K$. We can also count the number of vectors coming from the metric, by counting the number of invariant one-forms $\eta^a\in\{\Xi_i\}$, giving 
\begin{equation}
   \text{metric gauge fields}  \quad \Leftrightarrow \quad   \mathcal{A}^a\, \hat{\eta}_a ,
\end{equation}
where $\hat{\eta}_a$ are the dual singlet vectors. For singlet torsion, the torsion is completely determined by the Lie derivatives of the invariant tensors
\begin{equation}
   \label{eq:LXi}
   \mathcal{L}_{\hat{\eta}_a}\Xi_i = f_{ai}{}^j\,  \Xi_j \, , 
\end{equation}
where $f_{ai}{}^j$ are constants, fixed by the intrinsic torsion. For example, the gauging of the truncated theory depends on the Lie bracket
\begin{equation}
   \label{eq:Lie-alg}
   \BLie{\hat{\eta}_a}{\hat{\eta}_b} = f_{ab}{}^c \,\hat{\eta}_c\, , 
\end{equation}
and we see that the singlet intrinsic torsion determines the gauge algebra of the metric gauge fields.

To see how the construction works in practice consider the reduction on a Sasaki--Einstein manifold $M$ of dimension $d=2n+1$, which appeared in the context of reductions of M-theory and type IIB in~\cite{Gauntlett:2009zw} and~\cite{Cassani:2010uw,Gauntlett:2010vu} respectively. The invariant tensors $(\eta,\omega,\Omega)$, where $\eta$ is a real one-form, $\omega$ a real two-form and $\Omega$ a complex $n$-form on $M$, define an $\Gst=\SU(n)\subset\GL(d,\bbR)$ structure and satisfy
\begin{equation}
\label{eq:SE}
   \dd \eta = 2 \omega , \qquad
   \dd \Omega = \ii\,(n+1)\, \eta \wedge \Omega \,, 
\end{equation}
implying we indeed have constant singlet torsion, since only invariant tensors appear on the right-hand sides of these equations. In this case the metric scalar manifold is 
\begin{equation}
\frac{\Com{\SU(n)}{\GL(2n+1,\bbR)}}{\Com{\SU(n)}{\SO(2n+1)}} =  \frac{\bbR^+\times\bbC}{\U(1)} = \bbR^+ \times \bbR^+ , 
\end{equation}
where the first $\bbR^+$ comes from $\Com{\GL(2n)}{\GL(2n+1)}$ and $\bbC$ from $\Com{\SU(n)}{\GL(2n)}$. There is a single invariant one-form $\eta$ and so there will be a single gauge field $\mathcal{A}_\mu(x)$ coming from the metric. Concretely the consistent truncation on $M$ is defined by
\begin{equation}
   \dd s^2 = g_{\mu\nu}\dd x^\mu\dd x^\nu 
         + \rme^{2U} \dd s^2_{2n} + \rme^{2V} (\eta + \mathcal{A})^2 ,
\end{equation}
where $\dd s^2_{2n}$ is the (local) $2n$-dimensional K\"ahler--Einstein metric defined by $(\omega,\Omega)$. The scalars fields $U(x)$ and $V(x)$ parametrise the scalar manifold $H$.

The Scherk--Schwarz reduction $M=\mathcal{G}$ is of course itself also an example. The group structure picks out a preferred co-frame $\{e^a\}\in T^*M$ of (say) left-invariant one-forms. Geometrically the one-forms define an ``identity structure''  $\Gst=\id\supset \GL(d)$ (or parallelisation). Since $\Com{\id}{K}=K$, the scalar fields are in the coset 
\begin{equation}
\frac{\Com{\id}{\GL(d)}}{\Com{\id}{\SO(d)}} = \frac{\GL(d,\bbR)}{\SO(d)} \, . 
\end{equation}
The one-forms define $d$ gauge fields with a Lie algebra given by the Lie bracket~\eqref{eq:Lie-alg}. The consistent truncation ansatz for the metric is 
\begin{equation}
   \dd s^2 = g_{\mu\nu}\dd x^\mu\dd x^\nu 
        + h_{ab} \big(e^a + \mathcal{A}^a\big) \big(e^b + \mathcal{A}^b \big) \,,
\end{equation}
where $h_{ab}(x)$ is matrix of scalar fields and the $\mathcal{A}^a_\mu(x)$ are gauge fields in the adjoint of $\Gst$. 

Any number of other examples can be constructed. We note that the standard consistent truncation keeping a volume modulus on an orientable manifold can be thought of arising from the corresponding $\SL(n,\bbC)$-structure. Similarly the universal sector of type~II Calabi--Yau compactifications arises from keeping $\SU(n)$-singlet fields in the metric and form-field degrees of freedom.

\subsection{Generalised $G$-structure constructions}
\label{sec:gen-geo-const}

We can now extend this picture to generalised geometry to describe the consistent truncations of eleven-dimensional and type II supergravities on $d$- and $(d-1)$-dimensional manifolds $M$ respectively. The generalisation is straightforward: we replace the conventional $\Gst$-structures with generalised $G$-structure on the generalised tangent space $E$ associated to $M$. The generic structure group on $E$ is the exceptional group $\Edd$ which has a maximal compact R-symmetry subgroup $\Hd$ (see table~\ref{tab:groups}). If a $\Gst\subset\Hd$ structure is defined by a set of generalised invariant tensors the idea is then to  expand the supergravity fields in terms of the tensors used to define the consistent truncation. This is a generalisation of the construction given in~\cite{Lee:2014mla}, where it was shown that maximally supersymmetric consistent truncations corresponded to ``Leibniz parallelisations'', that is, identity structures $\Gst=\id$ with constant intrinsic torsion.

\subsubsection{Main theorem}
\label{sec:theorem}

Let us start by stating the result and then discuss more details of the generalised geometry and the proof of the statement. We claim
\begin{thm}
\label{thm:gen}
   Let $M$ be a $d$-dimensional (respectively $(d-1)$-dimensional) manifold with a generalised $\Gst$-structure defining a set of invariant tensors $\{Q_i\}$ with $\Gst\subset\Hd$ and only constant, singlet intrinsic torsion. Then there is a consistent truncation of eleven-dimensional (respectively type II) supergravity on $M$ defined by expanding all bosonic fields in terms of the invariant tensors. If $\dHd$ is the double cover of $\Hd$, acting on fermions the structure group lifts to $\tGst\subset\dHd$ and the truncation extends to the fermionic sector, provided again one expands the spinor fermion fields in terms of $\tGst$ singlets.  
\end{thm}

To see how this works, we start by summarising the generalised geometry reformulation of eleven-dimensional or type II supergravity on a product space $X\times M$ where $X$ is a $D$-dimensional Lorentzian space, and the internal manifold $M$ is $d$-dimensional, or, in the case of type II supergravity, $(d-1)$-dimensional. In generalised geometry, the $\GL(d,\bbR)$ or $\GL(d-1,\bbR)$ structure group of conventional geometry on $M$ is extended to $\Edd$ for $d\leq 7$~\cite{Hull:2007zu,Pacheco:2008ps}. This allows one to reformulate supergravity, so that the bosonic supergravity fields and their equations of motion are rearranged into generalised tensors transforming as representations of $\GL(D,\bbR)\times\Edd$. The $\GL(D,\bbR)$ scalar degrees of freedom are repackaged into a generalised metric, that is a symmetric generalised tensor $G\in\Gs{S^2E^*}$ which is invariant under the R-symmetry subgroup $\Hd\subset\Edd$. Thus geometrically the generalised metric defines an $\Hd$-structure~\cite{Coimbra:2011ky,Coimbra:2012af}. The $\GL(D,\bbR)$ one-form, vector degrees of freedom are sections of the generalised tangent space $E$, while the two-form tensor degrees of freedom are sections of a generalised tensor bundle here denoted $N$~\cite{Hohm:2013pua,Hohm:2013vpa,Hohm:2013uia,Abzalov:2015ega,Musaev:2015ces}. In summary we have
\begin{equation}
\begin{aligned}
   \label{eq:sugra-fields}
   \text{scalars:} &&& & G_{MN}(x,y) &\in \Gs{S^2 E^*}\, ,\\[1mm]
   \text{vectors:} &&& & \mathcal{A}_\mu{}^M(x,y) &\in \Gs{T^*X\otimes E}\, ,\\[1mm]
   \text{two-forms:} &&& & \mathcal{B}_{\mu\nu}{}^{MN}(x,y) &\in \Gs{\Lambda^2T^*X\otimes N}\, ,
\end{aligned}
\end{equation}
where $x$ and $y$ are coordinates on $X$ and $M$ respectively, the index $M$ denotes components of vectors in $E$ (or $E^*$ if lowered) and we are using the fact that $N\subset S^2 E$. One can also further introduce higher form-field degrees of freedom following the tensor hierarchy~\cite{deWit:2008ta,deWit:2005hv}. However, these do not introduce new degrees of freedom but are dual to the scalar, vector and two-forms\footnote{Note that for $D=4$ this means the $\mathcal{A}_\mu{}^M$ contain both the vectors and their duals, and in $D=6$ the $\mathcal{B}_{\mu\nu}{}^{MN}$ contain both the two-forms and their duals.}. The relevant groups and $\Edd$ representations are all listed in table~\ref{tab:groups}. Note that $\dHd$ is actually the double cover of $\Hd$. The dynamics of the supergravity is completely determined by the Levi--Civita connection on the external space and a generalised connection $\Dgen$ on the internal space. The latter is the generalised analogue of the Levi--Civita connection: it has vanishing generalised torsion and is compatible with the generalised metric. We also include in table~\ref{tab:groups} the $\Edd$ representation of the generalised tensor bundle $W$ in which the generalised torsion lies and the $\dHd$ representation of the spinor bundle $\mathcal{S}$ in which the supersymmetry parameter lies~\cite{Coimbra:2012af}.   
\begin{table}[!th]
\centering
\setlength{\tabcolsep}{0.45em}
\begin{tabular}{@{}llllll@{}}
  \toprule
  $\Edd$ & $E$ & $N$ & $W$ & $\dHd$ & $\mathcal{S} $ \\
  \midrule
  $\Ex{7}$ & $\rep{56}$ & $\rep{133}$ & $\rep{912}\oplus\rep{56}$ &
  $\SU(8)$ & $\rep{8}\oplus\rep{\bar{8}}$ \\
  $\Ex{6}$ & $\rep{27}$ & $\rep{27}'$ & $\rep{351}\oplus\rep{27}'$ &
  $\USp(8)$ & $\rep{8}$ \\
  $\Spin(5,5)$ & $\rep{16}^s$ & $\rep{10}$ 
     & $\rep{144}^s\oplus\rep{16}^c$ &
  $\USp(4)\times\USp(4)$ & $\repp{4}{1}\oplus\repp{1}{4}$ \\
  $\SL(5,\bbR)$ & $\rep{10}$ & $\rep{5}'$ &
     $\rep{40}\oplus\rep{15}'\oplus\rep{10}'$ &
  $\USp(4)$ & $\rep{4}$ \\
  \bottomrule
\end{tabular}
\caption{Generalised geometry groups, bundles and representations.}\label{tab:groups}
\end{table}

Now suppose we have a reduced structure group $\Gst\subset\Hd$ defined by a set of $\Gst$-invariant generalised tensors $\{Q_i\}$. As described in~\cite{Coimbra:2014uxa}, one can again define an intrinsic torsion $\Tint$ for the generalised $\Gst$-structure, and decompose it into representations of $\Gst$. The definition is as follows. Let $\tilde{D}$ be a generalised connection compatible with the $\Gst$-structure, that is, sastisfying $\tilde{D}Q_i=0$ for all $Q_i$. Formally, the generalised torsion $T$ of $\tilde{D}$ is defined by, acting on any generalised tensor $\alpha$, 
\begin{equation}
   \label{eq:gen-torsion}
   \big(\Lgen^{\tilde{D}}_V - \Lgen_V\big)\, \alpha = T(V) \cdot \alpha
\end{equation}
where $\Lgen$ is the generalised Lie derivative, $\Lgen^{\tilde{D}}$ is the generalised Lie derivative calculated using $\tilde{D}$ and we view the torsion as a map $T:\Gs{E}\to\Gs{\adj \tilde{F}}$ where $\adj\tilde{F}$ is the $\Edd\times\bbR^+$ adjoint bundle, so that $T(V)$ acts via the adjoint action on $\alpha$. The intrinsic torsion is then the component of $T$ that is independent of the choice of compatible connection $\tilde{D}$. We are interested in the case where only singlet representations appear in the intrinsic torsion. This means we can define a generalised Levi-Civita connection such that, in analogy with~\eqref{eq:nabla-Xi}, acting on any invariant generalised tensor $Q_i$,
\begin{equation}
   \label{eq:DQ}
   D_M Q_i = \Sigma_M \cdot Q_i
\end{equation}
where $\Sigma_M$ is a section of $E^*\otimes\adj P_{\Hd}$ that is completely determined in terms of the singlet torsion\footnote{Note there is a subtlety that the connection $D$ is not uniquely determined by the conditions of compatibility with the generalised metric and being torsion-free. However only certain projections of the action of $D$ appear in the supergravity and these are unique~\cite{Coimbra:2011ky}. In equation~\eqref{eq:DQ}, we are choosing a particular torsion-free compatible $D$. Equivalently, one can show that the unique projected operators, acting on $Q_i$, are completely determined by the singlet intrinsic torsion.}. Here we are using a notion where $\adj P_{\Hd}$ is the bundle of tensors transforming on the adjoint representation of $\Hd$.

The proof of consistency is just as before. By expanding in terms of invariant tensors, all the fields one keeps transform as singlets under the structure group, with the only dependence on the internal space coming from the $\{Q_i\}$. Furthermore from~\eqref{eq:DQ} the derivatives of all the truncated fields also have expansions in terms of singlets. So long as we keep all possible singlets and nothing else, the truncation is then necessarily consistent, since products of singlet representations can never source the non-singlet representations that were truncated away. 

\subsubsection{Structure of the truncated theory}
\label{sec:struc-trunc}

So far we have made a general argument that a $\Gst$-structure with singlet intrinsic torsion will lead to a consistent truncation of eleven-dimensional or type II supergravity. However, one can go further and deduce the structure of the truncated theory from the $\Gst$-structure and the torsion. We will find that in all cases, even when there is no preserved supersymmetry, it is described by a version of the embedding tensor formalism (see e.g.~\cite{Samtleben:2008pe,Trigiante:2016mnt} for a review of this formalism).

We start by identifying the $\Gst$-singlet truncated degrees of freedom. Since $\Gst\subset\Hd$ the structure encodes the generalised metric $G_{MN}$. In the truncation we want to keep singlet deformations of the structure, modulo those singlet deformations that do not deform the metric. At each point in $M$ the metric is an element of the coset $\Edd/\Hd$, thus we can generate the singlet deformations of the metric by acting on the structure by elements of $\Edd$ that commute with $\Gst$ modulo elements of $\Hd$ that commute with $\Gst$, since the latter will not change the metric. Thus we find the scalars parametrise the coset  
\begin{equation}
   \text{scalars:} \qquad h^{I}(x) \in
   \Mscal = \frac{\Com{\Gst}{\Edd}}{\Com{\Gst}{\Hd}}
      :=\frac{\mathcal{G}}{\mathcal{H}}\,.
\end{equation}
Recall that the vector fields are sections of $T^*X\otimes E$. If $\{K_\mathcal{A}\}$ is a basis for the $\Gst$-invariant generalised vectors, spanning a vector space $\mathcal{V}\subset\Gs{E}$, then we have
\begin{equation}
\label{vectors}
   \text{vectors:} \qquad \mathcal{A}_\mu{}^\mathcal{A}(x)\, K_\mathcal{A} \,\in\, \Gs{T^*M}\otimes \mathcal{V} . 
\end{equation}
If $\{J_\Sigma\}$ is a basis generating the $\Gst$-invariant vector space $\mathcal{B}\subset\Gs{N}$, we similarly have the two-form degrees of freedom 
\begin{equation}
\label{two-forms}
   \text{two-forms:} \qquad \mathcal{B}_{\mu\nu}{}^\Sigma(x) J_\Sigma  
      \in \Gs{\Lambda^2T^*X}\otimes\mathcal{B} .
\end{equation}
Note that by definition $\mathcal{V}$ and $\mathcal{B}$ are both representation spaces for the action of the commutant group $\mathcal{G}$. Note we also have $N\subset S^2E$ and so we can use the projection map $\times_N$ and embedding to define the constants $d_{\mathcal{A}\mathcal{B}}{}^\Sigma$ and $\tilde{d}_{\Sigma}{}^{\mathcal{A}\mathcal{B}}$
\begin{equation}
   \label{eq:const-d}
   K_\mathcal{A} \times_N K_\mathcal{B}
      = d_{\mathcal{A}\mathcal{B}}{}^\Sigma J_\Sigma , \qquad
   J_\Sigma = \tilde{d}_{\Sigma}{}^{\mathcal{A}\mathcal{B}}
       K_\mathcal{A}\otimes K_\mathcal{B}\, .
\end{equation}
intertwining the representation spaces.

Turning to the singlet intrinsic torsion, we note that, since $\tilde{D}K_{\mathcal{A}}=0$, in analogy with~\eqref{eq:LXi}, we have 
\begin{equation}
   \label{eq:LQ}
   \Lgen_{K_\mathcal{A}}Q_i = - \Tint(K_\mathcal{A}) \cdot Q_i  \,,
\end{equation}
where we recall that $L$ is the generalised Lie derivative.
Since $\Tint$ is a singlet, then $\Tint(K_{\mathcal{A}})$ must be a singlet of $\adj\tilde{F}$, but such singlets are precisely the Lie algebra of the commutant group $\mathcal{G}=\Com{\Gst}{\Edd}$. Thus $-\Tint$ defines an ``embedding tensor''~\cite{Samtleben:2008pe,Trigiante:2016mnt}, that is a linear map
\begin{equation}
   \label{eq:Theta}
   \Theta : \mathcal{V} \to \Lie\mathcal{G} \,.
\end{equation}
Acting on the $K_\mathcal{A}$, we get
\begin{equation}
\label{eq:K-alg}
   \Lgen_{K_\mathcal{A}}K_\mathcal{B}
   = \Theta_\mathcal{A}\cdot K_\mathcal{B}
   = \Theta_\mathcal{A}{}^{\hat{\alpha}}(t_{\hat{\alpha}})_\mathcal{B}
   {}^\mathcal{C}K_\mathcal{C}
   := X_{\mathcal{A}\mathcal{B}}{}^\mathcal{C} K_\mathcal{C}\, , 
\end{equation}
where $(t_{\hat{\alpha}})_\mathcal{B}{}^\mathcal{C}$ are the representations of the generators of $\Lie\mathcal{G}$ acting on $\mathcal{V}$. The Leibniz property of the generalised Lie derivative then implies~\cite{Coimbra:2011ky,Lee:2014mla} the standard quadratic condition on the embedding tensor
\begin{equation}
   \comm{X_\mathcal{A}}{X_\mathcal{B}}
       =  - X_{\mathcal{A}\mathcal{B}}{}^\mathcal{C} X_\mathcal{C}\, , 
\end{equation}
where we are viewing $(X_\mathcal{A})_\mathcal{B}{}^\mathcal{C}=X_{\mathcal{A}\mathcal{B}}{}^\mathcal{C}$ as a matrix. Thus we can view the $K_\mathcal{A}$ as generating a Lie algebra with structure constants $X_{[\mathcal{A}\mathcal{B}]}{}^\mathcal{C}$. Since the image of $\Theta$ may not be the whole of $\Lie\mathcal{G}$, we see that the vector fields describe a gauge group
\begin{equation}
   \label{eq:gauge}
   \text{gauge group:} \qquad \Ggauge\subseteq \mathcal{G} \,,
\end{equation}
where $\Lie\Ggauge=\im \mathcal{V}\subseteq\Lie\mathcal{G}$ under the embedding tensor map $\Theta$. The $X_\mathcal{A}$ then define the adjoint representation and $\Theta$ defines how the gauge action embeds as an action in $\mathcal{G}$. 

By reducing the generalised geometry/EFT reformulation of supergravity of~\cite{Coimbra:2011ky,Coimbra:2012af,Hohm:2013pua,Hohm:2013vpa,Hohm:2013uia,Abzalov:2015ega,Musaev:2015ces}, we can then summarise the structure and gauging of the truncated theory, which match the standard formulae for gauging of a tensor hierarchy via an embedding tensor~\cite{Samtleben:2008pe,Trigiante:2016mnt}:
\begin{itemize}
\item 
\emph{The fields in the truncated theory are as follows 
   \begin{equation}
   \label{eq:fields}
   \begin{aligned}
      \text{scalars:} & &&& &
      h^{I}(x) \in \Mscal
         = \frac{\Com{\Gst}{\Edd}}{\Com{\Gst}{\Hd}}
         :=\frac{\mathcal{G}}{\mathcal{H}} \,, \\[1mm]
      \text{vectors:} & &&& &
         \mathcal{A}_\mu^\mathcal{A}(x)\, K_\mathcal{A}
          \quad \in \Gs{T^*X}\otimes \mathcal{V}\,, \\[1mm]
       \text{two-forms:} & &&& &
          \mathcal{B}_{\mu\nu}^\Sigma(x)\, J_\Sigma
         \quad  \in \Gs{\Lambda^2T^*X}\otimes\mathcal{B}\,.
   \end{aligned}
\end{equation} }
\item 
\emph{The theory is gauged by $\Ggauge\subseteq\mathcal{G}$ with the scalar covariant derivatives 
   \begin{equation}
      \label{eq:gauging}
      \hat{D}_\mu h^{I}
      = \der_\mu h^{I}
           - \mathcal{A}^{\mathcal{A}}_\mu \,\Theta_{\mathcal{A}}{}^{\hat{\alpha}} k_{\hat{\alpha}}{}^{I}\,,
        \end{equation}
        where $k_{\hat{\alpha}}$ are the Killing vectors on $\Mscal$ generating the action of the $\Lie\mathcal{G}$\,.}
\item
\emph{The gauge transformations of the vectors and two-forms are 
   \begin{equation}
      \begin{aligned}
         \delta \mathcal{A}^\mathcal{A}_\mu
         &= \der_\mu \Lambda^\mathcal{A}
         + X_{\mathcal{B}\mathcal{C}}{}^\mathcal{A}\big(
         \mathcal{A}^\mathcal{B}_\mu\, \Lambda^\mathcal{C}
         - \Xi_\mu^{\mathcal{B}\mathcal{C}}
         \big)\,, \\[1mm]
         \delta \mathcal{B}^\Sigma_{\mu\nu}
         &= 2d_{\mathcal{A}\mathcal{B}}{}^\Sigma\Big(
         \der_{[\mu} \Xi_{\nu]}^{\mathcal{A}\mathcal{B}}
         + 2X_{\mathcal{C}\mathcal{D}}{}^\mathcal{A}
         \mathcal{A}^\mathcal{C}_{[\mu}\, \Xi_{\nu]}^{\mathcal{D}\mathcal{B}}
         - \Lambda^\mathcal{A}\mathcal{H}_{\mu\nu}^\mathcal{B}
         - \mathcal{A}^\mathcal{A}_{[\mu}\,\delta \mathcal{A}^\mathcal{B}_{\nu]} 
         \Big)\,, 
      \end{aligned}
   \end{equation}
where $\,\Xi^{\mathcal{A}\mathcal{B}}_\mu=\Xi_\mu{}^\Sigma \tilde{d}_\Sigma{}^{\mathcal{A}\mathcal{B}}$ and $\,\mathcal{H}^{\mathcal{A}}=\dd \mathcal{A}^\mathcal{A}
+X_{\mathcal{B}\mathcal{C}}{}^\mathcal{A}(\mathcal{A}^\mathcal{B}\wedge \mathcal{A}^\mathcal{C}
+ \mathcal{B}^\Sigma\tilde{d}_\Sigma{}^{\mathcal{B}\mathcal{C}})\,$.}  
\item
\emph{Given a lift $\tGst\subseteq\dHd$, the number of supersymmetries preserved by the truncated theory is given by the number of $\tGst$-singlets in the generalised spinor bundle~$\mathcal{S}$.}
\end{itemize}
\noindent

The key point here is that the geometrical data of the $\Gst$-structure and its singlet intrinsic torsion completely determine the truncated theory. The precise relationship between  these expressions and the uplifted supergravity fields depends on the normalisations of the basis vectors $K_\mathcal{A}$ and $J_\Sigma$ and the explicit expression for the generalised metric $G_{MN}$ in terms of the relevant normalised invariant tensors. We will turn to the details of these relationships in the explicit example of half-maximal truncations in the following sections.

\subsection{Maximal structure groups and pure supergravities}
\label{sec:max-min}

To see how the truncated theories arise for some specific structure groups and match known consistent truncations, in this and the next sub-section let us focus on truncations preserving a given amount of supersymmetry in $D=11-d$ dimensions. For $\mathcal{N}$ supersymmetries the generalised spinor bundle $\mathcal{S}$ must have $\mathcal{N}$ singlets when decomposed under the structure group $\Gst\subset\dHd$\footnote{Note that here and in the following subsection we will ignore discrete factors in the structure group and hence ignore the possible distinction between $\Gst$ and $\tGst$.}. Let $\GN$ be the \emph{maximal} subgroup of $\dHd$ for which this is true, that is the largest possible generalised structure group that preserves $\mathcal{N}$ supersymmetries. These groups are listed in table~\ref{tab:max-structure}.
\begin{table}[htb]
\centering
\setlength{\tabcolsep}{0.45em}
\begin{tabular}{@{}lll@{}}
  \toprule
   $\dHd$ &  $\GN$  \\ 
   \midrule
   $\SU(8)$ &  $\SU(8-\mathcal{N})$\\
   $\USp(8)$ & $\USp(8-2\mathcal{N})$ \\
   $\USp(4) \times \USp(4)$ &   
      $\USp(4-2\mathcal{N}_+) \times \USp(4-2\mathcal{N}_-)$ \\
   $\USp(4)$ & $\USp(4-2\mathcal{N}) $ \\
  \bottomrule    
\end{tabular}
\caption{Maximal generalised structure subgroups $\GN\subset \dHd$ preserving $\mathcal{N}$ supersymmetries in the truncated theory. Note that for $d = 5$ we have six-dimensional supergravity with $(\mathcal{N}_+,\mathcal{N}_-)$ supersymmetry.}
\label{tab:max-structure}
\end{table}

We can then use our formalism to determine the corresponding consistent truncations. In each case we need to find the commutant groups $\mathcal{G}$ and $\mathcal{H}$ and the spaces of vector and tensor multiplets. Both are fixed once one knows the embedding $\GN\subset\Edd$. The results are summarised in table~\ref{tab:commutants}. For the vector and two-form degrees of freedom we include only the minimum dynamical set. In particular, in $D=4$ and $D=5$, the two-forms are dual to scalars and vectors respectively, and so are not listed. For vectors in $D=4$ and two-forms in $D=6$ we include both the fields and their duals. In $D=6$ the self- and anti-self-dual two-forms are distinguished by their transformation under the two R-symmetry groups. Comparing with the standard literature (see for example the review in~\cite{Tanii:1998px}) we see that these theories are in one-to-one correspondence with the possible pure supergravity theories. This includes, in particular, the maximally supersymmetric cases of the sphere reductions. In each case, the gauging of the theory will depend on the singlet torsion, as described for the sphere cases in~\cite{Lee:2014mla}. 
\begin{table}[!th]
\centering
\setlength{\tabcolsep}{0.45em}
\begin{tabular}{@{}llllll@{}}
  \toprule
  $\Edd$ & $\mathcal{N}$ & $\mathcal{G}=\Com{\Gst}{\Edd}$
  & $\mathcal{H}=\Com{\Gst}{\dHd}$ & $\mathcal{V}$ & $\mathcal{B}$ \\
  \midrule
  $\Ex{7}$ & $1$ & $U(1)$ & $U(1)$ & -- & \\
  & $2$ & $\SU(2)\stimes U(1)$ & $\SU(2)\stimes U(1)$
  & $\rep{1}\splus\rep{1}$ & \\
  & $3$ & $\SU(3)\stimes U(1)$ & $\SU(3)\stimes U(1)$
  & $\rep{3}\splus\bar{\rep{3}}$ \\
  & $4$ & $\SU(4)\stimes\SL(2,\bbR)$ & $\SU(4)\stimes U(1)$ 
  & $\repp{6}{2}$ \\
  & $5$ & $\SU(5,1)$ & $\SU(5)\stimes U(1)$
  & $\rep{20}$ \\
  & $6$ & $\SO^*(12)$ & $\SU(6)\stimes U(1)$
  & $\rep{32}$ \\
  & $8$ & $E_{7(7)}$ & $\SU(8)$ & $\rep{56}$ \\
  \midrule
  $\Ex{6}$ & $1$ & $\USp(2)$ &  $\USp(2)$ & $\rep{1}$ \\
  & $2$ &  $\USp(4)\stimes\bbR^+$ & $\USp(4)$ & $\rep{5}+\rep{1}$ \\
  & $3$ & $\SU^*(6)$ & $\USp(6)$ & $\rep{15}$ \\
  & $4$ & $E_{6(6)}$ & $\USp(8)$ & $\rep{27}$ \\
  \midrule
  $\Spin(5,5)$ & $(1,0)$ & $\USp(2)$ & $\USp(2)$
  & -- & $\rep{1}$ \\
  & $(1,1)$ & $\USp(2)\stimes\USp(2)\stimes\bbR^+$
  &  $\USp(2)\stimes\USp(2)$
  & $\repp{2}{2}$ & $2\!\cdot\!\repp{1}{1}$ \\
  & $(2,0)$ & $\USp(4)$ & $\USp(4)$
  & -- & $\rep{5}$ \\
  & $(2,1)$ & $\SU^*(4)\stimes\USp(2)$ & $\USp(4)\stimes\USp(2)$  
  & $\repp{4}{2}$ & $\repp{6}{1}$ \\
  & $(2,2)$ & $\Spin(5,5)$ & $\USp(4)\stimes\USp(4)$ 
  & $\rep{16}$ & $\rep{10}$ \\
  \midrule
  $\SL(5,\bbR)$ & $1$ & $\USp(2)\stimes\bbR^+$ & $\USp(2)$ 
  & $\rep{3}$ & $\rep{1}$ \\
  & $2$ & $\SL(5,\bbR)$ & $\USp(4)$ 
  & $\rep{10}$ & $\rep{5}$ \\
  \bottomrule
\end{tabular}
\caption{Commutant groups and $\mathcal{G}$-representations of vectors and two-forms for $\GN$-structure consistent truncations.}\label{tab:commutants}
\end{table}

From one perspective, this is not surprising -- the representation theory is the same as that giving each pure supergravity theory as truncation of the maximally supersymmetric one in that dimensions. However, this analysis does allow us to give a proof of the conjecture in~\cite{Gauntlett:2007ma} (see also~\cite{Duff:1985jd,Pope:1987ad}):
\begin{cor}
Any supergravity solution with a $D$-dimensional AdS (or Minkowski) factor preserving $\mathcal{N}$ supersymmetries, defines a consistent truncation to the corresponding pure supergravity theory.     
\end{cor}
\noindent
The proof follows from the analysis of supersymmetric background in~\cite{Coimbra:2014uxa,Coimbra:2016ydd,Coimbra:2017fqv}. There it was showed that solutions with AdS (or Minkowski) factors with $\mathcal{N}$ supersymmetries correspond to $\GN$ generalised structures with singlet torsion. The corollary then follows as a direct application of Theorem~\ref{thm:gen}. For the Minkowski space case, the intrinsic torsion vanishes and the truncated theory is ungauged. 

\subsection{Supersymmetric truncations from conventional $G$-structures}
\label{sec:conv-trunc}

The more interesting case is when the structure group $\Gst$ is a subgroup of $\GN$ but one still has the same number of supersymmetries, that is, the same number of $\Gst$-singlets in the generalised spin bundle $\mathcal{S}$, since this can allow for truncated theories with non-trivial matter content. A simple way to achieve this situation is to consider the case of a conventional $G$-structure that corresponds to the appropriate number of supersymmetries. This analysis will allow us to connect to a number of known consistent truncations, including cases that require considerable calculation to derive the structure of the truncated theory. 

For definiteness we consider the cases of truncations of eleven-dimensional and type IIB supergravity to $D=4$ or $D=5$ on manifolds with $G_2$, $\SU(3)$ or $\SU(2)$ conventional $G$-structures. Calculating the commutant groups and the representation of the space of vector fields $\mathcal{V}$ we find the structure of the truncated theory is the same, independent of whether it came from eleven-dimensional or type IIB supergravity. We list the relevant groups and representations in table~\ref{tab:conv}. Note that for $D=4$ we give both the vectors and their duals, forming doublets of the $\SL(2,\bbR)$ subgroup of $\mathcal{G}$. 
\begin{table}[!th]
\centering
\setlength{\tabcolsep}{0.45em}
\begin{tabular}{@{}llllll@{}}
  \toprule
  $\Edd$ & $\mathcal{N}$ & $\Gst$ & $\mathcal{G}=\Com{\Gst}{\dHd}$
  & $\mathcal{H}=\Com{\Gst}{\Edd}$ & $\mathcal{V}$ \\
  \midrule
  $\Ex{7}$ & $1$ & $G_2$ & $\SL(2,\bbR)$ & $U(1)$ & --  \\
  & $2$ & $\SU(3)$ & $\SU(2,1)\times \SL(2,\bbR) $
  & $\SU(2)\times U(1)^2$ & $2\cdot\repp{1}{2}$ \\
  & $4$ & $\SU(2)$ & $\SO(6,3)\times \SL(2,\bbR)$
  & $\SO(6)\times\SO(3)\times U(1)$ & $\repp{9}{2}$ \\
  \midrule
  $\Ex{6}$ & $1$ & $\SU(3)$ &  $\SU(2,1)$ 
  & $\SU(2)\times U(1)$ & $\rep{1}$ \\
  & $2$ & $\SU(2)$ &  $\SO(5,2)\times\bbR^+$ 
  & $\SO(5)\times\SO(2)$ & $\rep{7}\oplus\rep{1}$ \\
  \bottomrule
\end{tabular}
\caption{Commutant groups and $\mathcal{G}$-representations of the vector fields for consistent truncations using conventional $G$-structures.}\label{tab:conv}
\end{table}

In each case we can identify the multiplet structure of the truncated theory and match to known examples of truncations, as follows:

\paragraph{$G_2\subset\Ex{7}$ structure:}
This case only refers to eleven-dimensional supergravity. Singlet intrinsic torsion implies a \emph{weak $G_2$ manifold}. The $D=4$ truncated theory is $\mathcal{N}=1$ supergravity coupled to a single chiral multiplet
\begin{equation}
   \Mscal = \frac{\SL(2,\bbR)}{U(1)} , 
\end{equation}
and there are no vector multiplets, matching the truncation first derived in~\cite{Gauntlett:2009zw}.

\paragraph{$\SU(3)\subset\Ex{7}$ structure:}
The $D=4$ truncated theory is $\mathcal{N}=2$ supergravity coupled to a single hypermultiplet and a single vector multiplet, with the scalar manifolds
\begin{equation}
   \Mscal = \mathcal{M}_{\text{hyper}} \times \mathcal{M}_{\text{vector}}
   = \frac{\SL(2,\bbR)}{U(1)} \times
      \frac{\SU(2,1)}{\SU(2)\times U(1)} . 
\end{equation}
For eleven-dimensional supergravity this includes the case of consistent truncation on a \emph{Sasaki--Einstein seven-manifold} first derived in~\cite{Gauntlett:2009zw}. For type IIB, it includes the case of the universal sector of \emph{nearly K\"ahler reductions}, the analogue of the IIA case considered in~\cite{KashaniPoor:2007tr,Cassani:2009ck}.

\paragraph{$\SU(2)\subset\Ex{7}$ structure:}
The $D=4$ truncated theory is $\mathcal{N}=4$ (half-maximal) supergravity coupled to three vector multiplets, with scalar manifold
\begin{equation}
   \Mscal = \frac{\SL(2,\bbR)}{U(1)}
       \times \frac{\SO(6,3)}{\SO(6)\times\SO(3)}\, . 
\end{equation}
For eleven-dimensional supergravity this includes the case consistent truncation on a \emph{tri-Sasaki seven-manifold} first derived  in~\cite{Cassani:2011fu}. 

\paragraph{$\SU(3)\subset\Ex{6}$ structure:}
This case only refers to eleven-dimensional supergravity. The $D=5$ truncated theory is minimal supergravity coupled to a single hypermultiplet
\begin{equation}
   \Mscal = \frac{\SU(2,1)}{\SU(2)\times U(1)}\, ,
\end{equation}
and has only the graviphoton with no extra gauge fields. For the case of vanishing intrinsic torsion the theory is just the universal sector of eleven-dimensional supergravity compactified on a Calabi--Yau manifold. 

\paragraph{$\SU(2)\subset\Ex{6}$ structure:}
The $D=5$ truncated theory is half-maximal supergravity coupled to two vector multiplets, with the scalar manifolds
\begin{equation}
   \Mscal =\bbR^+ \times \frac{\SO(5,2)}{\SO(5)\times\SO(2)}\, . 
\end{equation}
For type IIB supergravity this includes the case of consistent truncation on a \emph{Sasaki--Einstein five-manifold} derived in~\cite{Cassani:2010uw,Gauntlett:2010vu}. We will analyse this case in considerable detail in Section~\ref{sec:IIBonSE}.

\medskip

In each of these cases the gauging of the theory will depend on the particular intrinsic torsion, via the embedding tensor $\Theta$ defined by~\eqref{eq:LQ}. Rather than work through the details in each case here we will focus in the following sections on the particular class of half-maximal $D=5$ truncations. This will in particular include the details of the Sasaki--Einstein five-manifold example. We will also go further and discuss more involved examples. Finally, we note that we could also have considered cases above where $\Gst$ is a subgroup of the conventional $\SU(3)$ or $\SU(2)$ structure groups such that we still have the same amount of supersymmetry. These would be relevant for example, to the consistent truncation of type IIB on the $T^{1,1}$ coset space~\cite{Cassani:2010na,Bena:2010pr} (which admits a left-invariant $\U(1)\subset \SU(2)$ structure) and of eleven-dimensional supergravity on the various coset spaces considered in~\cite{Cassani:2012pj}.

\section{Half-maximal truncations to five dimensions}
\label{sec:half_max_str_5d}

In order to make the general formalism more explicit, in the following sections we will focus on the case of consistent truncations of type IIB and eleven-dimensional supergravity to five dimensions, preserving half-maximal supersymmetry. In this section we will give the details of the generic formalism, identifying the possible structure groups $\Gst$, the invariant generalised tensors and, in particular, how they determine the generalised metric. Concrete examples will be discussed in the following sections. We note that the case of half-maximal truncations to five (and other) dimensions using exceptional field theory was first considered in the general analysis of~\cite{Malek:2017njj}. Here we give a number of new results, both for how the generalised structure is defined and how the truncations are constructed. For the general structure of half-maximal supergravity in five dimensions we refer to~\cite{Schon:2006kz} (see also~\cite{DallAgata:2001wgl}).

\subsection{$\SO(5-n)$ generalised structures}
\label{gen1/2s}

Dimensional reductions of eleven-dimensional supergravity on a six-dimensional manifold or of type IIB supergravity on a five-dimensional manifold 
are described by $\Ex{6} \times \mathbb{R}^+$ generalised geometry. The R-symmetry group of five-dimensional supergravity is contained in
 $\USp(8)$,  the maximal compact subgroup of $\Ex{6}$. 
For  half-maximal supergravity, $\USp(8)$ must be broken to  
\be
\label{USp8inUSp4USp4}
\USp(8) \,\supset\, \USp(4)_R \times \USp(4)_S \,\supseteq\,  \USp(4)_R \times \tGst \ ,
\ee
where  the factor $\USp(4)_R$ is identified with the R-symmetry of half-maximal supergravity,  while the other $\USp(4)_S$ factor contains the (double cover of) the reduced structure group, $\tGst\subseteq \USp(4)$. 
Under the first embedding in~\eqref{USp8inUSp4USp4}, the spinorial representation of $\USp(8)$ decomposes as ${\bf 8} = {\bf (4,1)} \oplus {\bf (1,4)}$, and we can identify the four spinor parameters of half-maximal supergravity as those that transform in the ${\bf (4,1)}$ representation, in the ${\bf 4}$ of $\USp(4)_R$ and singlets of $\tGst$.  Since we are focussing on dimensional reductions that do not have more than half-maximal supersymmetry, we also require that there are no further $\tGst$-singlets in the ${\bf (1,4)}$ representation. This (essentially) restricts the possible structure groups\footnote{For spinorial representations we of course need the double cover $\tGst$. Thus, for instance $\USp(4)$ is the double cover of $\SO(5)$, but when discussing bosonic representations we can use $\SO(5)$ at the place of $\USp(4)$.} to be $\Gst = \SO(5-n)$, $n=0,\ldots,3$. (Here we are ignoring the possibility of finite structure groups, hence exclude $n=4$). Thus half-maximal truncations correspond to dimensional reductions on (the double cover of) $\Gst =\SO(5-n)$ generalised structures.
This structure group is embedded in $\Ex{6}$ as:
\be
\label{SO5dec}
\Gst = \SO(5-n) \subseteq \SO(5)_S\subset \SO(5,5) \subset \Ex{6} \, . 
\ee
There are two extra cases of $\Gst\subset\SO(5)_S$ not included in this sequence. These come from the embeddings
\begin{equation}
   \SO(5)_S \supset \SO(4) = \frac{\SU(2)\times\SU(2)}{\bbZ_2}
      \supset \SU(2)\times U(1) \supset U(1)\times U(1) .
\end{equation}
Choosing either $\Gst=\SU(2)\times U(1)$ or $\Gst=U(1)^2$ still gives a half-maximal truncation. However, it is easy to show that the commutant subgroups and $\Gst$-singlets are the same as the case of $\Gst=\SO(4)$. Thus although the structure is different the resulting truncated theory is the same, meaning we can restrict to the sequence~\eqref{SO5dec}.    
   
As discussed in the previous section, the  vector fields in the truncation are in one-to-one correspondence with the $\Gst$-singlets in the  fundamental representation of $\Ex{6}$, while the scalar fields parameterise the coset
\be\label{half_max_scalar_manifold}
\mathcal{M}_\text{scal} \,=\, \frac{\Comm_{\Ex{6}} (\Gst)}{\Comm_{\USp(8)}(\Gst)} \,=\,  
O(1,1)\times \frac{\SO(5,n)}{\SO(5)\times\SO(n)}
:= \frac{\mathcal{G}}{\mathcal{H}}
\ ,
\ee
where as before $\Comm_{\Ex{6}}(\Gst)$ and $\Comm_{\USp(8)}(\Gst)$ are the commutants of $\Gst$ in $\Ex{6}$ and $\USp(8)$, respectively. This matches the standard structure of the scalar manifold for half-maximal supergravity coupled to $n$ vector multiplets~\cite{Schon:2006kz}. The single scalar in the gravity multiplet parameterises the $O(1,1)$ factor\footnote{Previously we denoted such factor by $\mathbb{R}^+$, while here we use $O(1,1)$ to match the standard supergravity literature.} while the scalars in the vector multiplets parameterise the $ \frac{\SO(5,n)}{\SO(5)\times\SO(n)}$ coset space.

We can also identify the number of singlets in the generalised tangent space, which determines the number of vector fields in the truncation. They also form a representation of $\mathcal{G}$. Recall that the generalised tangent space $E$ transforms in the $\rep{27}$ of $\Ex{6}$. Under $\SO(1,1)\times\SO(5,5)\subset\Ex{6}$ we have the decomposition
\begin{align}\label{eq:ESO(5,5)}
   E &= E_0 \oplus E_{10} \oplus E_{16}\ , \nn\\[1mm]
   \rep{27} &= \rep{1}_{-4} \oplus \rep{10}_{2} \oplus \rep{16}_{-1} \ , 
\end{align}
where the subscripts denote the $\SO(1,1)$ weights. Under $\SO(5)\times\SO(5)$ we have $\rep{16}_{-1}=\repp{4}{4}$. By construction the $\rep{4}$ representation has no singlets under $\Gst$ and hence there are no singlets in the $\rep{16}_{-1}$ component. On the other hand, the $\rep{10}_{2}$ representation decomposes as
\begin{equation}
   \label{eq:10-decomp}
   \rep{10}_2 = \repp{5+n}{1}_2 \oplus \repp{1}{5-n}_2 , 
\end{equation}
under $O(1,1)\times\SO(5,n)\times\SO(5-n)\subset O(1,1)\times\SO(5,5)$. Thus we see that we get $6+n$ singlets, one from the $\rep{1}_{-4}$ representation and $5+n$ from $\rep{10}_2$. In summary, as a $\mathcal{G}=O(1,1)\times\SO(5,n)$ representation, we have the space of vector fields 
\begin{align}\label{eq:Kn}
   \mathcal{V} &= \rep{1}_{-4} \oplus \rep{(5+n)}_{2} , \nn \\
   \{ K_\mathcal{A} \} &= \{ K_0, K_A : A=1,\dots,5+n \} ,
\end{align}
where we are using the index $\mathcal{A}=0,1,\dots,5+n$. In terms of the half-maximal supergravity six of these vectors come from the gravity multiplet and $n$ of them from the additional vector multiplets. 

In generalised geometry, the $\Ex{6}$ cubic invariant, acting on the generalised tangent space $E$, gives a map $c:S^3E\to\det T^*M$, which can be used to choose a natural parametrisation of the invariant generalised vectors. From the decompositions~\eqref{eq:ESO(5,5)} and~\eqref{eq:10-decomp} we have 
\begin{align}\label{eq:SO55cond_SO5n}
   c(K_0,K_0,V) & = 0\ ,   \qquad \forall \,V\in\Gamma(E)\, , \nn\\
   c(K_A,K_B,K_C) &= 0\ ,   \qquad\forall \, A,B,C\, ,
\end{align}
and hence, independent of the choice of $K_A$, an $\SO(5,5)$ metric $\eta$ on $E_{10}$ given by
\begin{equation}
   \label{eq:eta}
   c(K_0,V,W) = \eta(V,W) \vol\, ,
\end{equation}
where $\vol$ is a volume form on $\det T^*M$. Since the $K_A$ are fixed up to $\SO(5,5)$ rotations, we can use this to fix an orthonormal basis, and hence also the volume form $\vol$, by 
\begin{equation}
\label{eq:ortho-cond_SO5n}
  \eta(K_A,K_B) = \eta_{AB} \, ,
\end{equation}
where 
\begin{equation}
   \eta_{AB}={\rm diag}(-1,-1,-1,-1,-1,+1,\dots,+1)
\end{equation} 
is the flat $\SO(5,n)$ metric\footnote{The overall sign in $\eta$ is chosen so as to allow a straightforward identification with the $\SO(5,n)$ metric normally used in half-maximal supergravity~\cite{Schon:2006kz}.}. Note that the freedom in the normalisation of $\eta$ in~\eqref{eq:eta} and hence of the $K_{\mathcal A}$ vectors via rescaling $K_0\mapsto\lambda^2K_0$ with $K_A\mapsto \lambda^{-1}K_A$ is just the action of the $O(1,1)$ subgroup of $\mathcal{G}$. Note that specifying a set of vectors $\{K_\mathcal{A}\}$ satisfying~\eqref{eq:SO55cond_SO5n} and~\eqref{eq:ortho-cond_SO5n} fixes an $\SO(5-n)\subset\Ex{6}$ structure. That is, the structure is completely determined by the vectors and no other generalised tensors are needed. 

Turning to the two-form fields, for $\Ex{6}$ generalised geometry we have
\begin{align}\label{Nbundlegen}
   N\simeq \det T^*M \otimes E^* &= N_0 \oplus N_{10} \oplus N_{16}\ , \nn\\
   \rep{27}' &= \rep{1}_{4} \oplus \rep{10}_{-2} \oplus \rep{16}'_{1} , 
\end{align}
where again we decompose under $\SO(1,1)\times\SO(5,5)\subset\Ex{6}$. The same argument as for $E$ then gives the space of singlet two-forms $J_\Sigma$
\begin{align}\label{eq:Kn}
   \mathcal{B} &= \rep{1}_{4} \oplus \rep{(5+n)}_{-2}\, ,\nn \\[1mm]
   \{ J^\mathcal{A} \} &= \{ J^0, J^A : A=1,\dots,5+n \}\, ,
\end{align}
where the isomorphism $ N\simeq \det T^*M \otimes E^*$ allows us to identify the usual $\Sigma$ index on the basis with the dual of the index on $K_\mathcal{A}$. It is natural to normalise
\begin{equation}
   \GM{J^\mathcal{A}}{K_\mathcal{B}} = \delta^\mathcal{A}{}_\mathcal{B}\vol \,,
\end{equation}
where  $\GM{W}{V} $ denotes the natural pairing between a vector and the (weighted) dual vector. The cubic invariant provides the intertwining maps~\eqref{eq:const-d} via
\begin{align}
   \label{K*fromK}
   J^0 &= \tfrac{1}{5+n}\,\eta^{AB}\,c(K_A,K_B,\cdot)\ , \nn\\[1mm]
   J^A &= \eta^{AB}c(K_0,K_B,\cdot) \ .
\end{align}
It will be helpful in what follows to also define 
\be\label{Jvol}
J^0=\vol\cdot K^*_0\,,\qquad  J^A=\eta^{AB}\vol\cdot K^*_B\,,
\ee
so that $\{K^*_\mathcal{A}\}$ are a set on $E^*$, dual to $\{K_\mathcal{A}\}$, satisfying 
 \begin{equation}
 \GM{K^*_0}{K_0} = 1\ , \qquad\GM{K^*_A}{K_B}= \eta_{AB}\ , \qquad \GM{K^*_0}{K_A}= \GM{K^*_A}{K_0}= 0\ .
\end{equation}

Having identified the matter content of the truncated theory, we now turn to its gauging. From the general discussion, this is determined by the intrinsic torsion of the structure, which encodes an embedding tensor. Since in this case, the generalised vectors determine the $\Gst$-structure, all the information of the intrinsic torsion should be encoded in~\eqref{eq:K-alg}, namely 
\begin{equation}\label{genLieKKisXK}
   L_{K_{\mathcal{A}}}K_{\mathcal{B}}
        = X_{\mathcal{A}\mathcal{B}}{}^{\mathcal{C}} K_{\mathcal{C}}\, . 
\end{equation}
The analysis of gaugings of half-maximal supergravity in five dimensions can be found in~\cite{Schon:2006kz}. The embedding tensor has components $f_{ABC}=f_{[ABC]}$, $\xi_{AB}=\xi_{[AB]}$ and $\xi_A$. For simplicity we will only discuss the case $\xi_A=0$, although it would be straightforward to include the general case $\xi_A\neq 0$. The remaining components have to satisfy  the conditions 
\begin{equation}
f_{[AB}{}^Ef_{CD]E}=0\,,\qquad \xi_{A}{}^Df_{DBC}=0\,,
\end{equation}
where the indices are raised/lowered using the $\SO(5,n)$ metric $\eta_{AB}$. Using the composite index $\mathcal{A}=\{0,A\}$, the components can be assembled into the gauge group generators $(X_{\mathcal{A}})_{\mathcal{B}}{}^{\mathcal{C}}=X_{\mathcal{A}\mathcal{B}}{}^{\mathcal{C}}$ as:
\begin{equation}\label{rel_X_fxi}
X_{AB}{}^C = -f_{AB}{}^C\,,\qquad  X_{0A}{}^B = -\xi_A{}^B\,,
\end{equation}
with the other components vanishing.
Then the $(X_{\mathcal{A}})_{\mathcal{B}}{}^{\mathcal{C}}$ generators satisfy the commutation relations: 
\be\label{commutator_XX}
[X_{\mathcal{A}},X_{\mathcal{B}}]=-X_{\mathcal{A}\mathcal{B}}{}^{\mathcal{C}} X_{\mathcal{C}}\,.
\ee
Thus, in general, we expect that any consistent truncation (leading to a gauging with $\xi_A=0$) should have a generalised Lie derivative algebra~\eqref{genLieKKisXK} with the components of $X_{\mathcal{A}\mathcal{B}}{}^\mathcal{C}$ given by~\eqref{rel_X_fxi}. Note that, in the generalised geometry, the algebraic conditions $f_{ABC}=f_{[ABC]}$, $\xi_{AB}=\xi_{[AB]}$ follow from consistency of the generalised algebra~\eqref{genLieKKisXK} with the conditions~\eqref{eq:SO55cond_SO5n} and~\eqref{eq:ortho-cond_SO5n}.

Having determined the number $n$ of vector multiplets and the embedding tensor from the generalised $\SO(5-n)$ structure, we have fully characterised the five-dimensional half-maximal supergravity theory that is obtained after  truncation. However we still need to provide the truncation ansatz, namely the embedding of the lower-dimensional fields into the higher-dimensional ones. This is necessary to uplift any solution of the lower-dimensional theory.  In order to be able to do this we need a further geometrical ingredient, that is the construction of the generalised metric on the exceptional tangent bundle starting from the generalised vectors defining the $\SO(5-n)$ structure. This will be instrumental to specifying the scalar truncation ansatz.

\subsection{The generalised metric}

Recall that, in the generalised geometry reformulation, the generalised metric $G_{MN}$ can be viewed as an element of the coset $\Ex{6}\times\bbR^+/(\USp(8)/\bbZ_2)$. Here we have a $\Gst=\SO(5-n)$ structure. Given the embedding~\eqref{USp8inUSp4USp4}, since $\tGst\subset\USp(8)$, the structure determines the metric. Since the structure is completely determined by the vectors $\{K_\mathcal{A}\}$ this means we should be able to use them to construct $G$ explicitly.

The easiest way to see how this construction works is to use the embedding~\eqref{SO5dec}. The choice of $K_0$ and $K^*_0$ fixes the $\SO(1,1)\times\SO(5,5)\subset\Ex{6}$ subgroup and gives a decomposition of the generalised tangent space~\eqref{eq:ESO(5,5)}. 
This in turn gives a decomposition of the metric into orthogonal metrics on $E_0$, $E_{10}$ and $E_{16}$ subspaces,
\begin{equation}
   G = G_0 + G_{10} + G_{16} \ .
\end{equation}
We can then use our knowledge of $\SO(5)\times\SO(5)\subset\SO(5,5)$ generalised structures to construct the three pieces of the metric as:
 \begin{align} 
 \label{G0}
G_0(V,V) &= \GM{K^*_0}{V}\GM{K^*_0}{V} \,,   \\
 \label{G10}
G_{10}(V,V) &= 2\,  \delta^{ab}  \langle K^*_a,V  \rangle \langle K^*_b, V \rangle   +  \eta(V,V)  \,, \\ 
 \label{G16}
G_{16}(V,V) &=  - 4 \sqrt{2}\, \langle K_1 \cdots  K_5 \cdot  V, V \rangle \, ,
\end{align}
where we have denoted the first five generalised vectors $\{K_a\}$ by an index $a=1,\dots,5$. Recall from~\eqref{eq:ortho-cond_SO5n} that these satisfy $\eta(K_a,K_b) = - \delta_{ab}$.

Let us explain these formulae. The metric $G_0$ is simply obtained by projecting onto  the singlet. 
For $G_{10}$,  we use the fact that  $E_{10}$ is the generalised tangent bundle for the $\SO(5,5)$ geometry and that the structure  $\SO(5)\times\SO(5)\subset\SO(5,5)$
induces a split of $E_{10}$  into positive- and negative-definite eigenspaces
\be
E_{10}=C_+\oplus C_- \, . 
\ee
Then the $\SO(5,5)$ invariant metric $\eta$ given in~\eqref{eq:eta} and the generalised metric $G_{10}$ can be written as
 \begin{align} 
   \eta(V,V)& = G_+ - G_- \,, \nn\\
 G_{10}&=   G_++G_- \, ,
\end{align}
where $G_\pm$ are metrics on $C_\pm$.  Since the $K_a$ form a basis for $C_-$, we have 
\begin{equation}
   G_-(V,V) =  \delta^{ab}\GM{K^*_a}{V}\GM{K^*_b}{V}\,.
\end{equation}
Hence $G_{10}=   G_++G_-  =  2G_- + \eta$, and we recover  \eqref{G10}.

For $G_{16}$ we recall  that, given  the $\SO(5)\times\SO(5)$ structure,  the positive definite inner product on $\SO(5,5)$ spinors is
\begin{equation}
   \GM{\Psi}{\Gamma^{(+)}\Psi}\,,
\end{equation}
where $\GM{\cdot}{\cdot}$ is the Mukai pairing and $\Gamma^{(+)}$ is the chirality operator on $C_+$, that is 
\begin{equation}
   \Gamma^{+}=\Gamma^+_1\cdots \Gamma^+_5\,,
\end{equation}
where we decompose the $\SO(5,5)$ gamma matrices into $\{\Gamma^+_a\}\cup\{\Gamma^-_{\hat{a}}\}$ spanning  $C_+$ and $C_-$. 
In this case, the Mukai pairing is just the natural pairing between $\Psi\in\Gamma(E)$ and $\Psi^*\in\Gamma(E^*)$. Thus we can write   $G_{16}$ as 
\begin{equation}
   G_{16}(V,V)
     = - 4 \sqrt{2} \, \GM{K_1\cdots K_5\cdot V}{V}\,,
\end{equation}
where the Clifford actions of $K_a$ map between $E$ and $E^*$ and are given by
\begin{equation}
\label{eq:Cliff2}
\begin{aligned}
  W \cdot \Psi  & : =\,W^M\Gamma_M \Psi 
     = \frac{c(W,\Psi,\cdot)}{\vol}
     \quad \in \Gamma(E^*) \,, \\ 
   W \cdot \Psi^* & : =\, W^M\Gamma_M \Psi^*
     = \vol \cdot \,c^*(W^*,\Psi^*,\cdot) \quad \in \Gs{E} \, .
\end{aligned}
\end{equation}
Here we define $V^*=\eta(V,\cdot)\in\Gs{E^*}$, and $c^*(\cdot,\cdot,\cdot)$ is the $\Ex{6}$ cubic invariant on $E^*$. Note that one does not need to project $V$ onto $E_{16}$ since as defined $G_{16}$ will vanish identically when acting on sections of $E_0$ or $E_{10}$. 

We will also need the inverse generalised metric $G^{-1}$, which acts on dual generalised vectors $Z\in \Gamma(E^*)$. 
Its expression is closely related to the one for $G$ and reads
 \begin{align} 
  \label{geninvmet}
& G^{-1}_0(Z,Z)\, =\,  \langle Z, K_0 \rangle \langle Z, K_0 \rangle \,,  \nonumber  \\
& G^{-1}_{10}(Z,Z) \, =\,  2  \delta^{ab}  \langle Z , K_a  \rangle \langle Z, K_b \rangle   +  \eta^{-1}(Z,Z)  \,,\nonumber \\ 
& G^{-1}_{16}(Z,Z) \, =\,  -4 \sqrt{2}\, \langle Z, K_1 \cdots  K_5 \cdot  Z \rangle \, ,
 \end{align}
where $\eta^{-1}(Z,Z)=\vol\cdot\, c^*(K^*_0,Z,Z)$ is the inverse of the $\SO(5,5)$ metric $\eta$.

\subsection{The truncation ansatz}\label{sec:E6_trunc_ansatz}

We provide here the main steps of the construction of the truncation ansatz, which is entirely based on the generalised vectors  $K_\mathcal{A}$ defining the $\SO(5-n)$ structure.
 More explicit formulae will be provided in the next sections, where we will specialise the formalism to both type IIB supergravity or M-theory, and discuss some concrete examples. 

We start from the ansatz for the vector fields. By taking the higher-dimensional supergravity fields with one external index we make a generalised vector $\mathcal{A}^M_\mu$, where we recall that  $\mu$ is an external spacetime index while $M$ labels the components of a generalised vector, which in $\Ex{6}$ generalised geometry transform in the ${\bf 27}$.
We expand this generalised vector as in \eqref{vectors} \be
\label{ansatz-vec}
\mathcal{A}^M_\mu (x, y) = \sum_{\mathcal{A} =0}^{5+n}  \mathcal{A}^\mathcal{A}_\mu (x)  K^M_\mathcal{A} (y)   \, , 
\ee
where  $\mathcal{A}_\mu^\mathcal{A}$ are  the five-dimensional supergravity vector fields.  Similarly, 
the supergravity fields with two antisymmetrised external indices can be arranged in a generalised tensor, as a section of the bundle $N$. Exploiting the isomorphism $ N\simeq \det T^*M \otimes E^*$, we can write this as a weighted dual vector $\mathcal{B}_{\mu\nu\,M}$, and express the truncation ansatz \eqref{two-forms} as
\be\label{ansatz_two-forms}
\mathcal{B}_{\mu\nu\,M}(x,y) \,=\, \sum_{\mathcal{A} =0}^{5+n} \mathcal{B}_{\mu \nu\,\mathcal{A}}(x) J^{\mathcal{A}}{}_{M}(y)\, . 
\ee

The ansatz for the scalar fields is more elaborated as it requires the generalised metric.
This is specified by choosing a metric on the coset space \eqref{half_max_scalar_manifold},
which is also the scalar manifold of half-maximal supergravity in five dimensions.
We parameterise the  $O(1,1)$ factor by a non-vanishing scalar $\Sigma$. 
 The $\frac{\SO(5,n)}{\SO(5)\times\SO(n)}$ factor is described by a coset representative $(\cV_{A}{}^a,\cV_A{}^{\ul a}) \in \SO(5,n)$ and its inverse $(\cV_{a}{}^A,\cV_{\ul a}{}^A)^T$, where $a=1,\ldots,5$ and $\ul{a}=1,\ldots,n$ are local $\SO(5)$ and $\SO(n)$ indices, respectively. The coset representative satisfies
\begin{align}
\eta_{AB} &= -\delta_{ab}\,\cV_A{}^a \cV_B{}^b + \delta_{\ul{ab}}\,\cV_A{}^{\ul a} \cV_B{}^{\ul b}\,,\nonumber\\[1mm]
M_{AB} &= \delta_{ab}\,\cV_A{}^a \cV_B{}^b + \delta_{\ul{ab}}\,\cV_A{}^{\ul a} \cV_B{}^{\ul b}\ .
\end{align}
Note that the matrix $M_{AB}$ is a metric on the coset, with inverse $M^{AB}=\eta^{AC} M_{CD}\, \eta^{DB}$.

The construction of the generalised metric now goes as follows.
We introduce the ``dressed'' generalised vectors
\begin{equation}
\tilde{K}_0 = \Sigma^{2} K_0\ , \qquad \tilde K_a = \Sigma^{-1}\,\cV_a{}^A K_A\ , \qquad \tilde K_{\ul a} = \Sigma^{-1}\, \cV_{\ul a}{}^A K_A\ ,
\end{equation}
and their duals
\begin{equation}
\tilde{K}_0^* = \Sigma^{-2} K_0^*\ , \qquad \tilde K^*_a = \Sigma\,\cV_a{}^A K^*_A\ , \qquad \tilde K^*_{\ul a} = \Sigma\, \cV_{\ul a}{}^A K^*_A\ .
\end{equation}

The generalised metric and its inverse are  defined as in \eqref{G0}--\eqref{G16}, this time using the dressed generalised vectors  $\tilde K_0$ and $\tilde K_a$, $a=1,\ldots, 5$. The generalised metric is then $G = G_0 + G_{10} + G_{16}$,  with 
\begin{align}
   G_0(V,V) &= \Sigma^{-4}\,\GM{K^*_0}{V}\GM{K^*_0}{V} \ ,\nn\\[1mm]
   G_{10}(V,V) &= \Sigma^{2} \left(2\,\delta^{ab} \mathcal{V}_a{}^A\mathcal{V}_b{}^B\, \GM{K^*_A}{V}\GM{K^*_B}{V}
     +  \eta(V,V) \right)\ ,\nn\\[1mm]
   G_{16}(V,V)
   &= -\tfrac{4 \sqrt{2}}{5!}\,\Sigma^{-1}\,\epsilon^{abcde}\cV_{a}{}^{A}\cV_{b}{}^{B}\cV_{c}{}^C{}\cV_{d}{}^{D}\cV_{e}{}^{E}\,
   \GM{K_{A}\cdots K_{E}\cdot V}{V}\ .
\end{align}
Similarly, the inverse generalised metric  $G^{-1} = G_0^{-1} + G_{10}^{-1} + G_{16}^{-1}$ is given by
\begin{align}
\label{Ginv10forSO5n}
 G^{-1}_0(Z,Z) &=  \Sigma^{4}\,\GM{Z}{K_0}\GM{Z}{K_0} \,  , \nn\\[1mm]
 G^{-1}_{10}(Z,Z) &= \Sigma^{-2}\left( 2 \,\delta^{ab}\mathcal{V}_a{}^A\mathcal{V}_b{}^B\GM{Z}{K_A}\GM{Z}{K_B} +  \eta^{-1}(Z,Z)\right)  \,  , \nn \\[1mm]
 G^{-1}_{16}(Z,Z) &= -\tfrac{4\sqrt2}{5!}\,\Sigma\,\epsilon^{abcde}\cV_{a}{}^{A}\cV_{b}{}^{B}\cV_{c}{}^{C}\cV_{d}{}^{D}\cV_{e}{}^{E}\,\GM{Z}{K_{A}\cdots K_{E}\cdot Z}\ .
\end{align}
Notice that the $\SO(5)\times \SO(n)$ invariant matrices $2\,\delta^{ab}\mathcal{V}_a{}^A\mathcal{V}_b{}^B = M^{AB}-\eta^{AB}$, $M^{ABCDE}= \epsilon^{abcde}\cV_{a}{}^{A}\cV_{b}{}^{B}\cV_{c}{}^{C}\cV_{d}{}^{D}\cV_{e}{}^{E}$ are familiar from the construction of half-maximal supergravity in five dimensions~\cite{Schon:2006kz}.
Also note that to get the correct power of $\Sigma$ in the $G_{16}$ and $G_{16}^{-1}$ expressions it is important to keep track of how many of the Clifford actions are with $\tilde{K}_a$ and how many with  $\tilde{K}^*_a$. 

The scalar ansatz is obtained by equating the inverse generalised metric with the one obtained from the split frame~\cite{Coimbra:2011ky,Lee:2014mla}, which encodes all supergravity fields with purely internal indices (including the warp factor of the external metric). By separating the different tensorial structures on the internal manifold $M$, we obtain the scalar ansatz for the individual higher-dimensional supergravity fields.

\section{Type IIB truncations}\label{sec:IIBtrunc}

In this section we specialise our formalism to dimensional reductions of type IIB supergravity on five-dimensional manifolds.  To this end, we first recall the details of type IIB $\Ex{6}$ geometry and present the truncation anzatz adapted to the type IIB fields.
Then we discuss concrete examples of consistent truncations.  The first is the truncation on squashed Sasaki--Einstein manifolds of \cite{Cassani:2010uw,Gauntlett:2010vu}, leading to half-maximal supergravity coupled to two vector multiplets.  Although this truncation is not new and can be understood based on ordinary $\SU(2)$ structure, it will serve to illustrate the validity of our approach in a relatively simple case. This will also make clear how generalised goemetry fully characterises the lower-dimensional theory even before the lower-dimensional Lagrangian is constructed from the truncation of the higher-dimensional equations of motion.
We will then consider a $\beta$-deformed Sasaki--Einstein manifold and will show that there is a consistent truncation on such manifolds leading to the same half-maximal supergravity obtaind from the Sasaki--Einstein truncation. This truncation includes 
the supersymmetric, $\beta$-deformed AdS$_5$ solution.

\subsection{$E_{6(6)}$ geometry for  type IIB }\label{sec:E66geometry_maintext}

We recall here some basic definitions of the $E_{6(6)}$ generalised geometry for type IIB supergravity on a five-dimensional manifold $M$. A more detailed account is given in Appendix~\ref{PreliminariesE66_IIB} following the conventions of~\cite[App.~E]{Ashmore:2015joa}.

It is convenient to decompose the generalised tangent bundle  $E$, whose fibers transform in the  ${\bf 27}$ of $\Ex{6}$, according to the  $\GL(5) \times \SL(2)$ subgroup of 
$\Ex{6}$ 
\begin{equation}\label{genbundle5}
   E \simeq TM \oplus (T^*M \oplus T^*M) \oplus \Lambda^3T^*M 
      \oplus (\Lambda^5T^*M \oplus \Lambda^5T^*M)\ ,
\end{equation}
where the two copies of $T^*M$ and the two copies of $\Lambda^5 T^*M$  transform as  $\SL(2)$ doublets.
A generalised vector can be written as
\be
V = v + \lambda^\alpha + \rho  + \sigma^\alpha \, , 
\ee
where $v$ is a vector, $\lambda^\alpha$ is an $\SL(2)$ doublet of one-forms, $\rho$ is a three-form and $\sigma^\alpha$ is an $\SL(2)$ doublet of  five-forms, $\alpha=\{+,-\}$ being the $\SL(2)$ index.
   The dual bundle decomposes accordingly as 
\begin{equation}
   E^* \simeq T^*M \oplus (TM \oplus TM) \oplus \Lambda^3TM 
      \oplus (\Lambda^5TM \oplus \Lambda^5TM)\ ,
\end{equation}
with sections
\be\label{dual_gen_vector}
Z = \hat{v} + \hat{\lambda}_\alpha + \hat{\rho}  + \hat{\sigma}_\alpha \, ,
\ee
where $\hat v$ is a one-form, $\hat{\lambda}_\alpha$ is an $\SL(2)$ doublet of vectors, $\hat{\rho}$ is a three-vector, and $\hat{\sigma}_\alpha$ is an $\SL(2)$ doublet of five-vectors.
The natural pairing between a generalised vector and a dual one is
\be
\label{pairing_vector_dualvector}
\GM{Z}{V} = \hat v_m v^m + \hat\lambda^{m}_\alpha\lambda_m^{\alpha} + \tfrac{1}{3!}\,\hat\rho^{mnp}\rho_{mnp} +  \tfrac{1}{5!}\,\hat\sigma^{mnpqr}_\alpha\sigma_{mnpqr}^\alpha \,.
\ee
The cubic invariant is defined on $E$ and $E^*$, respectively, as
\begin{align}
\label{cubic_inv}
 c(V, V, V)  &=  -3 \, \big(   \iota_v   \rho  \wedge \rho  +   \epsilon_{\alpha\beta}  \,\rho \wedge \lambda^\alpha  \wedge \lambda^\beta    - 2\,  \epsilon_{\alpha\beta} \, \iota_v \, \lambda^\alpha  \sigma^\beta \big) \, , \\[1mm]
\label{dual_cubic_inv}
c^*(Z, Z, Z) &=  -3\,  \big(   {\hat v} \lrcorner  \hat\rho  \wedge \hat\rho  +   \epsilon^{\alpha\beta}  \hat\rho \wedge \hat\lambda_\alpha  \wedge \hat\lambda_\beta    - 2 \, \epsilon^{\alpha\beta} \, \hat{v}\lrcorner  \, \hat\lambda_\alpha  \hat\sigma_\beta \big) \, .
\end{align}

The bosonic fields of type IIB supergravity are the metric, the dilaton $\phi$, the axion $C_0$,  an $\SL(2)$ doublet of two-form potentials $\hat B^\alpha$ ($\hat B^+$ being the NSNS two-form and $\hat B^-$ being the RR one), a self-dual four-form $\hat C$, and a doublet of six-form potentials $\hat{\tilde B}^\alpha$ that are on-shell dual to the two-forms.\footnote{In this subsection the symbol hat  denotes ten-dimensional fields.}
When dimensionally reducing on a five-dimensional manifold, the ten-dimensional fields are decomposed 
 according to the $\SO(1,9) \supset\SO(1,4)\times \SO(5)$ splitting of the  Lorentz group.  We will use coordinates $x^\mu$, $\mu = 0,\ldots, 4$ for the external spacetime and $y^m$, $m=1,\ldots,5$ for the internal manifold $M$. 
Then the type IIB metric takes the form
\be
\label{KK_decomp_metr}
g_{10}  \,= \,  \rme^{2\Delta}\, g_{\mu\nu} \,\dd{x}^\mu\dd{x}^\nu + g_{mn} Dy^m Dy^n\ ,
\ee
where  $Dy^m \,=\, \dd{y}^m - h_\mu{}^m \dd{x}^\mu$
and $\Delta(x,y)$ is the warp factor of the external metric $g_{\mu\nu}(x)$. The form fields decompose as
\begin{align}
\label{expand_10dfields}
\hat{B}^\alpha  &=  \tfrac{1}{2} \, B^\alpha_{m_1m_2} Dy^{m_1m_2} + \overline{B}{}^\alpha_{\mu m} \dd{x}^\mu \wedge Dy^m + \tfrac{1}{2}\,\overline{B}{}^\alpha_{\mu\nu} \dd{x}^{\mu\nu} \, ,\nn\\[1mm]
\hat{C} &= \tfrac{1}{4!} C_{m_1\ldots m_4 } D{y}^{m_1\ldots m_4 } + \tfrac{1}{3!}\overline{C}_{\mu m_1m_2 m_3 } \dd{x}^\mu\wedge D{y}^{m_1m_2 m_3 } + \tfrac{1}{4}\overline{C}_{\mu\nu m_1 m_2 }\dd{x}^{\mu\nu} \!\wedge\! Dy^{m_1 m_2}  + \ldots  ,\nn\\[1mm]
\hat{\tilde B}^\alpha &=  \tfrac{1}{5!}\,\overline{\tilde B}{}^{\,\alpha}_{\mu m_1\ldots m_5} \dd{x}^\mu\!\wedge\! D{y}^{m_1\ldots m_5} +\tfrac{1}{2\cdot 4!} \,\overline{\tilde B}{}^\alpha_{\mu\nu m_1\ldots m_4 }\dd{x}^{\mu\nu} \!\wedge\! Dy^{m_1\ldots m_4}+\ldots , 
\end{align}
where $\dd{x}^{\mu\nu} = \dd x^\mu\wedge \dd x^\nu$ and $Dy^{m_1\ldots m_p} = Dy^{m_1}\wedge \cdots \wedge Dy^{m_p}$.
The ellipsis denote forms with more than two external indices which we will not need. 
The expansion in $Dy$ instead of $\dd y$ ensures covariance of the components under internal diffeomorphisms. 

As discussed in e.g.~\cite{Cassani:2016ncu,Baguet:2015xha}, covariance under generalised diffeomorphisms also requires a redefinition of the  barred fields in the expansion above.  We adopt a notation such that $B_{\mu, p}$  indicates the components of a one-form in the external spacetime which are $p$-forms in the internal manifold. Similarly,  $B_{\mu\nu, p}$ are the components of a two-form in the external spacetime that are $p$-forms in the internal manifold. 
We perform the following field redefinitions of the one-forms in the external spacetime:
\begin{align}\label{redef_one_forms}
& \overline{B}{}^\alpha_{\mu,1} = B^\alpha_{\mu,1} \,,\nn \\[1mm]
& C_{\mu,3} =   \overline{C}_{\mu,3}  + \frac{1}{2}\,  \epsilon_{\alpha \beta} B^\alpha_{\mu,1} \wedge B^\beta\,, \nn \\[1mm]
& \tilde{B}_{\mu,5}^{\alpha}=\overline{ \tilde{B}}{}_{\mu,5}^{\alpha}-\tfrac{1}{2}\,\overline{B}{}_{\mu,1}^{\alpha}\wedge C -\tfrac{1}{2}\,\overline{C}_{\mu,3}\wedge B^{\alpha}\,,
\end{align}
where $B^\alpha$, $C$ are just internal.
The external two-forms are redefined as
\begin{align}
& B^\alpha_{\mu \nu}= \overline{B}{}^\alpha_{\mu \nu}+  h_{[\mu} \lrcorner  \overline{B}{}^\alpha_{\nu] } \,, \nn \\[1mm]
& C_{\mu \nu,2} =   \overline{C}{}_{\mu \nu,2}  + \tfrac{1}{2}\,  \epsilon_{\alpha \beta} B^\alpha_{\mu \nu}\, B^\beta\,, \nn \\[1mm]
& \tilde{B}{}_{\mu\nu,4}^{\alpha}=\overline{\tilde{B}}{}_{\mu\nu,4}^{\alpha}+\tfrac{1}{2}\,\overline{B}{}_{\mu\nu}^{\alpha}\, C +\tfrac{1}{2}\,\overline{C}_{\mu\nu,2}\wedge B^{\alpha} -  \overline{B}{}^\alpha_{[\mu,1} \wedge \overline{C}_{\nu],3}\,\, .
\end{align}
The new (unbarred) fields transform covariantly both under internal diffeomorphisms and form gauge transformations, that is under generalised diffeomorphisms.

The next step is to arrange the redefined fields into the inverse generalised metric $G^{MN}$, the generalised vectors $\mathcal{A}_\mu^M$ and  the tensors $\mathcal{B}_{\mu\nu M}$.
The generalised metric is made by all the type IIB supergravity fields with only internal indices, including the warp factor~$\Delta$,
\be
G^{MN} \ \leftrightarrow\ \{\Delta,\,g_{mn},\,  \phi , \, C_0, \, B^\alpha_{m_1m_2}, \, C_{m_1\ldots m_4}  \}\ .
\ee
Its precise expression is given in \eqref{generalised_metric_IIB}.   
The fields with one external index can be arranged into the generalised vector $\mathcal{A}_\mu^M \in \Gamma(E)$, 
\be
\label{def_calA_mu^M}
\mathcal{A}_\mu{}^M \,=\, \{ h_\mu{}^m ,\, B^\alpha_{\mu m} , \, C_{\mu m_1m_2 m_3 },\,  \tilde{B}^\alpha_{\mu m_1\ldots m_5}    \}\ .
\ee
 Similarly 
 the fields with  two external indices form the generalised tensor $\mathcal{B}_{\mu\nu\,M}$, that is a section of $ N\simeq \det T^*M \otimes E^*$ (see \eqref{NbIIB}, \eqref{sectionNIIB} for its $\GL(5)\times \SL(2)$ decomposition), 
 \be\label{gen_ten_BmunuMN}
\mathcal{B}_{\mu\nu\,M} \,  =  \,   \{ B_{\mu\nu\,\alpha}, \,   C_{\mu\nu m_1m_2}  ,\,  \tilde{B}_{\mu\nu m_1\ldots m_4\,\alpha},\, \tilde g_{\mu\nu m_1\ldots m_5,n} \}\, .
\ee
Here, the $\SL(2)$ index $\alpha$ on the type IIB fields has been lowered with $\epsilon_{\alpha\beta}$, and $\tilde g \in \Gamma ( \Lambda^7 T^*M_{10} \otimes T^*M_{10} )$ is a tensor related to the dual graviton in ten dimensions. The latter is not part of type IIB supergravity in its standard form and will not play a role in the specific truncations we will discuss below.

We have thus decomposed the ten-dimensional tensors according to their external or internal legs and repackaged the components into generalised geometry objects. We can then specify the dependence of these fields on the internal coordinates
by making the consistent truncation ansatz described in Section~\ref{sec:E6_trunc_ansatz}.

\subsection{Truncation from generalised $\SU(2)$  structure on Sasaki--Einstein manifolds}
\label{sec:IIBonSE}

We discuss type IIB supergravity on a five-dimensional a Sasaki--Einstein manifold $M$, which admits a consistent truncation to half-maximal gauged supergravity with two vector multiplets~\cite{Cassani:2010uw}, see also \cite{Gauntlett:2010vu,Liu:2010sa,Skenderis:2010vz}.

\subsubsection{Generalised $\SU(2)$ structure}
 
 Five-dimensional Sasaki--Einstein ($\rm{SE}_5$) structures are examples of ordinary $\SU(2)$ structures, whose torsion is also an $\SU(2)$-singlet. 
 The $\SU(2)$ structure is defined by a vector $\xi$, a one-form $\eta$ and a triplet of real two-forms $j_i$, $i=1,2,3$, satisfying the compatibility conditions\footnote{The $j_i$ are identified with the forms used in eq.~\eqref{eq:SE} as $j_3 = \omega$ and $j_1+\ii\, j_2=\Omega$.}
\begin{equation}
\begin{aligned}
\label{SU2SE}
\xi \lrcorner j_i &= 0\, &&& \xi \lrcorner \eta &= 1    \, , \\[1mm]
j_i \wedge j_j &= 0\ \ {\rm for}\ i\neq j\,, &&& \tfrac{1}{2}\, j_i  \wedge j_j \wedge \eta &= \delta_{ij}\, {\rm vol}\,,
\end{aligned}
\end{equation}
where ${\rm vol}$ is the volume form compatible with the ${\rm SE}_5$ metric
\begin{equation}
\label{SEmetric}
g_{\rm SE_5} = g_{\rm KE}  +  \eta^2 \,,
 \end{equation}
which is taken with canonical normalization $R_{mn}=4g_{mn}$.
Locally this metric describes a fibration over a four-dimensional K\"ahler-Einstein 
base with metric $g_{\rm KE}$. In a neighbourhood, the fibre direction corresponds to the orbit of the vector $\xi$, which is also an isometry and is called the Reeb vector.
In addition the $\SU(2)$ invariant forms satisfy the differential conditions
\be
{\rm d} \eta = 2 j_3\,, \qquad  {\rm d}  j_3 =0\,,   \qquad      {\rm d} (j_1+\ii\, j_2) = 3\, \ii\, \eta \wedge (j_1+\ii\, j_2)\,.
\ee
 The AdS$_5\times {\rm SE}_5$ supersymmetric solution of type IIB supergravity has string frame metric 
\be\label{AdS5xSE5metric}
g_{10} = \ell^2\left( g_{\rm AdS_5} + g_{\rm SE_5}\right)\,,
\ee
where $g_{\rm AdS_5}$ is the unit ${\rm AdS}_5$ metric and $\ell$ sets the overall scale. The solution also contains a non-trivial self-dual five-form flux whose internal part is proportional to the SE$_5$ volume, 
 \be\label{F5_SE}
 F^{\rm fl} = \dd C = \kappa\, {\rm vol} \, ,\qquad\qquad \xi\lrcorner C = 0\,,
 \ee
 where $\kappa$ is a constant related to the overall scale as $\ell^4 = \frac{\kappa}{4} \,\rme^{\phi_0}\,$.\footnote{The parameter $\kappa$ is related to the $N$ units of five-form flux as $\kappa=27\pi N$.}
The second expression in \eqref{F5_SE} is just a convenient gauge choice for the four-form potential.
 
 The consistent truncation was originally constructed by expanding the type IIB supergravity fields in the most general way possible in the basis of $\SU(2)$-singlets given above~\cite{Cassani:2010uw,Gauntlett:2010vu}. 
We now show how this truncation is easily derived from $\Ex{6}$ generalised geometry; 
this will also give the opportunity to illustrate the general statements made in Section~\ref{sec:half_max_str_5d} in a concrete example.
We thus lift the Sasaki--Einstein $\SU(2)$ structure
to the generalised tangent bundle, and take $G_S =  \SU(2) \subset \USp(4)$. Under 
\be\label{USp8toSU2S}
\USp(8) \,\supset\, \USp(4)_R \times \USp(4) \,\supset\, \USp(4)_R\times \U(1) \times \SU(2)_S\ ,
\ee
the spinorial representation decomposes as
\be
{\bf 8} \to {\bf (4,1)} \oplus {\bf (1,4)} \to  {\bf (4,1)} \oplus {\bf (1,2_1)} \oplus {\bf (1,2_{-1})}\,,
\ee
so we have precisely four $G_S$-singlets and the truncation preserves half-maximal supersymmetry. 
In order to count the vector fields in the truncation,  we embed $\SU(2) \sim \SO(3)$  in $\Ex{6}$ 
\be\label{E6toSO3S}
\Ex{6} \,\supset\, \SO(5,5) \times \SO(1,1) \,\supset\,   \SO(1,1) \times \SO(5,2) \times \SO(3)_S \ ,
\ee
and decompose the fundamental representation of $\Ex{6}$, 
\be
{\bf 27} \to {\bf 10}_2 \oplus {\bf 16}_{-1} \oplus {\bf 1}_{-4} \to {\bf (7,1)}_2 \oplus {\bf (1,3)}_2 \oplus {\bf (8,2)}_{-1} \oplus {\bf (1,1)}_{-4}\ .
\ee
We find  8 singlets of  $\SO(3)$, 7   transforming in the fundamental of $\SO(5,2)$  and one neutral. 
This matches the vector field content of half-maximal supergravity coupled to two vector multiplets.

From  \eqref{USp8toSU2S}, \eqref{E6toSO3S}, we see that the scalar manifold of the truncated theory is 
\be\label{MscalIIBonSE}
\mathcal{M}_{\rm scal}\,=\,\frac{\Comm_{\Ex{6}} (\SU(2)_S)}{\Comm_{\USp(8)}(\SU(2)_S)} \, = \, \SO(1,1)\times \frac{\SO(5,2)}{ \SO(5)\times\SO(2)} \,,
\ee
that is the scalar manifold of half-maximal supergravity coupled to two vector multiplets.\footnote{Precisely the same group-theoretical arguments described here were used in \cite{Khavaev:1998fb,Freedman:1999gp} to identify a consistent truncation of maximal $\SO(6)$ supergravity to half-maximal supergravity with two vector multiplets. Although the matter content of the five-dimensional theory is the same, the gauging is different.}

The eight generalised vectors $K_{\mathcal{A}}$, with $\mathcal{A}=\{0,A\}=0,1, \dots, 7$, defining the generalised $\SU(2)$ structure are constructed from the tensors defining the ordinary $\SU(2)$ structure on the Sasaki--Einstein manifold.
For the generalised vectors to contain all the information about the background, we should also include a twist by the four-form $C$ satisfying \eqref{F5_SE},
\be
\label{twistgenvectSE5}
K_{\mathcal{A}} = \rme^{C} \check{K}_{\mathcal{A}} \, ,
\ee
where $\check{K}_{\mathcal{A}}$ denotes the untwisted vectors and the adjoint action of $C$ on a generalised vector is given in~\eqref{twisted_vector}.
We find that the generalised $\SU(2)$ structure is defined by
\begin{align}
\label{SEgenv}
& K_0 =   \xi \ ,\nonumber \\[1mm]
& K_i =   \tfrac{1}{\sqrt{2}}  \, \eta \wedge j_i  \qquad \quad i=1,2,3 \ , \nonumber \\[1mm]
& K_4 =  \tfrac{1}{\sqrt{2}}  \left(   \np \,  \eta  -  \nm\,   {\rm vol}  -  \np\, \eta \wedge C  \right) \, , \nonumber \\[1mm]
& K_5 =  \tfrac{1}{\sqrt{2}} \left(  - \nm\,  \eta  -   \np\,   {\rm vol}  +    \nm\, \eta \wedge C \right) \, , \nonumber \\[1mm]  
& K_6 =  \tfrac{1}{\sqrt{2}}  \left( \np\,   \eta  +  \nm\,  {\rm vol}  -   \np\, \eta \wedge C \right) \, , \nonumber \\[1mm]
& K_7 = \tfrac{1}{\sqrt{2}}  \left( - \nm\,  \eta  + \np\,   {\rm vol} +   \nm\, \eta \wedge C  \right)   \, , 
 \end{align}
where \be\label{SL2states}
\np^\alpha  = {1\choose 0}^\alpha\,,\qquad \nm^\alpha = {0\choose 1}^\alpha
\ee 
are a basis for the $SL(2)$ doublets.
Using  \eqref{cubic_inv} for the cubic invariant, it is straightforward to verify that the 
compatibility  relations \eqref{eq:SO55cond_SO5n}, \eqref{eq:ortho-cond_SO5n} are satisfied, with $n=2$.

We will also need the dual vectors $K^*_{\mathcal{A}}$. 
Evaluating \eqref{K*fromK}, \eqref{Jvol}, we find that these are
\begin{align}
\label{SE_dual_vec}
& K^\ast_0 = \eta \, ,\nonumber \\
& K_i^\ast = -\tfrac{1}{\sqrt{2}}\: \hat j_i \wedge \xi \ , \qquad \quad i=1,2,3\, ,\nonumber \\[1mm]
& K_4^\ast =  \tfrac{1}{\sqrt{2}} \,\big(- \hat{\nm}\, \xi  + \hat{\np}\,  C \lrcorner  \hat{\rm vol}   + \hat{\np}\, \hat{\rm vol}  \big) \, , \nonumber \\[1mm]
& K_5^\ast =  \tfrac{1}{\sqrt{2}}\, \big( \hat{\np}\, \xi + \hat{\nm}\,  C \lrcorner  \hat{\rm vol}   + \hat{\nm}\,  \hat{\rm vol}\big) \, , \nonumber \\[1mm]
& K_6^\ast =  \tfrac{1}{\sqrt{2}}\, \big( \hat{\nm}\, \xi  + \hat{\np}\,  C \lrcorner  \hat{\rm vol}   + \hat{\np}\, \hat{\rm vol}\big) \, , \nonumber \\[1mm]
& K_7^\ast =  \tfrac{1}{\sqrt{2}}\, \big(  -\hat{\np}\, \xi  + \hat{\nm}\,  C \lrcorner  \hat{\rm vol}  + \hat{\nm}\, \hat{\rm vol}\big) \, , 
\end{align}
where  $\hat j_i,$ are the two-vectors dual to the two forms $j_i$, $\hat{\rm vol}$ is the five-vector dual to the volume form, and
\be
\hat{\nm}_\alpha =\epsilon_{\alpha\beta} \,\nm^\beta = {1\choose 0}_\alpha\ ,\qquad \hat{\np}_\alpha = -\epsilon_{\alpha\beta}\, \np^\beta = {0\choose 1}_\alpha\ .
\ee

The gauging  of  the five-dimensional theory is obtained by computing  the generalised Lie derivative between the set of generalised vectors, as in \eqref{genLieKKisXK}. The definition of the type IIB generalised Lie derivative can be found in \eqref{genLieE66second}. We find that the algebra closes into the non-vanishing structure constants
\begin{align}\label{algebraSE5}
&X_{01}{}^2 =- X_{02}{}^1= 3\, ,\nn\\[1mm]
&X_{04}{}^5 =-X_{05}{}^4 =-X_{04}{}^7 = -X_{07}{}^4 = X_{05}{}^6 = X_{06}{}^5 = -X_{06}{}^7 = X_{07}{}^6   = \frac{\kappa}{2} ,\nn \\[1mm]
& X_{34}{}^5= -X_{34}{}^7 =-X_{35}{}^4 = X_{35}{}^6= X_{36}{}^5 = -X_{36}{}^7 = -X_{37}{}^4 =  X_{37}{}^6  =  \sqrt 2\, ,\nn\\[1mm]
&X_{45}{}^3= X_{47}{}^3=  -X_{56}{}^3 = X_{67}{}^3  = \sqrt 2\,,
\end{align}
where the terms in the last two lines are antisymmetric in the lower indices.
From \eqref{rel_X_fxi} we conclude that the embedding tensor components are
\begin{align}
&\xi_{12} = 3\,,\qquad \xi_{45} = \xi_{47}= -\xi_{56} = \xi_{67} = \frac{\kappa}{2} \,,\nn\\[1mm]
& f_{345} = f_{347} = - f_{356}= f_{367} = \sqrt 2\,.\label{embedding_tensor_IIBonSE}
\end{align}
This is fully consistent with the embedding tensor found in \cite{Cassani:2010uw}.\footnote{The precise matching between the embedding tensor components in \eqref{embedding_tensor_IIBonSE} and those in \hbox{\cite[eq.~(4.20)]{Cassani:2010uw}} is obtained upon renaming the indices $(1234567)_{\rm here} = (3451267)_{\rm there}$ (which can be achieved by a trivial $\SO(5)$ transformation), multiplying all components in \eqref{embedding_tensor_IIBonSE} by $-\sqrt{2}$ (which is a harmless rescaling of the gauge group generators) and noticing from comparing the five-form fluxes  that $\kappa_{\rm here} = 2k_{\rm there}$.}
 As discussed there, the corresponding gauge algebra is $\Heis_3 \times \U(1)$, where $\Heis_3$ is the three-dimensional Heisenberg algebra. The remaining four generators, that transform in a non-adjoint representation of the gauge algebra, determine the vector fields that are eaten-up by two-form fields via a St\"uckelberg mechanism.

\subsubsection{Generalised metric}

In order to recover the scalar truncation ansatz we need to construct the generalised metric evaluating the formulae \eqref{Ginv10forSO5n}.
We first derive the generalised metric for the background solution AdS$_5\times$SE$_5$ using \eqref{geninvmet}, since this is simpler and it allows one to see how the construction works.
 Then in the next subsection we will discuss the generalised metric for the dressed generalised vectors, allowing for general $\Sigma$, $\mathcal{V}$, and extract the scalar ansatz.
For simplicity, we also momentarily set  the four-form $C$ to zero, that is we work with the untwisted vectors, and reintroduce it
in a second step.  

Recalling the decomposition \eqref{dual_gen_vector} of the arbitrary dual generalised vector $Z$, we find that $G_0^{-1}$ in \eqref{Ginv10forSO5n} is 
\begin{align}
G_0^{-1}(Z,Z) = (\xi^m \hat{v}_m)^2\,,
\end{align}
while the two terms defining $G_{10}^{-1}$ evaluate to
\begin{equation}
2  \delta^{ab}  \langle Z , K_a  \rangle \langle Z, K_b \rangle =  \frac{1}{4}\sum_{i=1,2,3}\Big(\eta_m j_{i\, np}\,\hat{\rho}^{mnp}\Big)^{2} +  \sum_{\alpha=1,2} (\eta_m \hat{\lambda}^{m}_{\alpha})^2 + \frac{1}{5!}\sum_{\alpha=1,2} (\hat{\sigma}_{\alpha}^{mnpqr})^2\,,
\end{equation}
and
\begin{equation}
\frac{c^*(K_0^*,Z,Z)}{\hat \vol} =  -\frac{1}{12} \, \eta_m \hat\rho^{mnp} \hat\rho^{qrs} \epsilon_{npqrs} + 2\, \epsilon^{\alpha\beta}\,\eta_m \hat\lambda^m_\alpha *\!\hat\sigma_\beta \ .\end{equation}
The term involving the $j_i$ projects $\eta_m \hat\rho^{mnp}$ on its anti-self-dual part on the K\"ahler-Einstein basis, hence it can be written as
\begin{align}
\frac{1}{4}\sum_{i}\Big(\eta_m j_{i\, np}\hat{\rho}^{mnp}\Big)^{2} &=\frac14  \Big(\eta_m\hat{\rho}^{mnp} - \frac12\eta_m\epsilon^{mnpqr}\hat\rho_{qrs}\eta^s\Big)^2 \nonumber\\[1mm]
&=\frac12  (\eta_m\hat{\rho}^{mnp})^2 + \frac{1}{12} \, \eta_m \hat\rho^{mnp} \hat\rho^{qrs} \epsilon_{npqrs}\,.
\end{align}
Adding up the two contributions we obtain 
\begin{align}
G_{10}^{-1}(Z,Z) 
&= \sum_{\alpha=1,2} (\eta_m \hat{\lambda}^{m}_{\alpha})^2  +\frac{1}{2}\,(\eta_{p}\,\hat{\rho}^{pmn})^2 + \frac{1}{5!}\sum_{\alpha=1,2} (\hat{\sigma}_{\alpha}^{mnpqr})^2 \,.
\end{align}
We see that the tensor structure of $G_0^{-1}$ and $G_{10}^{-1}$ is such that at least one index is along the fiber of the Sasaki--Einstein manifold.
It remains to evaluate $G_{16}^{-1}$: as explained in the general discussion of Section \ref{sec:half_max_str_5d}, this is obtained by the recursive Clifford action  of $K_5$, $K_4^*, K_3, K_2^*, K_1$ on a dual vector $Z$, and by  finally pairing up the resulting  vector with $Z$ itself. After a long but relatively straightforward computation, we find
\be
\label{G16inv} 
 G_{16}^{-1}(Z,Z)  =     g^{mn}_{\rm KE}\, \hat{v}_m\hat{v}_n  + \delta^{\alpha\beta} g^{\rm KE}_{mn}\, \hat \lambda^m_\alpha \hat\lambda^m_\beta + \tfrac{1}{6}\,g^{\rm KE}_{mq}g^{\rm KE}_{nr}g^{\rm KE}_{ps}\,\hat{\rho}^{mnp}\hat{\rho}^{qrs} \,.
\ee
Hence $G_{16}^{-1}$ is just a generalised metric on the four-dimensional K\"ahler-Einstein base.
Adding up the three contributions, we arrive at 
\begin{equation}\label{gen_metric_SE_untwisted}
 G^{-1}(Z,Z)  =  g^{mn}\hat v_m \hat v_n + \delta^{\alpha\beta} g_{mn} \hat \lambda^m_\alpha \hat\lambda^m_\beta + \tfrac{1}{6}\,\hat{\rho}^{mnp}\hat{\rho}_{mnp} + \delta^{\alpha\beta} \hat{\sigma}_{\alpha}^{mnpqr}\hat{\sigma}_{\beta,mnpqr}\,,
\end{equation}
where $g_{mn}$ is  the Sasaki--Einstein metric \eqref{SEmetric}, which is also used to lower the curved indices in the last two terms.

The metric associated with the twisted generalised vectors $K_{\mathcal{A}} = \rme^{C}\check K_{\mathcal{A}}$ is easily obtained by recalling that the $\Ex{6}$ cubic invariant is preserved by the twist, 
\begin{equation}
c({\rm e}^C V,{\rm e}^C V',{\rm e}^C V'') = c(V,V',V'')\ .
\end{equation}
This means that the generalised metric with non-trivial four-form potential can be computed using the untwisted $K$'s (\eqref{SEgenv} with $C=0$) and  $\rme^{-C}Z$. Thus,
 to reintroduce $C$,  it is sufficient to consider  \eqref{gen_metric_SE_untwisted} and to make 
the following substitutions
\begin{align}
\hat v &\to \hat v + \hat\rho \lrcorner C\, ,\nonumber\\[1mm]
\hat\lambda_\alpha &\to \hat\lambda_\alpha - C \lrcorner \hat\sigma_\alpha \,.
\end{align}
Comparing the generalised metric and  \eqref{generalised_metric_IIB} with only non-zero  $g_{mn}$ and four-form $C$, we recover the metric and four-form potential  of the AdS$_5\times$SE$_5$ solution of type IIB supergravity.

\subsubsection{Recovering the truncation ansatz}

In \cite{Cassani:2010uw}, the  scalar truncation ansatz based on the Sasaki--Einstein structure is given in the Einstein frame by\footnote{Compared to \cite{Cassani:2010uw} we have renamed $b_1 = {\rm Re} b^\Omega$, $b_2 = {\rm Im} b^\Omega$, $b_3 = b^J$, and similarly for $c_i$.}
\begin{align}\label{scalar_ansatz_IIBonSE}
g_{10} &= \rme^{-\frac{2}{3}(4U+V)} g_{\mu\nu} \diff x^\mu \diff x^\nu + \rme^{2U} g_{\rm KE} + \rme^{2V} \eta^2\nn\\[1mm]
B^+ &=   b_i\, j_i \,,\quad B^- =  c_i\,j_i\ , \qquad C = \Cfl -a \,j_3 \wedge j_3\ ,
\end{align}
where  $\{U,V,b_i,c_i,a\}$, with $i=1,2,3$, together with the axion $C_0$ and the dilaton $\phi$, are eleven scalar fields depending just on the external coordinates, and $\Cfl$ is the background four-form potential that we called $C$ in the previous subsection, satisfying \eqref{F5_SE}. 
 These  eleven scalars parameterise the coset manifold \eqref{MscalIIBonSE}. Specifically, the $\SO(1,1)$ factor is parameterised by the combination $\Sigma = \rme^{-\frac{2}{3}(U+V)}$. For the $\frac{\SO(5,2)}{ \SO(5)\times\SO(2)}$ coset representative, it is convenient to use a solvable parametrization, which is obtained exponentiating the Cartan and positive root generators of the coset. 
The explicit form of  $\{\mathcal{V}_A{}^{b},\mathcal{V}_A{}^{\underline{b}}\}$ (with $b=1,\ldots,5$ and $\underline{b}=1,2$) chosen in \cite{Cassani:2010uw} reads\footnote{Compared to \cite{Cassani:2010uw}, we have renamed the indices $(1234567)_{\rm here} = (3451267)_{\rm there}$ via an $\SO(5)$ transformation.}
\begin{equation}\label{cosetrep}
\!\!\!{\small\left(\begin{matrix} 
1 & 0 & 0 &\rme^{-\frac{\phi_1}{2}}(-c_1+C_0b_1) & -\rme^{-\frac{\phi_2}{2}}b_1 &  \rme^{-\frac{\phi_1}{2}}(c_1-C_0b_1) & \rme^{-\frac{\phi_2}{2}}
b_1 \\[2mm]
0 & 1 & 0 &\rme^{-\frac{\phi_1}{2}}(-c_2+C_0b_2) & -\rme^{-\frac{\phi_2}{2}}b_2 &  \rme^{-\frac{\phi_1}{2}}(c_2-C_0b_2) & \rme^{-\frac{\phi_2}{2}}b_2 \\[2mm]
0 & 0 & 1 &\rme^{-\frac{\phi_1}{2}}(-c_3+C_0b_3) & -\rme^{-\frac{\phi_2}{2}}b_3 &  \rme^{-\frac{\phi_1}{2}}(c_3-C_0b_3) & \rme^{-\frac{\phi_2}{2}}b_3 \\[2mm]
c_1 & c_2 & c_3 & \frac{\rme^{-\frac{\phi_1}{2}}}{2}(\rme^{\phi_1}+\mathfrak{c_-}+ C_0\mathfrak{a}_+) & \frac{  \rme^{\frac{\phi_2}{2}} }{2}C_0 -\frac{\rme^{-\frac{\phi_2}{2}}}{2}\mathfrak{a}_+ &  \frac{\rme^{-\frac{\phi_1}{2}}}{2}(\rme^{\phi_1}-\mathfrak{c}_- -C_0\mathfrak{a}_+ ) & \frac{  \rme^{\frac{\phi_2}{2}}}{2}C_0 + \frac{\rme^{-\frac{\phi_2}{2}}}{2}\mathfrak{a}_+ \\[2mm]
b_1 & b_2 & b_3 & \frac{\rme^{-\frac{\phi_1}{2}}}{2}(\mathfrak{a}_- - C_0\mathfrak{b}_- ) & \frac{\rme^{-\frac{\phi_2}{2}}}{2}(\rme^{\phi_2}+\mathfrak{b}_-) &  \frac{\rme^{-\frac{\phi_1}{2}}}{2}(-\mathfrak{a}_-+ C_0\mathfrak{b}_- ) & \frac{\rme^{-\frac{\phi_2}{2}}}{2}(\rme^{\phi_2}-\mathfrak{b}_-)\\[2mm]
c_1 & c_2 & c_3 & \frac{\rme^{-\frac{\phi_1}{2}}}{2}(\rme^{\phi_1}-\mathfrak{c}_+ + C_0\mathfrak{a}_+)  & \frac{  \rme^{\frac{\phi_2}{2}} }{2}C_0 -\frac{\rme^{-\frac{\phi_2}{2}}}{2}\mathfrak{a}_+ &   \frac{\rme^{-\frac{\phi_1}{2}}}{2}(\rme^{\phi_1}+\mathfrak{c}_+-C_0\mathfrak{a}_+ ) & \frac{  \rme^{\frac{\phi_2}{2}}}{2}C_0 + \frac{\rme^{-\frac{\phi_2}{2}}}{2}\mathfrak{a}_+  \\[2mm]
b_1 & b_2 & b_3 & \frac{\rme^{-\frac{\phi_1}{2}}}{2}(\mathfrak{a}_- + C_0\mathfrak{b}_+ ) & \frac{\rme^{-\frac{\phi_2}{2}}}{2}(\rme^{\phi_2}-\mathfrak{b}_+) &  \frac{\rme^{-\frac{\phi_1}{2}}}{2}(-\mathfrak{a}_-- C_0\mathfrak{b}_+ ) & \frac{\rme^{-\frac{\phi_2}{2}}}{2}(\rme^{\phi_2}+\mathfrak{b}_+)
 \end{matrix}\right)}
\end{equation}
where we defined  $\phi_1 = 4U-\phi$, $\phi_2=4U+\phi$, and
\begin{align}
\mathfrak{a}_+  &=2a+b_ic_i\,,\,\ \quad \mathfrak{a}_-=2a-b_ic_i\,, \nn\\[1mm]
\mathfrak{b}_+  &= 1+b_ib_i\, ,\qquad \mathfrak{b}_- = 1- b_ib_i\,,\nn\\[1mm]
 \mathfrak{c}_+ &= 1+c_ic_i \,,\qquad \mathfrak{c}_- = 1-c_ic_i\,.
\end{align}

Note that the solvable parameterisation has a nice interpretation in terms of $E_{6(6)}$ adjoint action (recall \eqref{eq:IIB_adjoint})  
\be\label{action_sugra_fields}
{\tilde K_a \choose \tilde K_{\underline{a}} }= \rme^{-(B^++B^- + C)} \cdot m \cdot r \cdot \rme^{-l} \cdot  {K_a \choose  K_{\underline{a}} } \,=\, \Sigma^{-1}\, {\mathcal{V}{}_a{}^B \choose \mathcal{V}{}_{\underline{a}}{}^B} K_B \, ,
\ee
where 
\be
B^+ =  b_i \,j_i\,,\qquad B^- =  c_i \, j_i \,, \qquad C = -a\, j_3 \wedge j_3\ ,
\qquad
m^\alpha{}_\beta = \left(\begin{matrix}  \rme^{\frac{\phi}{2}} \,&\, 0 \\ \rme^{\frac{\phi}{2}}C_0 \,&\, \rme^{-\frac{\phi}{2}} \end{matrix}\right) \ ,
\ee
\be
r = {\rm diag}\left(\rme^{V},\,\rme^{U},\,\rme^{U},\,\rme^{U},\,\rme^{U} \right) \ ,\qquad l = \tfrac{1}{3}\,{\rm tr}(r) = \tfrac{1}{3}\left(4U+V\right)\ 
\ee
so that the action is by only supergravity fields, with no need to introduce the poly-vector components in the  $E_{6(6)}$ adjoint.\footnote{The $\GL(5)$ matrix $r$ is given in the basis of vielbeine that makes the metric diagonal. This should not be confused with the $\SL(2)$ doublet $r^\alpha$.}

Having chosen an explicit parameterisation of the coset representative $\mathcal{V}$, we can compute the full generalised metric using formula \eqref{Ginv10forSO5n}. This will depend on the eleven scalars $\{U,V,C_0,\phi,b_i,c_i,a\}$. Comparing the expression obtained in this way with  form \eqref{generalised_metric_IIB} of the generalised metric, we can extract the truncation ansatz for the supergravity fields $g_{mn}$, $C_0$, $\phi$, $B^\alpha_{mn}$, $C_{mnpq}$, as well as the warp factor $\Delta$.\footnote{
A minor subtlety is that the truncation of \cite{Cassani:2010uw} was derived in the Einstein frame of type IIB supergravity, while the generalised metric in \eqref{generalised_metric_IIB} is adapted to the string frame; however \eqref{generalised_metric_IIB} can be turned to the Einstein frame by simply ignoring the explicit factors of $\rme^{-\phi}$ appearing there, and we do so in our computation.}

Although straightforward in principle, the computations are  lengthy  and we just discuss the final result. The warp factor is easily extracted using \eqref{eq:GB-metric}, \eqref{detGB} and reads\footnote{In this case $\Delta $ is not really a warp factor as it is independent of the internal coordinates. It is just a Weyl rescaling setting the external metric in the Einstein frame.}
\begin{equation}
\rme^{2\Delta}=\rme^{-\frac{2}{3}(4U+V)}\ , 
\end{equation}
while the internal metric is given by
\be
 \rme^{-2\Delta} (G^{-1})^{mn}  = g^{mn} = \rme^{-2U} g^{mn}_{\rm KE} + \rme^{-2V}\xi^m\xi^n \, . 
\ee
Proceeding  in a similar way for the other supergravity fields, we recover  precisely the scalar  ansatz \eqref{scalar_ansatz_IIBonSE}.

The ansatz for the  five-dimensional vectors follows  straightforwardly from  \eqref{ansatz-vec}. We construct the linear combination of generalised vectors   
$ \mathcal{A}^\mathcal{A}_\mu K_{\mathcal{A}}$,  
where the coefficients $ \mathcal{A}^\mathcal{A}_\mu$ are vectors in five dimensions, and we equate it to the  generalised vector
 \eqref{def_calA_mu^M}, with the fields $B^\alpha_{\mu,1}$, $C_{\mu,3}$, and $\tilde{B}^\alpha_{\mu,5}$ being defined as in \eqref{redef_one_forms}.  Separating the fields transforming in different representations of $\GL(5)$, we find:
\begin{align}
h_\mu  &=  \mathcal{A}_\mu^0\, \xi \,, \nn\\[1mm]
B^+_{\mu,1} & =   \tfrac{1}{\sqrt{2}}  \left(   \mathcal{A}_\mu^4 +  \mathcal{A}_\mu^6 \right)    \eta\,, \nn\\[1mm]
B^-_{\mu,1} & =  - \tfrac{1}{\sqrt{2}}    \left(  \mathcal{A}_\mu^5  + \mathcal{A}_\mu^7  \right) \eta\,, \nn\\[1mm]
C_{\mu,3} & =  \tfrac{ 1}{\sqrt{2}}  \,
  \mathcal{A}_\mu^i  \, j_i  \wedge \eta \,,\nn\\[1mm]
\tilde{B}^+_{\mu,5} & =   - \tfrac{1}{\sqrt{2}}    \left( \mathcal{A}^5_\mu  -  \mathcal{A}^7_\mu \right)   \vol 
+   \tfrac{1}{\sqrt{2}}   \left( \mathcal{A}_\mu^4 +  \mathcal{A}_\mu^6 \right)   \Cfl \wedge \eta  \,, \nn \\[1mm]
\tilde{B}^-_{\mu,5} & =  -  \tfrac{1}{\sqrt{2}}  \left(\mathcal{A}^4_\mu - \mathcal{A}^6_\mu \right)   \vol 
+  \tfrac{1}{ \sqrt{2} }   \left(\mathcal{A}^5_\mu + \mathcal{A}^7_\mu \right)   \Cfl \wedge \eta\,.
\end{align}   
 
The ansatz for the two-form fields follows from \eqref{ansatz_two-forms}. The weighted dual vectors $J^\mathcal{A}$ can be computed by multiplying the dual vectors $K^*_{\mathcal{A}}$ in \eqref{SE_dual_vec} by the internal volume form as in \eqref{Jvol}. Doing so we find: 
\begin{align}
J^0 & = \vol \otimes\, \eta \,,\nn\\
J^i& =   \tfrac{1}{\sqrt{2}} \, j_i  \,, \nn\\ 
J^4& =  \tfrac{1}{\sqrt{2}}  (-\hat{n}  + \hat{r} \vol_{4} - \hat{n}\, \Cfl ) \, ,\nn \\ 
J^5& =   \tfrac{1}{\sqrt{2}}  (-\hat{r}  - \hat{n} \vol_4 - \hat{r}\, \Cfl ) \,, \nn \\ 
J^6& =  \tfrac{1}{\sqrt{2}}  (\hat{n}  + \hat{r} \vol_4 + \hat{n}\, \Cfl ) \, ,\nn \\ 
J^7& =   \tfrac{1}{\sqrt{2}}  (\hat{r}  - \hat{n} \vol_4 + \hat{r}\, \Cfl ) \, , 
\end{align} 
where we defined ${\rm vol}_4 = \xi \lrcorner \vol$. 
 Equating $\mathcal{B}_{\mu \nu\,\mathcal{A}} J^{\mathcal{A}}{}_{M}$ to the weighted dual vector \eqref{gen_ten_BmunuMN} and separating the terms in different $\GL(5)$ representations, we find 
\begin{align}
B_{\mu \nu, 0\,+}&  =   \tfrac{1}{\sqrt{2}} \left( \mathcal{B}_{\mu \nu\,7}  -\mathcal{B}_{\mu \nu\,5}  \right)\,, \nn \\[1mm]
B_{\mu \nu, 0\,-}&  =   \tfrac{1}{\sqrt{2}} \left(  \mathcal{B}_{\mu \nu\,6} - \mathcal{B}_{\mu \nu\,4} \right)\,, \nn \\[1mm]
C_{\mu \nu,2} & =  \tfrac{1}{\sqrt{2}}   \, \mathcal{B}_{\mu \nu\, i}\, j_i \,, \nn \\[1mm] 
\tilde{B}_{\mu \nu,4\,+} & =  \tfrac{1}{\sqrt{2}} \left( \mathcal{B}_{\mu \nu\,4} + \mathcal{B}_{\mu \nu\,6} \right)  \vol_4  +  \tfrac{1}{\sqrt{2}}  \left( \mathcal{B}_{\mu \nu\,7} - \mathcal{B}_{\mu \nu\,5} \right) \Cfl    \,,\nn\\[1mm]
\tilde{B}_{\mu \nu,4\,-} & =  -\tfrac{1}{\sqrt{2}} \left( \mathcal{B}_{\mu \nu\,5} + \mathcal{B}_{\mu \nu\,7}\right)  \vol_4 +  \tfrac{1}{\sqrt{2}} \, \left( \mathcal{B}_{\mu \nu\,6} - \mathcal{B}_{\mu \nu\,4} \right) \Cfl    \, . 
\end{align} 
The tensor $\tilde g$ associated with the dual graviton would be expanded as $ \mathcal{B}_{\mu\nu\,0} \vol\otimes \eta$, but we will not need this. 
 
This ansatz for the one-form and two-form fields agrees with the one of \cite{Cassani:2010uw}.
We have thus shown how the full bosonic truncation ansatz for type IIB supergravity on Sasaki--Einstein manifolds can be derived from our general approach to half-maximal truncations.

We observe that the particular Sasaki--Einstein manifold given by the $T^{1,1} = \frac{\SU(2)\times\SU(2)}{\U(1)}$ coset space admits a further reduced $\U(1)\subset \SU(2)$ structure. In the generalised geometry, this introduces an additional singlet  vector $K_8 = \eta \wedge \Phi$, where $\Phi$ in the only harmonic two-form in the Sasaki--Einstein metric on $T^{1,1}$. On $T^{1,1}$ one can also twist the generalised tangent bundle by NSNS and RR three-form fluxes proportional to the cohomologically non-trivial three-form $\eta \wedge \Phi$. Following the same steps as above including the extra vector, we would retrieve the larger consistent truncation of  \cite{Cassani:2010na,Bena:2010pr}, yielding half-maximal gauged supergravity coupled to three vector multiplets.

\subsection{Truncations for $\beta$-deformed backgrounds}

It was shown in~\cite{Gauntlett:2007ma} that for any AdS$_5$ solution to type IIB supergravity preserving minimal supersymmetry, and hence dual to an $\mathcal{N}=1$ SCFT$_4$, there is 
a consistent truncation to pure gauged supergravity in five dimensions containing that AdS$_5$ solution.  A class of such backgrounds is provided by the $\beta$-deformation of Lunin and Maldacena \cite{Lunin:2005jy}. 
For the case where the internal manifold is $S^5$, the explicit truncation ansatz of type IIB supergravity on the  $\beta$-deformed geometry to pure gauged supergravity has been given very recently in \cite{Liu:2019cea}. Here we show that if one starts from a toric Sasaki--Einstein manifold, the generalised $\SU(2)$ structure of the $\beta$-deformed background allows for a much larger truncation. The resulting five-dimensional supergravity is in fact just the same half-maximal supergravity with two vector multiplets that arises from type IIB supergravity on squashed Sasaki--Einstein manifolds.  One way to see this is to observe that the full truncation ansatz on toric Sasaki--Einstein manifolds can be $\beta$-deformed.

\subsubsection{The $\beta$-deformed  $T^{1,1}$ background}

In \cite{Lunin:2005jy}, Lunin and Maldacena  showed that,  given an $\mathcal{N}=1$ background with two $\U(1)$ isometries commuting with the $R$-symmetry, a new supersymmetric solution can be obtained by applying a TsT transformation, namely a sequence of T-duality along one of the $\U(1)$, a shift along the second $\U(1)$ and another T-duality along the first one. 
Any toric Sasaki--Einstein manifold can be deformed in this way.  
We will present explicit formulae for the $T^{1,1}$ manifold, however our results apply to any toric Sasaki--Einstein five-manifold.

The canonically normalised Sasaki--Einstein metric on $T^{1,1}$ is
\begin{align}
\label{T11met}
g_{\rm SE_5} \, =\, \frac{1}{6}\sum_{i=1,2} \left( \diff \theta_i^2 + \sin^2\theta_i \diff\phi_i^2  \right) + \frac{1}{9}\left( \diff\psi 
+\cos\theta_1 \diff\phi_1 + \cos\theta_2 \diff\phi_2 \right)^2 \,,
\end{align}
and for the internal part of the four-form potential satisfying \eqref{F5_SE} we choose the gauge
\be
\label{T11C4}
C  = -\frac{\kappa}{108}\, \psi \sin\theta_1\sin\theta_2 \,\diff \theta_1\wedge \diff \phi_1\wedge \diff \theta_2\wedge \diff \phi_2\,. 
\ee
The dilaton is constant and all other fields vanish, $ \phi = \phi_0 = {\rm const}$, $C_0=B^\alpha=0$.\footnote{The axion $C_0$ is set to zero for simplicity, although any constant value would be allowed.}

The $\beta$-deformed solution\footnote{This is a solution for a real deformation $\beta$. The generalisation to a complex deformation is straightforward and amounts to an $SL(2,\bbR)$ rotation.}
 given in \cite{Lunin:2005jy} reads
\begin{align}\label{LMsolutionT11}
g_{10} &= \ell^2 \bigg\{ g_{\rm AdS_5} + \frac{\sin^2\theta_1\sin^2\theta_2}{324 f} \, \diff \psi^2 + \frac{1}{6}\left(\diff \theta_1^2 + \diff\theta_2^2\right) \nn \\[1mm]
\,&\qquad \quad + \mathcal{G} \left[ h \left( \diff \phi_1 + \frac{\cos\theta_1\cos\theta_2}{9h}\diff \phi_2 + \frac{\cos\theta_1}{9h}\diff\psi \right)^2 + \frac{f}{h}\Big( \diff \phi_2 + \frac{\cos\theta_2\sin^2\theta_1}{54f}\,\diff\psi \Big)^2\right] \bigg\}\,,\nn\\[1mm]
\rme^{2\phi} \,&=\, \rme^{2\phi_0} \mathcal{G}\,,\nn\\[1mm]
B^+  &= 2 \gamma \ell^4\, \mathcal{G} f \left( \diff \phi_1 + \frac{\cos\theta_1\cos\theta_2}{9h}\diff \phi_2 + \frac{\cos\theta_1}{9h}\diff\psi \right) \wedge \left( \diff\phi_2 + \frac{\cos\theta_2\sin^2\theta_1}{54f}\,\diff\psi \right)\,,\nn\\[1mm]
B^- &=   \frac{\kappa\,\gamma}{54} \,\cos\theta_1\sin\theta_2\,\diff\theta_2\wedge\diff\psi\,,\nn\\[1mm]
F^{\rm fl} \,&=\, \kappa\, \mathcal{G} \, {\rm vol}_{\rm SE} \,,
\end{align}
where $\ell^4 = \frac{\kappa}{4} \,\rme^{\phi_0}\,$ and $\gamma$ is a real parameter. Moreover one has the functions
\begin{align}\label{defs_LM}
& \mathcal{G}^{-1} = 1+ 4\gamma^2 \ell^4f\,,
 \nn\\[1mm]
& h = \frac{\cos\theta_1^2}{9} + \frac{\sin^2\theta_1}{6}\,,\qquad f= \frac{1}{54}\left( \cos^2\theta_2\sin^2\theta_1 + \cos^2\theta_1\sin^2\theta_2 \right) + \frac{1}{36}\sin^2\theta_1\sin^2\theta_2 \,.
\end{align}

\subsubsection{The $\beta$-deformation in generalised geometry}
\label{betadefgg}

We next show that the type IIB $\beta$-deformation has a very simple interpretation in generalised geometry as the $E_{6(6)}$ action by a bi-vector with components along  the two $U(1)$ isometries commuting with the Reeb vector.\footnote{Similarly, the $\beta$-deformation of AdS$_4$ solutions to M-theory  is generated by a tri-vector in $\Ex{7}$ generalised geometry, see \cite{Ashmore:2018npi}.}  For the $T^{1,1}$ metric \eqref{T11met}, these correspond to the rotations by angles $\phi_1$ and $\phi_2$. Then the $\beta$ deformed solution is generated by the bivector
\be
\label{betadef}
\beta^\alpha\, = \,  { 0 \choose \beta } =  { 0  \choose - 2\gamma\, \partial_{\phi_1}\wedge  \partial_{\phi_2} } \, ,
\ee
where $\gamma$ is a real constant. This acts on a  generalised vector $V= v+ \lambda^\alpha + \rho + \sigma^\alpha$  in the adjoint of  $E_{6(6)}$ as (see \eqref{eq:IIB_adjoint}): 
\begin{align}
\label{E66adjoint_act_beta}
V^\prime  & =  \rme^\beta \cdot V =  V + \beta \cdot V  \nn \\
& = (v -   \epsilon_{\alpha\beta} \beta^\alpha \lrcorner \lambda^\beta ) + (\lambda^\alpha + \beta^\alpha \lrcorner\rho ) + (\rho +  \epsilon_{\alpha\beta} \beta^\alpha \lrcorner \sigma^\beta ) + \sigma^\alpha \,.
\end{align}
In particular it is easy to show that the deformation \eqref{betadef}  maps the  generalised vector $K_\mathcal{A}$,  \eqref{SEgenv}, defining the generalised $SU(2)$ structure
into new generalised vectors 
\begin{align}
\label{betaSEgenv}
& K^\prime_0 =   \xi \ ,\nonumber \\
& K^\prime_i =   \tfrac{1}{\sqrt{2}} \left[  \nm  \beta \lrcorner \left( \eta \wedge j_i \right) +  \eta \wedge j_i \right]   \qquad \quad i=1,2,3 \ , \nonumber \\
& K^\prime_4 =  \tfrac{1}{\sqrt{2}}  \left[  -\beta \lrcorner  \eta    + \np \,  \eta  + \beta \lrcorner \left( \eta \wedge C \right)   -  \nm\,   {\rm vol}  -  \np\, \eta \wedge C  \right] \ , \nonumber \\  
& K^\prime_5 =  \tfrac{1}{\sqrt{2}} \left[ \beta \lrcorner  \eta  - \nm\,  \eta   + \beta \lrcorner  {\rm vol}   -   \np\,   {\rm vol}  +    \nm\, \eta \wedge C \right] \ , \nonumber \\  
& K^\prime_6 =  \tfrac{1}{\sqrt{2}}  \left[ -\beta \lrcorner  \eta +   \np\,     \eta  + \beta \lrcorner \left( \eta \wedge C \right)  +  \nm\,  {\rm vol}  -   \np\, \eta \wedge C \right] \ , \nonumber \\  
& K^\prime_7 = \tfrac{1}{\sqrt{2}}  \left[ \beta \lrcorner  \eta- \nm\,  \eta  -  \beta \lrcorner  {\rm vol} + \np\,   {\rm vol} +   \nm\, \eta \wedge C  \right]   \, , 
 \end{align}
that are still globally defined. Since the new $K_{\mathcal{A}}$, $\mathcal{A}=0,\ldots,7$, are obtained from the original ones by an $\Ex{6}$ transformation, they still satisfy the conditions \eqref{eq:SO55cond_SO5n}, \eqref{eq:ortho-cond_SO5n} with $n=2$, and therefore  define a generalised $\SU(2)$ structure. Moreover, evaluating the generalised Lie derivative between them, one can check that they satisfy exactly the same algebra~\eqref{algebraSE5} as the original generalised vectors associated with the Sasaki--Einstein structure. 

We conclude that there exists a consistent truncation on the $\beta$-deformed geometry, which leads to the very same five-dimensional half-maximal gauged supergravity obtained via reduction on Sasaki--Einstein manifolds.

To compute the algebra for the deformed generalised vectors it is helpful to make an explicit choice of parametrisation for the  
$SU(2)$ structure on $T^{1,1}$. We introduced  the coframe one-forms
\begin{align}
\nonumber e^1 &= \frac{1}{3} \left(\diff \psi + \cos \theta_1 \diff \phi_1 +  \cos \theta_2 \diff \phi_2\right) \,, & 
\nonumber e^2 &= \frac{1}{\sqrt{6}} \left( \cos \tfrac{\psi}{2} \, \sin \theta_1 \diff \phi_1  - \sin \tfrac{\psi}{2}\,   \diff \theta_1 \right)  \, ,\\[1mm]
\nonumber e^3 &=  \frac{1}{\sqrt{6}} \left( \sin \tfrac{\psi}{2} \, \sin \theta_2 \diff \phi_2 +  \cos \tfrac{\psi}{2} \,  \diff \theta_2 \right)  \, ,& 
\nonumber e^4 &=  \frac{1}{\sqrt{6}} \left( \cos \tfrac{\psi}{2} \, \sin \theta_2 \diff \phi_2  -  \sin \tfrac{\psi}{2} \,  \diff \theta_2 \right)  \, ,\\[1mm]
 e^5 &= \frac{1}{\sqrt{6}} \left( \sin \tfrac{\psi}{2} \, \sin \theta_1 \diff \phi_1  + \cos \tfrac{\psi}{2} \,  \diff \theta_1 \right)  \,, 
\label{vielbeine}
\end{align}
such that the Sasaki--Einstein metric~\eqref{T11met} is 
$g_{\rm SE} \,=\, \sum_{a=1}^5 (e^a)^2$, and the $SU(2)$ structure  \eqref{SU2SE}  is given by 
\be
\begin{gathered}
\xi = -3\, \partial_\psi \, , \qquad
 \eta =  -e^1 \, , \\[1mm]
 j_1=  e^{24} + e^{35} \,, \qquad
j_2 =  e^{23} - e^{45} \,, \qquad
 j_3 =  e^{25} -e^{34} \,.
 \label{leftinvforms}
\end{gathered}
\ee
The RR four-form potential satisfying \eqref{F5_SE} can be written as
\be
C = -\tfrac{1}{6}\kappa\, \psi\, j_3  \wedge j_3 \, .
\ee
For completeness we can also list the  $\beta$-deformed generalised dual vectors
\begin{align}
\label{betaSEsgenv}
& K^{\ast \prime}_0 = \eta  - \np \, \beta \lrcorner \eta \, ,\nonumber \\[1mm]
& K_i^{\ast \prime} = -\tfrac{1}{\sqrt{2}} \left(  \hat j_i \wedge \xi -  \beta  \wedge   \hat j_i \wedge \xi \right)  \ , \qquad \quad i=1,2,3\, ,\nonumber \\[1mm]
& K_4^{\ast \prime} =  \tfrac{1}{\sqrt{2}} \left(- \hat{\nm}\, \xi  + \hat{\np}\,  C \lrcorner  \hat{\rm vol}  -\beta \wedge   C \lrcorner  \hat{\rm vol}  + \hat{\np}\, \hat{\rm vol}  \right) \, , \nonumber \\[1mm]
& K_5^{\ast \prime} =  \tfrac{1}{\sqrt{2}} \left( \hat{\np}\, \xi + \hat{\nm}\,  C \lrcorner  \hat{\rm vol}  - \beta \wedge \xi   + \hat{\nm}\,  \hat{\rm vol}\right) \, , \nonumber \\[1mm]
& K_6^{\ast \prime} =  \tfrac{1}{\sqrt{2}} \left( \hat{\nm}\, \xi  + \hat{\np}\,  C \lrcorner  \hat{\rm vol}    - \beta \wedge   C \lrcorner  \hat{\rm vol}   + \hat{\np}\, \hat{\rm vol}\right) \, , \nonumber \\[1mm]
& K_7^{\ast \prime} =  \tfrac{1}{\sqrt{2}} \left(  -\hat{\np}\, \xi  + \hat{\nm}\,  C \lrcorner  \hat{\rm vol} + \beta \wedge \xi  + \hat{\nm}\, \hat{\rm vol}\right) \, . 
\end{align}

As for the Sasaki--Einstein case, the inverse generalised metric is computed by plugging the $\beta$-deformed generalised vectors and their duals in \eqref{geninvmet}. The computation is long but relatively straightforward. Comparing the result with \eqref{generalised_metric_IIB}, we can then extract the supergravity fields describing the $\beta$-deformed solution. We illustrate here the main steps. From \eqref{detGB} one finds that the deformed solution has trivial warp factor
\be
\rme^{\Delta^\prime} \, =\, (\det \mathcal{H})^{-1/20} = 1 \,. 
\ee
 The inverse metric  $(G^{-1})^{mn} = g^{mn}$ reproduces the   metric in \eqref{LMsolutionT11},
\begin{align}\label{internal_metric_LM_simpler_form}
g_5'   =   \frac{1}{6}   \sum_{i=1,2} (  {\rm d} \theta^2_i  +   \mathcal{G}  \sin \theta_i {\rm d} \phi_i^2 )  + \frac{1}{9}     \mathcal{G}
({\rm d}\psi +   \cos \theta_1 {\rm d} \phi_1  +   \cos \theta_2 {\rm d} \phi_2)^2     + \frac{ \gamma^2}{81}      \mathcal{G}   \sin^2 \theta_1 \sin^2  \theta_2 \diff \psi^2 \, ,
 \end{align}
 which we have written in a way that will make the comparison with the truncation ansatz easier.
The relation
\be
B'^\alpha_{mn} =  G_{m[p} (G^{-1})^{p}{}^\alpha_{n]}  \, , 
\ee
gives the  NS and RR two-form potentials
\begin{align}
B^{\prime+} & =   \gamma\, \mathcal{G} \left[ 2 f  \diff \phi_1 \wedge \diff \phi_2 + \tfrac{1}{27} ( \sin^2 \theta_1 \cos \theta_2\, \diff \phi_1  -   \sin^2 \theta_2 \cos \theta_1\, \diff \phi_2)  \wedge \diff \psi \right]   \,,\nn \\[1mm] 
B^{\prime-} &=-  \frac{\kappa\, \gamma}{54} \sin \theta_1 \sin  \theta_2\, \diff \theta_1 \wedge \diff \theta_2 \, . 
\end{align}
While the NS two-form is exactly the same as in \eqref{LMsolutionT11}, the RR two-form  is related to the one of \cite{Lunin:2005jy} by a  gauge transformation 
$ B^-_{{\rm LM} } = B^{\prime-} + \diff \Lambda$ with $\Lambda = -   \frac{\gamma \kappa}{54}  \psi  \cos \theta_1 \sin \theta_2  \diff \theta_2$. 
Next we use the component $(G^{-1})_{m \, n}^{\alpha\,\,\beta} $ in \eqref{generalised_metric_IIB} to extract the axio-dilaton
\be
\rme^{- \phi'} h^{\alpha \beta} =  \frac{1}{5}\, \left[ (G^{-1})^{mn} (G^{-1})_{n \, m}^{\alpha\,\,\beta} + (G^{-1})^{m}{}^\alpha_{n}  (G^{-1})^{n}{}^\beta_{m} \right] 
= \begin{pmatrix} 1 & 0 \\ 0 & \mathcal{G}^{-1} \end{pmatrix} \, .
\ee
From \eqref{habi} we see that $C_0$ is zero (as we set it to zero in the undeformed solution) and the dilaton reproduces the one in  the  solution of \cite{Lunin:2005jy}. 
Finally, from the component $ (G^{-1})^m{}_{npq} $ we find the four-form potential 
 \begin{align}
  C^\prime & = \frac{ \kappa \,\psi}{108}   \mathcal{G} \, \Big[ - \frac{\gamma^2}{54}   \sin \theta_1 \sin \theta_2   \left( \cos \theta_2 \sin^2 \theta_1  \diff \phi_1  - \cos \theta_1 \sin^2 \theta_2 \, \diff \phi_2 \right) \wedge  \diff \theta_1 \wedge \diff \theta_2 \wedge \diff \psi  \nn \\[1mm]
 & \quad \qquad \quad  +    \left(1+ 2 \gamma^2 f \right) \diff \theta_1 \wedge \diff \theta_2 \wedge  \diff \phi_1 \wedge \diff \phi_2  \Big]\,,
 \end{align}
which gives the five-form of \cite{Lunin:2005jy},
\be
F'_5 \,\equiv\, \diff C' + \frac{1}{2}\left(  B'^+\wedge \diff B'^- - B'^-\wedge \diff B'^+\right)\,=\, \mathcal{G}\, \kappa \vol_{\rm SE}\,,
\ee
where $\vol_{\rm SE}$ is the $T^{1,1}$ volume form in the undeformed solution.

An equivalent way to compute the generalised metric for the deformed background is to act with a $\beta$-deformation on the generalised metric of the Sasaki--Einstein solution.  We consider 
 the action of a nilpotent bivector,  $\beta \wedge \beta =0$. This is not the most general bivector deformation, but it is enough to describe the $\beta$-deformation of Lunin and Maldacena. The transformed metric is
\be
G'^{-1} \,=\, \rme^\beta \cdot G^{-1} \cdot \rme^{-\beta} =  G^{-1} +  \beta \cdot  G^{-1} -  G^{-1}  \cdot \beta  - \beta \cdot   G^{-1} \cdot \beta  \, .
\ee 
For the purpose of extracting the type IIB supergravity fields, we will only need the following components of the $\beta$-transformed generalised metric 
\begin{align} 
\label{compgenmet}
(G'^{-1})^{mn} \,&=\,  (G^{-1})^{mn} - \beta_\alpha^{mp} (G^{-1})_{p}^{\alpha\, n} + (G^{-1})^{m\,}{}^\gamma_{p}\beta^{pn}_\gamma  - \beta_\alpha^{mp} (G^{-1})_{p\,q}^{\alpha\,\gamma}\beta^{qn}_\gamma  \,,\nn \\[1mm]
(G'^{-1})^m{}^\gamma_n \,&=\, (G^{-1})^m{}^\gamma_n + \tfrac{1}{2} (G^{-1})^m{}_{npq}\beta^{\gamma\,pq} - \beta_\alpha^{mp}(G^{-1})_{p\,n}^{\alpha\,\gamma}  \,,\nn \\[1mm]
(G'^{-1})^{\alpha\,\gamma}_{m\,n} \,&=\,  (G^{-1})^{\alpha\,\gamma}_{m\,n} + \tfrac{1}{2}\beta^{\alpha\,pq}(G^{-1})_{mpq\,}{}_n^\gamma + \tfrac{1}{2}(G^{-1})_{m\,}^\alpha{}_{npq}\beta^{\gamma\,pq}  + \tfrac{1}{4}\beta^{\alpha\,pq}(G^{-1})_{mpq\,nrs}\beta^{\gamma\,rs}    \,,\nn \\[1mm]
(G'^{-1})^m{}_{npq} \,&=\, (G^{-1})^m{}_{npq} - \beta_\alpha^{rs} (G^{-1})^{m\,}{}^\alpha_{npqrs}\,.
\end{align}
Plugging in the formulae above the metric, dilaton and four-form potential of the AdS$_5 \times T^{1,1}$ solution and the form \eqref{betadef} of the bivector $\beta$, we recover exactly the expressions for the different fields of the $\beta$-deformed solution discussed above.

In the specific example of $T^{1,1}$,  one can also include  the $\beta$ deformation of the generalised vector  $K_8$ introduced in the previous section 
\be
 K^\prime_8 = \Phi \wedge \eta +  \beta\lrcorner(\eta \wedge \Phi )\, n \, . 
\ee
The $\beta$-transformed vector  should still preserve the algebra and, after also introducing three-form fluxes, the corresponding  enhanced truncation contains the $\beta$-transformed Klebanov-Strassler solution discussed in the Appendix of~\cite{Lunin:2005jy}.

\subsubsection{The truncation ansatz}

The truncation ansatz for the vectors  is obtained substituting in \eqref{ansatz-vec} the generalised vectors defining the generalised $SU(2)$ structure on the beta-deformed  $T^{1,1}$ are given in \eqref{betaSEgenv}, \eqref{betaSEsgenv}
\begin{align}
h_\mu  &=   \mathcal{A}_\mu^0\, \xi  - \tfrac{1}{\sqrt{2}}  \left(  \mathcal{A}_\mu^4 +   \mathcal{A}_\mu^6\right)\,  \beta\lrcorner \eta  \, , \nn\\[1mm]
B^+_{\mu,1} & =   \tfrac{1}{\sqrt{2}}  \left(  \mathcal{A}_\mu^4  +  \mathcal{A}_\mu^6  \right)   \eta   \, , \nn\\[1mm]
B^-_{\mu,1} & =    -\tfrac{1}{\sqrt{2}}  \left(   \mathcal{A}_\mu^5 +  \mathcal{A}_\mu^7 \right)   \eta  +    \tfrac{1}{\sqrt{2}}\,    
 \mathcal{A}_\mu^i\, \beta \lrcorner \left(j_i\wedge \eta\right)  \, , \nn\\[1mm]
C_{\mu,3} & =  \tfrac{1}{\sqrt{2}} \,
   \mathcal{A}_\mu^i \,  j_i \wedge \eta   - \tfrac{1}{\sqrt{2}} \left( \mathcal{A}_\mu^4 +  \mathcal{A}_\mu^6\right)   \beta \lrcorner  \big(\eta \wedge \Cfl\big) 
+  \tfrac{1}{\sqrt{2}}  \left( \mathcal{A}_\mu^5-  \mathcal{A}_\mu^7\right)   \beta \lrcorner    \vol  \, ,  \nn\\[1mm]
\tilde{B}^+_{\mu,5} & =   
-  \tfrac{1}{\sqrt{2}}  \left(   \mathcal{A}_\mu^5 -  \mathcal{A}_\mu^7 \right) {\rm vol} 
 -  \tfrac{1}{\sqrt{2}}  \left( \mathcal{A}_\mu^4 +  \mathcal{A}_\mu^6  \right) \Cfl \wedge \eta \,,\nn\\[1mm]
 \tilde{B}^-_{\mu,5} & =   
-  \tfrac{1}{\sqrt{2}}  \left(   \mathcal{A}_\mu^4-  \mathcal{A}_\mu^6 \right)  {\rm vol} 
 +  \tfrac{1}{\sqrt{2}}  \left(  \mathcal{A}_\mu^5 +  \mathcal{A}_\mu^7 \right) \Cfl \wedge \eta 
\,  .
\end{align}   
To give the  ansatz for the two-forms one has to compute the tensors $J^\mathcal{A}$ in the bundle  $N \simeq \det T^*M \otimes E^*$. As for the Sasaki--Einstein truncation, these are obtained acting on the dual generalised vectors $K^*$ with the internal volume, as in \eqref{Jvol},
\begin{align}
J^0 & = \eta \otimes\vol + \, \hat{r}\,   \beta \lrcorner  ( \eta \otimes\vol)  \, , \nn  \\[1mm]
J^i& =   \tfrac{1}{\sqrt{2}}  \left( \hat{r}  \,  \beta \lrcorner j_i  +   j_i  \right)  \,,\nn \\[1mm]
J^4& =     \tfrac{1}{\sqrt{2}} \left( - \hat{n}  + \beta \lrcorner \Cfl  + \hat{r} \,{\rm vol}_4  -  \hat{n}\,  \Cfl \right) \, ,  \nn  \\[1mm] 
J^5& =     \tfrac{1}{\sqrt{2}} \left( - \hat{r}  + \beta \lrcorner {\rm vol}_4   - \hat{n}\,{\rm vol}_4 -  \hat{r}\,  \Cfl \right)  \, , \nn  \\[1mm] 
J^6& =     \tfrac{1}{\sqrt{2}} \left(  \hat{n}  -  \beta \lrcorner \Cfl  + \hat{r} \,{\rm vol}_4  +  \hat{n}\,  \Cfl \right) \, ,  \nn  \\[1mm] 
J^7& =     \tfrac{1}{\sqrt{2}} \left(  \hat{r}  + \beta \lrcorner  {\rm vol}_4   - \hat{n}\,{\rm vol}_4 +  \hat{r}\,  \Cfl \right)  \, , 
\end{align} 
where again we are using ${\rm vol}_4 = \xi \lrcorner \vol$. 
Then equating the components of the generalised tensor  \eqref{gen_ten_BmunuMN} with   the linear combination  $\mathcal{B}_{\mu\nu\, \mathcal{A} } J^{\mathcal{A}}$, we find
\begin{align}
B_{\mu\nu,0 \,+} & =  \tfrac{1}{\sqrt{2}} \left( \mathcal{B}_{\mu\nu\, 7} -  \mathcal{B}_{\mu\nu\, 5}  \right)  - \tfrac{1}{\sqrt{2}}    
 \left( \mathcal{B}_{\mu\nu \,1}  \,\beta \lrcorner j_1 + \mathcal{B}_{\mu\nu \,2}\,  \beta \lrcorner j_2  \right) \, ,  \nn \\[1mm] 
B_{\mu\nu,0 \,-} & =  \tfrac{1}{\sqrt{2}} \left(  \mathcal{B}_{\mu\nu\, 6} -  \mathcal{B}_{\mu\nu\, 4} \right)   \, ,  \nn \\[1mm] 
C_{\mu\nu ,\,2 }&=   \tfrac{1}{\sqrt{2}} \big[  
 \mathcal{B}_{\mu\nu\, i}  \, j_i  +  \big(  \mathcal{B}_{\mu\nu\, 4}  - \mathcal{B}_{\mu\nu\, 6} \big)  \beta \lrcorner \Cfl + 
 \big(  \mathcal{B}_{\mu\nu\, 5} +  \mathcal{B}_{\mu\nu\, 7}  \big)   \beta \lrcorner {\rm vol}_4 \big]  \, , \nn\\[1mm]
 \tilde{B}_{\mu\nu, 4\,+}  & =      -  \mathcal{B}_{\mu\nu\, 0} \,  \beta \lrcorner (  \eta \otimes \vol )   +   \tfrac{1}{\sqrt{2}}  \left( \mathcal{B}_{\mu\nu\, 4} +  \mathcal{B}_{\mu\nu\, 6} \right) {\rm vol}_4 + 
   \tfrac{1}{\sqrt{2}}  \left( \mathcal{B}_{\mu\nu\, 7} -  \mathcal{B}_{\mu\nu\, 5} 
 \right)  \Cfl     \,,\nn\\[1mm]
 \tilde{B}_{\mu\nu, 4\,-}  & =        -   \tfrac{1}{\sqrt{2}}  \left(  
  \mathcal{B}_{\mu\nu\, 5} +  \mathcal{B}_{\mu\nu\, 7} \right)    {\rm vol}_4 + 
   \tfrac{1}{\sqrt{2}}  \left(  \mathcal{B}_{\mu\nu\, 6} -  \mathcal{B}_{\mu\nu\, 4} \right)  \Cfl     \, .
\end{align}  

The generalised metric contains the ansatz for the internal fields, metric and form potential, the dilaton and the warp factor. Here we give the final result for the internal and mixed components of the ten-dimensional metric \eqref{KK_decomp_metr}:
\begin{align}
g_{mn}Dy^mDy^n  &=  \mathcal{F}^{-1}  \Big[ \tfrac{1}{3}\, \rme^{2 A_1 }   \left( \gamma f_1  \,\diff  \theta_1 + \gamma f_2 \,  \diff \theta_2 + f_0  D  \psi\right)^2  +
 \tfrac{1}{3}\, \rme^{A_1} E_1 D  \psi^2 \nn\\[1mm]
&\qquad  +  \tfrac{1}{6}\, \rme^{A_2} E_2  \left(\diff \theta_1^2 +\diff \theta_2^2 \right)  + \tfrac{9}{2} \,\rme^{A_1 + A_2}    \left( \sin^2 \theta_1D  \phi_1^2 + \sin^2 \theta_2  D \phi_2^2 \right)  \nn \\[1mm]
&\qquad + \tfrac{1}{3}\,\rme^{2 A_1 }   \left( 3 \cos \theta_1D  \phi_1 + 3 \cos \theta_2 D \phi_2 + f_0 D \psi\right)^2   \nn \\[1mm]
&\qquad  +   \gamma\,   \rme^{A_1} \sin^2\theta_1\sin^2\theta_2\,  \Big(\sum_{i=1,2}  g_{i}  \sin \theta_i  D  \phi_i\,\diff\theta_1 +  \sum_{i=1,2} h_{i}  \sin \theta_i D  \phi_i\,  \diff\theta_2 \Big)    \Big]\,,
\end{align}  
where  the differentials $D$ contain the shift by the five-dimensional vectors
\begin{align} 
D \psi  &= \diff \psi  + 3 \mathcal{A}^0  \, , \nn \\
D \phi_1  &= \diff \phi_1  +\tfrac{\sqrt{2}}{3}  \gamma \cos \theta_2 (  \mathcal{A}^4 + \mathcal{A}^6)  \, ,   \nn \\
D \phi_2  &= \diff \phi_2  - \tfrac{\sqrt{2}}{3}  \gamma \cos \theta_1 (  \mathcal{A}^4 + \mathcal{A}^6)   \, . 
\end{align}
For simplicity of notation we defined  $A_1 = \frac{8}{3} (U+V)$, $A_2 = \frac{2}{3} (7 U+V)$, $A_3 = \frac{1}{3} (3 \phi  -8 U+ 4V)$ as well as  the functions
\begin{align} 
b^{+}_{12} &= b_1 \cos \psi + b_2 \sin \psi  \,, \nn  \\[1mm]
b^{-}_{12} &= b_2 \cos \psi - b_1 \sin \psi  \,, \nn  \\[1mm]
 f_0   & = 3 - \gamma\,  b_{12}^+   \sin \theta_1  \sin \theta_2  \,, \nn  \\[1mm]
 f_1 & =    b_{12}^- \cos \theta_1\sin \theta_2 + b_3 \sin \theta_1 \cos \theta_2 \,, \nn \\[1mm]
 f_2 & =    b_{12}^- \sin \theta_1\cos \theta_2 - b_3 \cos \theta_1 \sin \theta_2\,,  \nn \\[1mm] 
 \mathcal{F}  &=   3 \, \rme^{A_1}  f_0^2  +  \gamma^2  \rme^{A_2+A_3}  \left[  2\, \rme^{A_1}  \left(\cos^2 \theta_1\sin^2\theta_2 +  \sin^2\theta_1\cos^2 \theta_2 \right) +3\, \sin^2 \theta_1  \sin^2 \theta_2\, \rme^{ A_2} \right]\,,
\end{align}  
and 
\begin{align} 
g_{1} & = ( 3\, \rme^{A_2}   + 2\,  \rme^{A_1} \cot^2 \theta_1)   b_{12}^-   +  2 \,\rme^{A_1}  b_3  \cot \theta_1  \cot \theta_2 \,,     \nn \\[1mm]
g_{2}  & = ( 3\, \rme^{A_2}   + 2\,  \rme^{A_1} \cot^2 \theta_2)   b_{3} +  2 \,\rme^{A_1}  b_{12}^-  \cot \theta_1  \cot \theta_2 \,,     \nn \\[1mm]
h_{1} & =  ( 3\, \rme^{A_2}   + 2\,  \rme^{A_1} \cot^2 \theta_1)   b_{3} -  2 \,\rme^{A_1}  b_{12}^-  \cot \theta_1  \cot \theta_2 \,,     \nn \\[1mm]
h_{2}  & = ( 3\, \rme^{A_2}   + 2\,  \rme^{A_1} \cot^2 \theta_2)   b_{12}^-   -  2 \,\rme^{A_1}  b_3  \cot \theta_1  \cot \theta_2 \,,     \nn \\[1mm]
E_1 & = \rme^{ 2 A_2 + A_3} \gamma^2  \sin^2 \theta_1  \sin^2 \theta_2  -\rme^{A_1}\, f_0^2   \,,  \nn \\[1mm]
E_2 & =   \mathcal{F} + 3 \gamma^2   \rme^{A_1} \left(  (b^-_{12} )^2 + b_3^2 \right)\sin^2 \theta_1  \sin^2 \theta_2 \,.
\end{align}  
%

In line with the results of \cite{Gauntlett:2007ma}, there exists a subtruncation to minimal five-dimensional gauged supergravity, that has recently been made explicit in \cite{Liu:2019cea}.
The bosonic sector of minimal five-dimensional  supergravity is made of the metric, a single vector (the graviphoton) and no scalars.  It is obtained from  the truncation derived here, by 
setting all two-form and scalar fields to zero except for $\rme^U=\rme^V= \ell$, taking
\be
 \mathcal{A}^1 =  \mathcal{A}^2 =  \mathcal{A}^4 =  \mathcal{A}^5 =  \mathcal{A}^6 =  \mathcal{A}^7 = 0 \,,
\ee
and identifying the other two gauge fields with the graviphoton $A$ as
\be
A =  3  \mathcal{A}^0 =  -  \mathcal{A}^3 \, . 
\ee
In this case it is easy to see that the generalised metric is the same as for the background solution, so that the internal fields are not modified.    
The ansatz for the full ten-dimensional metric becomes
\begin{align}
\label{10dmmin}
g_{10}  & \,= \,    g_{\mu\nu} \,\dd{x}^\mu\dd{x}^\nu  
+      \frac{1}{6}  \left(\diff \theta_1^2 +\diff \theta_2^2 \right) +  \frac{\mathcal{G}}{6}     \left(   \sin^2 \theta_1   \diff  \phi_1^2 +    \sin^2 \theta_2   \diff  \phi_2^2\right)   \nn \\
&\ + \frac{ \mathcal{G} }{9}  
\left({\rm d}\psi +   \cos \theta_1 {\rm d} \phi_1  +   \cos \theta_2 {\rm d} \phi_2 + 3\mathcal{A}^0 \right)^2   
+ \frac{ \gamma^2}{81}   \mathcal{G}   \sin^2 \theta_1 \sin^2  \theta_2  \left(\diff \psi + 3 \mathcal{A}^0 \right)^2 \,,
\end{align}
where we have set $\ell=1$. Note that the purely internal part coincides with~\eqref{internal_metric_LM_simpler_form}. 

\section{M-theory truncations including a Maldacena-N\'u\~nez AdS$_{\mathbf{5}}$ solution}\label{sec:MNsection}

In this section, we construct a generalised $\U(1)\subset \USp(4)$ structure on a manifold $M_6$ given by a fibration of $S^4$ over $\mathbb{\Sigma}$, where $\mathbb{\Sigma}$ is a constant curvature Riemann surface. Specifically, $\mathbb{\Sigma}$ can be the hyperbolic plane $H^2$, the flat space $\mathbb{R}^2$, a sphere $S^2$, or a quotient thereof. We argue that in each case the generalised structure provides a consistent truncation to five-dimensional half-maximal gauged supergravity coupled to three vector multiplets and with a $\U(1)\times \mathit{ISO}(3)$ gauging. The embedding tensor depends on the curvature of $\mathbb{\Sigma}$. When $\mathbb{\Sigma}$ is negatively curved, there is a fully supersymmetric AdS$_5$ solution which uplifts to the AdS$_5\times_{\rm w} M_6$ solution of \cite{Maldacena:2000mw} preserving 16 supercharges.\footnote{The symbol $\times_{\rm w}$ denotes the warped product.} This describes the low-energy limit of M5 branes wrapped on $\mathbb{\Sigma}$, which is an $\mathcal{N}=2$ SCFT$_4$, and our truncation captures some deformations of such theory.

Generic AdS$_5\times_{\rm w} M_6$ solutions of eleven-dimensional supergravity preserving half-maximal supersymmetry were classified in \cite{Lin:2004nb}.  It was shown in \cite{Gauntlett:2007sm} that for all such  solutions, there is a consistent truncation to pure half-maximal supergravity with $\U(1)\times\SU(2)$ gauging, such that the supersymmetric AdS$_5$ vacuum uplifts to the AdS$_5\times_{\rm w} M_6$ solution. In  Section~\ref{sec:gen-formalism} we discussed how this statement follows from restricting to the singlet sector of the $\USp(4)$ generalised structure on $M_6$. The results of this section show that, at least for the specific $M_6$ geometry of \cite{Maldacena:2000mw}, the generalised structure is further reduced to $\U(1)$ and correspondingly the truncation can be enlarged to half-maximal supergravity with three vector multiplets.

We note that the existence of such a consistent truncation, as well as a detailed analysis of its sub-truncations and vacua, was very recently proven using a complementary approach in~\cite{Cheung:2019pge}. These authors considered an explicit truncation directly from seven-dimensional maximal gauged supergravity. As we will see, the generalised structure we find is indeed built using the generalised parallelisation on $S^4$ that defines the seven-dimensional maximal gauged supergravity, thus giving a direct connection to the construction in~\cite{Cheung:2019pge}. 


\subsection{$E_{6(6)}$ generalised geometry for M-theory}

We start by recalling some basic notions of $E_{6(6)}$ generalised geometry for M-theory, which is relevant for dimensional reductions of eleven-dimensional supergravity on a six-dimensional manifold $M$. Again we follow the conventions of \cite[app.~E]{Ashmore:2015joa}.

Under $\GL(6)$, the exceptional tangent bundle on $M$ decomposes as:
\begin{equation}
   E \,\simeq \, TM \oplus \Lambda^2T^*M \oplus \Lambda^5T^*M \, ,
\end{equation}
so that a generalised vector reads
\begin{equation}
V = v + \omega + \sigma\,,
\end{equation}
where $v \in TM$, $\omega\in \Lambda^2 T^*M$ and $\sigma\in\Lambda^5T^*M$.
The $E_{6(6)}$ cubic invariant is defined as\footnote{This is 6 times the cubic invariant given in \cite{Ashmore:2015joa}.}
\begin{equation}
c(V,V,V) = - \, 6\,\iota_v \omega \wedge \sigma -  \omega\wedge \omega\wedge\omega\,.
\end{equation}
The bundle $N\simeq \det T^*M \otimes E^*$ similarly decomposes as:
\be
N \simeq T^*M \oplus \Lambda^4T^*M \oplus (T^*M \otimes \Lambda^6T^*M)\,,
\ee
so the sections are the sum of a one-form, a four-form and a tensor made of the product of a one-form and a volume form.

The eleven-dimensional supergravity fields, that is the metric $g_{11}$, the three-form potential $\hat{A}$ and its six-form dual $\hat{\tilde A}$, can be decomposed according to the $\SO(1,10) \to \SO(1,4)\times \SO(6)$ splitting of the Lorentz group similarly to the discussion in Subsection~\ref{sec:E66geometry_maintext} for type IIB supergravity:
\begin{align}
\label{11d_metric_general}
g_{11}  &= \rme^{2\Delta}\, g_{\mu\nu} \,\dd{x}^\mu\dd{x}^\nu + g_{mn} Dy^m Dy^n\ ,\nn \\[1mm]
\hat{A} &= \tfrac{1}{3!} A_{m_1m_2m_3} D{y}^{m_1m_2m_3} + \tfrac{1}{2}{A}_{\mu m_1m_2} \dd{x}^\mu\wedge D{y}^{m_1m_2} + \tfrac{1}{2}{A}_{\mu\nu m}\dd{x}^{\mu\nu} \wedge Dy^m + \ldots \, ,\nn\\[1mm]
\hat{\tilde{A}} &= \tfrac{1}{6!} \tilde{A}_{m_1\ldots m_6} Dy^{m_1\ldots m_6} + \tfrac{1}{5!} {\tilde A}_{\mu m_1\ldots m_5} \dd{x}^\mu \!\wedge\! Dy^{m_1\ldots m_5} \nn\\
& \quad +  \tfrac{1}{2\cdot 4!} {\tilde A}_{\mu\nu m_1\ldots m_4} \dd{x}^{\mu\nu} \!\wedge\! Dy^{m_1\ldots m_4} + \ldots  ,
\end{align}
where  $Dy^m \,=\, \dd{y}^m - h_\mu{}^m \dd{x}^\mu$ ensures covariance under internal diffeomorphisms,
and $\Delta(x,y)$ is the warp factor of the external metric $g_{\mu\nu}(x)$.
We can organise the eleven-dimensional supergravity fields into the inverse generalised metric on $M$\footnote{The precise expression for the inverse generalised metric in terms of the eleven-dimensional supergravity fields is easily obtained from the conformal split frame given in \cite{Coimbra:2011ky}.}
\be
G^{MN} \ \leftrightarrow\ \{\Delta,\,g_{mn}, \, A_{m_1m_2m_3}, \, \tilde A_{m_1\ldots m_6}  \}\ ,
\ee
the generalised vectors
\be
\mathcal{A}_\mu{}^M = \{ h_\mu{}^m ,\, A_{\mu mn} , \tilde{A}_{\mu m_1\ldots m_5}\,     \}\,,
\ee
and the weighted dual vectors
\be
\mathcal{B}_{\mu\nu\,M} =\{  A_{\mu\nu m}, \,   \tilde{A}_{\mu\nu m_1\ldots m_4}  ,\, \tilde g_{\mu\nu m_1\ldots m_6,n} \} \,,
\ee
where as in type IIB we will not need the last term, related to the dual graviton. The bosonic truncation ansatz is obtained by equating these generalised geometry objects to the corresponding terms given in Section~\ref{sec:half_max_str_5d}.

\subsection{Generalised $\U(1)$ structure}

The internal geometry of the half-maximal AdS$_5\times_{\rm w} M_6$ solution of \cite{Maldacena:2000mw} is constructed as a fibration of $S^4$ over $\mathbb{\Sigma}$, where $\mathbb{\Sigma}$ is a negatively curved Riemann surface. This $M_6$ has a $U(1)\subset\GL(6)$ structure in conventional geometry. As we will see below, this defines a  consistent truncation to half-maximal supergravity coupled to three vector multiplets.
 Explicitly under the embedding $\SO(2)\times\SO(5,3)\subset\SO(5,5)\subset\Ex{6}$  of~\eqref{SO5dec}, the generalised tangent space $E$ decomposes as
\begin{equation}
   \rep{27} = \rep{1}_0 + \rep{8}^v_0 + \rep{1}_+ + \rep{1}_- + \rep{8}^s_+ + \rep{8}^s_- \,,
\end{equation}
where $\rep{8}^v$ and $\rep{8}^s$ are vector and spinor representations and the subscript denotes the $\SO(2)\simeq U(1)$ charge. Thus we have nine singlets under $\U(1)$, which correspond to the generalised vectors $K_{\mathcal{A}}$, $\mathcal{A}=0,\ldots,8$. Under $\SO(5)\times\SO(3)$ these decompose as 
\begin{equation}
   \begin{aligned}
   \rep{1} + \rep{8}^v \ &= \repp{1}{1} + \quad\ \,\repp{5}{1}\; \quad \ +\;\  \quad \repp{1}{3} \, , \\[1mm]
      \Gamma(E) \ &\ni \ \ K_0\ \,\cup \{K_1,\dots,K_5\} \cup \{K_6,K_7,K_8\}\, .
\end{aligned}
\end{equation}

The explicit form of these vectors is determined by the $S^4$ fibration structure of the $M_6$ geometry. To see how they arise, we will first consider the direct product $\mathbb{\Sigma}\times S^4$ and recall some generalised geometry on $S^4$, and then implement the twist of $S^4$ over $\mathbb{\Sigma}$. On $\mathbb{\Sigma}\times S^4$ we can decomposes the generalised tangent space under $\GL(2,\bbR)\times\SL(5,\bbR)\subset\Ex{6}$ where $\GL(2,\bbR)$ is the structure group of the conventional tangent space on $\mathbb{\Sigma}$ and $\SL(5,\bbR)\simeq\Ex{4}$ is the structure group of the generalised tangent space on $S^4$. Explicitly we have 
\begin{equation}
\label{eq:SL5-decomp}
\begin{aligned}
   E &\simeq T\mathbb{\Sigma} \oplus (T^*\mathbb{\Sigma} \otimes N_4)
      \oplus (\Lambda^2 T^*\mathbb{\Sigma} \otimes N_4') \oplus E_4\,, \\
   \rep{27} &= \repp{2}{1} \oplus \repp{2}{5'} \oplus \repp{1}{5}
         \oplus \repp{1}{10} \,, 
\end{aligned}
\end{equation}
where in the second line we denote the $\SL(2,\bbR)\times\SL(5,\bbR)$ representations, and where we have introduced the generalised bundles on $S^4$ 
\begin{equation}
\begin{aligned}
   E_4 &\,\simeq\,  TS^4 \oplus \Lambda^2T^*S^4\,, \\
   N_4 &\,\simeq\, T^*S^4 \oplus \Lambda^4 T^*S^4\,, \\
   N_4' &\,\simeq\, \mathbb{R} \oplus \Lambda^3 T^*S^4\,,
\end{aligned}
\end{equation}
$E_4$ being the generalised tangent space on $S^4$. 

As discussed in~\cite{Lee:2014mla}, on $S^4$ these bundles are parallelisable, that is, they admit global frames, constructed as follows. Let us parameterise the round four-sphere $S^4$ of radius $R$ with coordinates $R\, y^i$, $i=1,\ldots,5$, constrained by the condition $\delta_{ij}y^i y^j=1$. The metric and the volume form induced from $\mathbb{R}^5$ are
\begin{equation}
\label{metricvolS4_ycoords}
g_{4} = R^2 \,\delta_{ij} \dd y^i \dd y^j\,,\qquad
\vol_{4} = \tfrac{1}{4!}R^4\,\epsilon_{i_1i_2i_3i_4 i_5} \,y^{i_1} \dd y^{i_2}\wedge \dd y^{i_3}\wedge \dd y^{i_4} \wedge \dd y^{i_5}\,.
\end{equation}
We can define the generalised frames
\begin{equation}
\label{frames_S4}
\begin{aligned}
   E_{ij} &= v_{ij} + R^2*_4\! (\dd y_i\wedge\dd y_j) + \iota_{v_{ij}} A  
      \qquad\in\, \Gamma(E_4) \,, \\
   E_i &= R\,\dd y_i - y_i \vol_{4} + R\,\dd y_i\wedge A 
      \qquad\quad\ \; \in\, \Gamma(N_4) \,, \\
   E'_i &= y_i + R\, *_4\! \dd y_i + y_i A 
     \qquad\qquad\qquad\; \  \in\, \Gamma(N_4')\, , 
\end{aligned}
\end{equation}
where $v_{ij}\in \Gamma(TS^4)$ are the Killing vector fields generating the $\SO(5)$ isometries, the Hodge star $*_4$ is computed using \eqref{metricvolS4_ycoords},  and the M-theory three-form $A$ is  chosen such that 
\be\label{flux_MN_background}
F = \dd A = 3R^{-1} \vol_{4}\,.
\ee
The frames~\eqref{frames_S4} are globally-defined and therefore parallelise the respective bundles. Furthermore, under the generalised Lie derivative, the $E_{ij}$ generate an $\so(5)$ algebra
\begin{equation}
   L_{E_{ij}} E_{kl} = - R^{-1} \left(
      \delta_{ik}E_{jl} -\delta_{il}E_{jk} + \delta_{jl} E_{ik}
      - \delta_{jk} E_{il} \right) \, . 
\end{equation}
This parallelisation is the basis of the generalised Scherk--Schwarz reduction of eleven-dimensional supergravity on $S^4$ \cite{Lee:2014mla}, which reproduces the well-known consistent truncation to maximal $\SO(5)$ supergravity in seven dimensions \cite{Nastase:1999kf}. In the generalised Scherk--Schwarz reduction, the $E_{ij}$ define the truncation ansatz for the seven-dimensional scalar and vector fields, while the $E_i$ and $E_i'$ define the ansatz for the two-form and three-form potentials.

In the solutions of~\cite{Maldacena:2000mw}, the internal space is a fibration
\begin{equation}
\label{eq:S4-fibration}
   \begin{tikzcd} 
      S^4 \arrow[r,"i"] & M_6 \arrow[d,"\pi"] \\
      & \mathbb{\Sigma}
   \end{tikzcd} 
\end{equation}
where topologically the sphere is twisted by a $U(1)$ subgroup of the $\SO(5)$ isometry group. Here $\mathbb{\Sigma}$ can be a negatively curved Riemann surface as in~\cite{Maldacena:2000mw}, but we can also allow it to be a torus $T^2$, or a sphere $S^2$. Let the one-forms $e_1,e_2$ be an orthonormal co-frame such that
\be\label{gSigmavolSigma}
g_{\mathbb{\Sigma}} = (e_1)^2+(e_2)^2\,,\qquad \vol_\mathbb{\Sigma} = e_1\wedge e_2
\ee
are the constant curvature metric and compatible volume form on $\mathbb{\Sigma}$,
all of which can be pulled back to $M_6$ using the projection map $\pi$. The twisting of the co-tangent space $T^*\mathbb{\Sigma}$ defines a $\U(1)$ spin-connection $\upsilon$ on $\mathbb{\Sigma}$ given by 
\begin{equation}
   \label{d_frame_Sigma}
\dd (e_1+\ii\,e_2) = \ii\, \upsilon \wedge (e_1+\ii\,e_2)\,, \qquad \dd \upsilon = \frac{\kappa}{R^2} \,\vol_\mathbb{\Sigma}\,,
\end{equation}
where $\kappa =-1$ for $H^2$, $\kappa =0$ for $\mathbb{R}^2$ and $\kappa = +1$ for $S^2$ (and quotients thereof), and for convenience we are identifying the overall scale of $\mathbb{\Sigma}$ with the radius $R$ of $S^4$. To preserve supersymmetry one needs to choose the $U(1)$ twisting of the sphere so that it cancels the $U(1)$ twisting of the cotangent space. For the half-maximal case one can choose conventions such that the twisting is the $U(1)$ generated by, for example, the $v_{12}$ Killing vector that appears in generalised frame $E_{12}$. In terms of the embedding in $\Ex{6}$ we thus have the breaking pattern
\begin{equation}
   \Ex{6} \supset \SL(2,\bbR)\times\SL(5,\bbR)
   \supset \SO(2) \times \SO(5)
   \supset \SO(2) \times \SO(2)\times\SO(3)
   \supset U(1) \, , 
\end{equation}
where the final $U(1)$ is the diagonal subgroup of $\SO(2)\times\SO(2)\simeq U(1)^2$. By calculating the commutants one can  see that this structure indeed corresponds to the case of half-maximal supersymmetry with $n=3$ vector multiplets, as claimed. 

Having identified the $U(1)$ structure we can now directly construct the singlet vectors in the generalised tangent space. Given the decomposition~\eqref{eq:SL5-decomp}, we note that these should come from $E_{12}\in\Gs{E_4}$, $E_{\alpha\beta}\in\Gs{E_4}$ and $\vol_{\mathbb{\Sigma}}\wedge E'_\alpha\in\Gs{\Lambda^2T^*\mathbb{\Sigma}\otimes N'_4}$ with $\alpha,\beta\in\{3,4,5\}$, since these are neutral under the  $U(1)$ action generated by $v_{12}$. In addition we get a complex generalised vector of the form $(e_1+\ii e_2)\wedge(E_1+\ii E_2)\in\Gs{T^*\mathbb{\Sigma}\otimes N_4}$ since the twisting means that the $U(1)$ action on the first term is cancelled by the $U(1)$ action on the second term. Concretely we find the nine globally defined generalised vectors $K_{\mathcal{A}}$ on $M_6$ with $\mathcal{A}=0,\ldots,8$: 
\begin{equation}
\label{gen_vec_S4xSigma}
\begin{aligned}
   K_0 &= \tfrac{1}{2}\,\me^\Upsilon \cdot E_{12} \,, \\[1mm]
   K_1 +\ii K_2 &= \left(e_1+\ii e_2\right) \wedge
       \,\me^\Upsilon\cdot \left(E_1 +\ii E_2 \right)\,, \\[1mm]
       K_\alpha &= \tfrac12 \epsilon_{\alpha\beta\gamma}
          \,\me^\Upsilon\cdot E_{\beta\gamma}
      + \vol_\mathbb{\Sigma}\wedge \; \me^\Upsilon\cdot E'_\alpha \,, \\[1mm]
      K_{3+\alpha} &=  \tfrac12 \epsilon_{\alpha\beta\gamma}
          \,\me^\Upsilon\cdot E_{\beta\gamma}
      - \vol_\mathbb{\Sigma}\wedge \; \me^\Upsilon\cdot E'_\alpha \,,
\end{aligned}
\end{equation}
with $\alpha=3,4,5$ and $\epsilon_{\alpha\beta\gamma}$ the antisymmetric symbol such that $\epsilon_{345}=1$. Note that each of the frame vectors on $S^4$ is twisted by the exponentiated $\SO(5)$ adjoint action (defined in \cite[app.~E.1]{Ashmore:2015joa}) of an element of the $\mathfrak{e}_{6(6)}\oplus \mathbb{R}$ algebra:\footnote{This is reminiscent of the construction in the context of $O(d,d)$ generalised geometry in \cite{Andriot:2009fp}.}
\begin{equation}
\label{twist_connection}
\begin{aligned}
\Upsilon &= -R\, \upsilon \times_{\rm ad} E_{12} \\[1mm]
 &=  -R\left[  v_{12} \otimes \upsilon +  \upsilon \wedge *_4 (R^2 \,\dd y_1\wedge\dd y_2) + \upsilon \wedge \iota_{v_{12}} A \right]\,,
\end{aligned}
\end{equation}
where the tensor product $\times_{\rm ad}$ contains a projection on the adjoint representation and $\upsilon$ is the spin-connection on $\mathbb{\Sigma}$. Concretely one finds
\begin{equation}
\label{twistedKs}
\begin{aligned}
   \me^\Upsilon\cdot E_{12} &= E_{12} \,, \\
   \me^\Upsilon\cdot ( E_1 + \ii E_2 )
      &= (E_1 + \ii E_2) + \ii R\, \upsilon
      \wedge (E_1'+\ii\, E_2')\,, \\
   \tfrac12 \epsilon_{\alpha\beta\gamma}
      \,\me^\Upsilon\cdot E_{\beta\gamma}
      &= \tfrac12 \epsilon_{\alpha\beta\gamma} E_{\beta\gamma}
      + R\, \upsilon \wedge  E_\alpha\,,\\
   \,\me^\Upsilon\cdot E'_\alpha &= E'_\alpha + \tfrac{1}{2}R^2\,\epsilon_{\alpha\beta\gamma}\, \upsilon \wedge  \diff y_\beta \wedge \diff y_\gamma\,.
\end{aligned}
\end{equation}
Note that the last term in the fourth line drops out when  wedged with $\vol_\mathbb{\Sigma}$ in $K_\alpha$ and $K_{3+\alpha}$ above.
One can check that these $K_{\mathcal{A}}$ do satisfy the conditions \eqref{eq:SO55cond_SO5n}, \eqref{eq:ortho-cond_SO5n} for a generalised $\U(1)$ structure, where $K_1,\dots K_5$ are the negative norm vectors transforming in the fundamental representation of $\SO(5)$, while the $K_{3+\alpha}$ are the positive norm ones forming an $\SO(3)$ triplet. Since the frame vectors on $S^4$ have been twisted by the same element $\Upsilon$ of $\Ex{6}$, one can actually check the \eqref{eq:SO55cond_SO5n}, \eqref{eq:ortho-cond_SO5n} using the untwisted basis. In particular, the twisting implies that, since $\dd E_i=0$ and $\dd E_i'=\frac{1}{R}\, E_i$,
\begin{equation}
\label{property_twisting_MN2}
\dd \left[\me^\Upsilon\cdot (E_1 + \ii \, E_2)\right]
    = - \ii \,\upsilon \wedge \me^\Upsilon\cdot (E_1+ \ii\, E_2)
     + \vol_\mathbb{\Sigma} \wedge (\ldots) \ ,
\end{equation}
which just cancels the exterior derivative of $e_1+\ii e_2$ in~\eqref{d_frame_Sigma} giving
\begin{equation}
\label{closure_K1K2}
   \dd(K_1+\ii \, K_2)=0\,.
\end{equation}

The reason for the twisting by $\Upsilon$ is straightforward. Given the fibration~\eqref{eq:S4-fibration}, although vectors on $S^4$ push forward via the inclusion map $i:S^4\to M_6$, we need a choice of $U(1)$ connection in order to push forward forms on $S^4$ to globally defined forms on $M_6$. If $\psi$ is a coordinate on $S^5$ such that $v_{12}=R^{-1}\der/\der\psi$ this means replacing $\dd\psi$ in any form on $S^4$  with $\dd\psi+\upsilon$. This is exactly what the action of the first term in~\eqref{twist_connection} does. However, in  the seven-dimensional consistent truncation on $S^4$ the $U(1)$ gauging actually comes from $E_{12}$ not just the leading isometry term $v_{12}$. Thus to match with the construction in~\cite{Maldacena:2000mw}, we should actually twist by the connection in~\eqref{twist_connection}, where the effect of the extra terms is to turn on additional $F$ flux.  This is the generalised geometry counterpart of the topological twist of the M5-brane $(2,0)$ theory on~$\mathbb{\Sigma}$. Our construction should also make it clear that the truncation we are going to define can equivalently be seen as a truncation of seven-dimensional maximal supergravity on~$\mathbb{\Sigma}$.

\subsection{The gauge algebra and the embedding tensor}

We now compute the algebra generated by the twisted generalised vectors $K_{\mathcal{A}}$ in \eqref{twistedKs}.
 The M-theory generalised Lie derivative on $M_6$ is defined as:
\begin{align}
L_V V' \, &= \, \mathcal{L}_v v' + \left(\mathcal{L}_v \omega' - \iota_{v'}\dd\omega 
\right) + \left( \mathcal{L}_v \sigma' - \iota_{v'}\dd\sigma 
 - \omega' \wedge \dd\omega  
 \right)\,.
\end{align}
In order to perform the computations, we find it convenient to use a parameterisation of the generalised vectors in terms of angular coordinates on $S^4$. This is given in Appendix~\ref{app:angular_coords_M6}.
We find that the only non-vanishing generalised Lie derivatives are:
\begin{equation}
L_{K_{0}} (K_1+\ii\, K_2) \ = \  \tfrac{\ii}{2R}\, (K_1+\ii\, K_2) \ ,
\end{equation}
where we crucially used \eqref{closure_K1K2},
and
\begin{align}
L_{R_\alpha} R_{\beta} \, = \, -\tfrac{1}{R}\epsilon_{\alpha\beta\gamma}R_\gamma\, , \qquad L_{R_\alpha} T_{\beta} \, = \, L_{T_\alpha} R_{\beta}\, = \, -\tfrac{1}{R}\epsilon_{\alpha\beta\gamma}T_\gamma\, , \qquad L_{T_\alpha}T_\beta \, = \, 0\, , 
\end{align}
where we introduced the combinations
\begin{align}
R_\alpha \ &:=\ \tfrac12\left[(1-\kappa)K_\alpha+ (1+\kappa)K_{3+\alpha}\right] 
\ , \nonumber\\[1mm]
T_\alpha \ &:=\ \tfrac12 \left[  K_\alpha-K_{3+\alpha}   \right]\,,\qquad \alpha=1,2,3\,.  
\end{align}
It follows that $K_0$ generates a $U(1)$ under which $K_1+\ii K_2$ is charged, and $R_\alpha,T_\alpha$ generate the $\mathit{ISO}(3)$ algebra, with $R_\alpha$ generating the $\SO(3)$ rotations and $T_\alpha$ generating the $\mathbb R^3$ translations. As is apparent from the form of the $R_\alpha$, the way the $\SO(3)$ subgroup of $\mathit{ISO}(3)$ is embedded in $\SO(5,3)$ depends on the value of $\kappa$. If $\kappa=-1$ then $\SO(3)\subset\SO(5)\subset\SO(5,3)$, if $\kappa=0$ then $\SO(3)$ is the diagonal subgroup of $\SO(3,3)\subset\SO(5,3)$, and if $\kappa=+1$ then $\SO(3)$ is the commutant of  $\SO(5)$ in $\SO(5,3)$.

Since all generalised Lie derivatives yield a combination of the $K_\mathcal{A}$ with constant coefficients, the consistent truncation will go through, giving half-maximal gauged supergravity in five dimensions coupled to three vector multiplets. Recalling \eqref{genLieKKisXK}, \eqref{rel_X_fxi}, we can determine the embedding tensor. 
 We find that the non-trivial embedding tensor components are:
\begin{equation}
\label{emb_tensor_MN}
\begin{aligned}
&\xi_{12} = -\frac{1}{2R} \,,\\[1mm]
&f_{\alpha\beta\gamma} = -\frac{3+\kappa}{2R}\,\epsilon_{\alpha\beta\gamma}\,,\qquad\,\ \qquad f_{\alpha\beta(\gamma+3)} = -\frac{1+\kappa}{2R}\,\epsilon_{\alpha\beta\gamma}\,, \\[1mm]
&f_{\alpha(\beta+3)(\gamma+3)} = \frac{1-\kappa}{2R}\,\epsilon_{\alpha\beta\gamma}\,,\qquad 
f_{(\alpha+3)(\beta+3)(\gamma+3)} = \frac{3-\kappa}{2R}\,\epsilon_{\alpha\beta\gamma}\,.
\end{aligned}
\end{equation}
We note that these indeed agree with the embedding tensor derived in~\cite{Cheung:2019pge}.

When $\kappa=-1$, the gauging satisfies the conditions for a half-maximal AdS$_5$ vacuum spelled out in \cite{Louis:2015dca}. This supersymmetric AdS$_5$ vacuum uplifts to the AdS$_5\times_{\rm w} M_6$ solution of \cite{Maldacena:2000mw}. 
In \cite{Bobev:2018sgr} the general conditions for five-dimensional half-maximal supergravity to admit supersymmetric flows between AdS fixed points were given.
Inspection of the gauging \eqref{emb_tensor_MN} shows that the consistent truncation cannot admit such a flow, the basic reason being that the way $S^4$ is twisted over the Riemann surface is fixed in our truncation ansatz. It follows that the truncation cannot describe a flow from the AdS$_5$ vacuum preserving 16 supercharges to another supersymmetric vacuum. Nevertheless, it may contain other interesting solutions that it might be worth exploring.

\subsection{Recovering the truncation to pure half-maximal supergravity}

The general formulae of Section~\ref{sec:E6_trunc_ansatz} provide an algorithmic construction of the full bosonic truncation ansatz for eleven-dimensional supergravity on $M_6$, leading to the five-dimensional half-maximal supergravity coupled to three vector multiplets described above. In the following we make this completely explicit for a sub-truncation of that theory: we recover the truncation to pure half-maximal supergravity given in  \cite{Gauntlett:2007sm}. 
This is only possible when $\mathbb{\Sigma}$ is a negatively curved Riemann surface. Indeed in order to be able to throw away the three vector multiplets consistently and be left with just the gravity multiplet we need 
the gauge algebra to close on the first six generalised vectors, $K_0,\ldots, K_5$, so that we have a $\USp(4)$ generalisd structure with singlet torsion. From \eqref{emb_tensor_MN} we see that this requires $\kappa=-1$. The gauging thus obtained is $\SU(2)\times U(1)$ and the half-maximal supergravity is the one dubbed $\mathcal{N}=4^+$ in \cite{Romans:1985ps}.

In order to determine how the only scalar field $\Sigma$ of pure half-maximal supergravity embeds in the eleven-dimensional fields we evaluate the inverse generalised metric \eqref{Ginv10forSO5n}, where we set $\mathcal{V}_a{}^A=\delta_a{}^A$ as we are now truncating all other scalar fields. In particular, from
\begin{align}\label{inv_gen_metricMN}
(G^{-1})^{mn}\,&= \rme^{2\Delta} g^{mn}\,,\nn\\[1mm]
(G^{-1})^{m}{}_{np}  &= \rme^{2\Delta}\,g^{mq} A_{qnp}\,,
\end{align}
we can extract the internal metric and the internal part of the three-form potential, after having computed the warp factor $\Delta$. The latter is given by the general formula \cite{Coimbra:2012af}
\begin{equation}
{\rm vol}_G \equiv (\det G_{MN})^{-\frac{9-d}{2\,{\rm dim}\,E}} = \sqrt{\det g_{mn}}\,\rme^{(9-d)\Delta}\,,
\end{equation}
where we need to take $d=6$ and ${\rm dim}\,E = 27$.\footnote{We correct a typo in footnote 3 of \cite{Coimbra:2012af}: $\det G$ appearing there should actually be $(\det G)^{1/2}$.} Equivalently we can write:
\be\label{gen_formula_Delta}
\rme^{9\Delta} = (\det \,G^{-1\,MN})^{\frac{1}{18}}(\det G^{-1\,mn})^{\frac{1}{2}} \,.
\ee

We explicitly evaluate the inverse generalised metric and express it in terms of the  $M_6$ coordinates introduced in Appendix~\ref{app:angular_coords_M6}.
In this way we find that \eqref{gen_formula_Delta} gives for the warp factor:
\be
\rme^{6\Delta} = \bar\Delta\,,
\ee
where we introduced the function
\be
\bar\Delta = \cos^2\theta + \frac{\Sigma^3}{2\sqrt 2} \sin^2\theta\,.
\ee
%
%
Inverting $(G^{-1})^{mn}$, we obtain the internal metric
$g_{mn} = \rme^{2\Delta} G_{mn} $,
which reads
\be
g_6 = R^2\bar\Delta^{1/3}\bigg [ \frac{\sqrt2}{\Sigma} \left( \diff \theta^2 + g_{\mathbb{\Sigma}} \right) + \frac{\sqrt2}{\Sigma \bar\Delta} \sin^2\theta\left(\diff\psi+\upsilon\right)^2 + \frac{\Sigma^2}{2 \bar\Delta}\cos^2\theta \:g_{S^2} \bigg]\,,
\ee
where $g_{\mathbb{\Sigma}}$ is the uniform metric on $\mathbb{\Sigma}$ and $g_{S^2}$ is the unit metric on the 2-sphere inside $S^4$.
The second line of \eqref{inv_gen_metricMN} gives for the internal part of the three-form potential:
\begin{align}
A &=\frac{R^3}{2 \bar{\Delta}} \cos^3\theta   \, \left[ 
  -2 \bar{\Delta} \, \upsilon +
   \left(\frac{\Sigma^{3}}{\sqrt2} -2 \right) \sin^2
   \theta   \, (\dd\psi + \upsilon) -6\,\psi \,\tan \theta\,
   \bar{\Delta}\,  \dd\theta 
   \right]  \wedge \vol_{S^2}\,,
\end{align}
whose field strength is
\begin{align}\label{internal_4form_MN}
\diff A &=\frac{R^3 \cos^3\theta }{2 \, \bar{\Delta}^2} \left[ 
 \left( \big( \sqrt2\, \Sigma^{3} + 2\bar{\Delta}\big)\tan\theta\,\dd\theta   + \tfrac{1}{\sqrt2}\sin^2\theta\,\dd \big(\Sigma^{3}\big) \right)\! \wedge(\dd\psi+\upsilon)
 + \frac{2\bar{\Delta}}{R^2}  \vol_{\mathbb{\Sigma}}    \right]\!\wedge \vol_{S^2}\! .
\end{align}
In this way we have obtained the embedding of the five-dimensional scalar $\Sigma$ into the eleven-dimensional supergravity fields. We note that the value of $\Sigma$ giving the AdS$_5\times_{\rm w}M_6$ solution of~\cite{Maldacena:2000mw} is $\Sigma=2^{1/6}$.

 We can go on and use our general formulae to determine the embedding of the five-dimensional vector and two-form fields.
For the mixed components of the eleven-dimensional metric we get
\be
h_\mu{}^m = \tfrac{1}{2}\mathcal{A}^0_\mu\, v^m_{12} +  \tfrac{1}{2} \,\epsilon_{\alpha\beta\gamma} \mathcal{A}_\mu^\alpha \, v^m_{\beta\gamma} \,, 
\ee
where we recall that $\alpha=3,4,5$. Then using \eqref{11d_metric_general} we  reconstruct the full eleven-dimensional metric:
\begin{align}\label{11d_metric_MN}
g_{11} \! &=   \bar{\Delta}^{1/3} \,g_5  + R^{2}\bar{\Delta}^{1/3}\left[\frac{\sqrt2}{\Sigma} (\dd \theta^{2} + g_{\mathbb{\Sigma}})+ \frac{\sqrt2}{\Sigma\bar{\Delta}}\sin^2\theta\,(\dd\psi+\upsilon - \tfrac{1}{2}\mathcal{A}^0)^2 + \frac{\Sigma^2}{2\bar{\Delta}}\cos^2\theta\, \hat g_{S^2} \right]
\end{align}
where $\mathcal{A}^0$ gauges the shifts of the angle $\psi$, while
$\hat g_{S^2}$ denotes the metric on $S^2$ where the $\SO(3)$ isometries are gauged by $\mathcal{A}^3,\mathcal{A}^4,\mathcal{A}^5$. When $S^2$ is described by constrained coordinates such that $\mu^\alpha\mu^\alpha =1$, this reads 
\be
\hat g_{S^2} = D\mu^\alpha D\mu^\alpha\,,
\ee
 with
\be
D\mu^\alpha= \dd\mu^\alpha -\tfrac{1}{2} \,\epsilon_{\beta\gamma\delta} \mathcal{A}^\delta \, v_{\beta\gamma}^\alpha\,,
\ee
$v_{\beta\gamma}^\alpha$ being the $S^2$ Killing vectors $v_{45}, v_{53}, v_{34}$ expressed in the $\mu^\alpha$ coordinates.

In order to determine the remaining part of the three-form potential we compute
\begin{align}\label{pieces_11d_3form}
 \mathcal{A}^{\mathcal{A}} \wedge K_{\mathcal{A}}|_2 &=   \tfrac{1}{2} R^2 \cos^3\theta\,  \mathcal{A}^0 \wedge\vol_{S^2}  \nn\\[1mm]
 &\quad + {\rm Re} \left[ R\,\rme^{\ii \,\psi}(\mathcal{A}^1- \ii \mathcal{A}^2)\wedge   \left(e_1+\ii \, e_2 \right)\wedge \left(\cos\theta \,\dd \theta + \ii \,\sin\theta \,(\dd \psi + \upsilon)\right) \right]    \nn \\[1mm]
& \quad+ \mathcal{A}^\alpha\wedge\left[  - R^2\, \dd (\cos\theta \,\mu^\alpha)\wedge (\dd\psi+\upsilon) + R^2\, \dd\left[ \mu^\alpha\, \dd\left(\psi\, \cos^3\theta\right) \right]+ \cos\theta\, \mu^\alpha \vol_{\mathbb{\Sigma}}  \right]  ,\nn\\[1mm] 
 \mathcal{B}_{\mathcal{A}}\wedge J^{\mathcal{A}}|_1 
&=\frac{R}{2} \left[   {\rm Re}\left( \ii\,(\mathcal{B}_{1}-\ii\,\mathcal{B}_{2}) \wedge  (e_1+\ii\, e_2)\,\sin\theta\,\rme^{\ii\, \psi}   \right) + \mathcal{B}_{\alpha}\wedge \dd (\cos\theta\, \mu^\alpha) \right]\,,
\end{align}
where $K_{\mathcal{A}}|_2$ and $J^{\mathcal{A}}|_1$ denote the 2-form and 1-form parts of 
$K_{\mathcal{A}}$ and $J^{\mathcal{A}}$, respectively (cf.~Appendix~\ref{app:angular_coords_M6}).
Then the full eleven-dimensional three-form potential is
\be\label{full_11d_3form}
\hat A = A +  \mathcal{A}^{\mathcal{A}} \wedge K_{\mathcal{A}}|_2 +  \mathcal{B}_{\mathcal{A}}\wedge J^{\mathcal{A}}|_1 \,,
\ee
where we also need to implement the shifts $\dd\psi \to \dd\psi - \tfrac{1}{2}\mathcal{A}^0$ and $\dd\mu^\alpha\to D\mu^\alpha$ so as to achieve covariance under internal diffeomorphisms.

We can now compare with the consistent truncation ansatz given in \cite{Gauntlett:2007sm}. To this extent, we redefine our scalar $\Sigma$ into the scalar $X$ appearing there as
$\Sigma = 2^{1/6}X^{-1},$ and fix the scale of $M_6$ as $R=1$. 
Then the eleven-dimensional metric \eqref{11d_metric_MN} precisely  matches the one given in \cite[eq.$\:$(3.1)]{Gauntlett:2007sm}\footnote{
After making the obvious identifications of the supergravity gauge fields and of the connection one-form on $\mathbb{\Sigma}$, as well as a trivial, constant rescaling of the external metric.}.
We also checked that the eleven-dimensional four-form field strength matches the corresponding one given in \cite{Gauntlett:2007sm} after the five-dimensional one-form and two-form potentials are set to zero.
Checking agreement of the remaining part of the eleven-dimensional four-form requires a little further work. Indeed our four-form, being constructed from the three-form potential, automatically satisfies the Bianchi identity, while the Bianchi identity of the four-form given in \cite{Gauntlett:2007sm} is not automatic and defines part of the the lower-dimensional equations of motion. Moreover in the embedding tensor formalism adopted in this paper one keeps the vector fields as propagating degrees of freedom, while the two-form potentials are auxiliary, non-propagating fields introduced just to ensure closure of the gauge algebra; on the other hand, in \cite{Gauntlett:2007sm} two of the six vector fields in the half-maximal gravity multiplet are dualised into propagating two-forms and do not appear in the five-dimensional Lagrangian. 
These two descriptions are related by dualisation of some of the fields.\footnote{See also \cite[Sect.$\:$3.2]{Cassani:2011fu} for a discussion of the procedure leading to select the relevant degrees of freedom from dual pairs in a related context.}
One starts from the on-shell duality between the eleven-dimensional three-form  and six-form potentials $\hat A$ and $\hat{\tilde A}$,
\be
*_{11} \diff \hat A + \tfrac{1}{2}\,\hat A \wedge\diff \hat A = \diff \hat{\tilde A}\,.
\ee
Plugging our truncation ansatz in, this yields a set of duality relations between five-dimensional fields, in particular between one- and two-form potentials. Using these relations we can trade some of the fields appearing in our three-form potential for those appearing in the dual six-form. In particular, it is possible to remove the $\diff \mathcal{B}_\alpha$, $\alpha=3,4,5$, from $\diff\hat A$ and replace them by $*\,\dd \mathcal{A}^\alpha$
(the reason being that in the expression for $\hat A$ given by \eqref{pieces_11d_3form}, \eqref{full_11d_3form}, $\mathcal{B}_\alpha$ wedges a closed one-form, implying that in $\dd \hat A$ only $\dd\mathcal{B}_\alpha$, and not $\mathcal{B}_\alpha$, appears).   
On the other hand, in $\dd\hat{A}$ the two-forms $\mathcal{B}_{1}, \mathcal{B}_{2}$ are St\"uckelberg-coupled to the one-forms $\mathcal{A}^1,\mathcal{A}^2$ as $\diff (\mathcal{A}^1-\ii\mathcal{A}^2)-\frac{\ii}{2}(\mathcal{B}_1-\ii\, \mathcal{B}_2)$
and cannot be removed. If desired, one could instead dualise $\mathcal{A}^1-\ii\mathcal{A}^2$ into $\mathcal{B}_1-\ii\, \mathcal{B}_2$ so that the latter becomes propagating in the five-dimensional theory, matching in this way the description of \cite{Gauntlett:2007sm}.


\section{Conclusions}\label{sec:Conclusions}

In this paper we have discussed how generalised geometry provides a formalism to understand consistent truncations of string and M-theory preserving varying amounts of supersymmetry, including non-supersymmetric cases.

When the generalised structure group $G_S$ is just the identity, and the generalised intrinsic torsion is a $G_S$-singlet, one has a generalised Leibniz parallelisation \cite{Lee:2014mla} and can perform a generalised Scherk--Schwarz reduction; this is a consistent truncation preserving maximal supersymmetry. When instead $G_S$ is non-trivial, and the intrinsic torsion is still a $G_S$-singlet, one obtains a consistent truncation preserving only a fraction of supersymmetry. As we discussed in Section~\ref{sec:gen-formalism}, the matter content of the reduced theory is obtained by evaluating the commutant of $G_S$ in $\Ex{d}$ in the relevant representations, while the gauging follows from the algebra of  $G_S$-singlets under generalised diffeomorphisms. In this way the lower-dimensional theory is completely determined. Our formalism is completely general, extending to less intuitive examples than the case where the consistent truncation is based on an ordinary $G_S$-structure. For instance we can allow for a non-trivial warp factor, or use generalised tensor fields whose fixed-rank components can vanish at points on the internal manifold, but the full generalised tensor is nowhere vanishing. 

After illustrating the general principles, in Section~\ref{sec:half_max_str_5d} we have discussed in detail truncations to five dimensions preserving half-maximal supersymmetry. These are based on $\SO(5-n) \subseteq \USp(4) \subset \Ex{6}$ structures. In this case, the generalised structure is entirely characterised by a set of generalised vectors $K_{\mathcal{A}}$, $\mathcal{A}=0,1,\ldots,5+n$, and the truncation contains $n$ vector multiplets. The sub-algebra of generalised diffeomorphisms generated by the $K_{\mathcal{A}}$ determines the gauging of the five-dimensional supergravity. We have given an algorithmic prescription to construct the full bosonic truncation ansatz. In particular, we provided an expression for the generalised metric on the internal manifold in terms of the $K_{\mathcal{A}}$, and using this we specified the scalar field ansatz for the truncated theory. 

We gave evidence for two new consistent truncations preserving half-maximal supersymmetry: the first is obtained from type IIB supergravity on $\beta$-deformed toric Sasaki--Einstein five-manifolds, and the second from eleven-dimensional supergravity on half-maximal Maldacena-N\'u\~nez geometries \cite{Maldacena:2000mw} (the latter recently independently derived using the truncation from seven-dimensional maximal supergravity in~\cite{Cheung:2019pge}). In both cases, we showed how the generalised geometry  completely characterises the truncated theory. For the type IIB reduction we also discussed the bosonic truncation ansatz, while for the M-theory one we recovered the ansatz for the sub-truncation to pure half-maximal supergravity previously given in~\cite{Gauntlett:2007sm}.

There are many other possible truncations that it would be intriguing to explore using our formalism. We sketch here some possibilities directly related to the cases we have studied. In type IIB $\Ex{6}$ geometry, it would be interesting to construct a generalised $\U(1)$ structure on the $Y^{p,q}$ family \cite{Gauntlett:2004yd} of Sasaki--Einstein manifolds, and check if it admits a $\U(1)$-singlet intrinsic torsion. If so, this would give a half-maximal consistent truncation on  $Y^{p,q}$ manifolds extending the one based on  generic Sasaki--Einstein $\SU(2)$ structure by one Betti vector multiplet, as in the $Y^{1,0}\simeq T^{1,1}$ truncation of~\cite{Cassani:2010na,Bena:2010pr}. For this to go through, one would need the full flexibility of generalised geometry in order to circumvent the issue pointed out in~\cite{Liu:2011dw} relevant working with ordinary $G$-structures.

In M-theory, it would be nice to extend the construction presented in Section \ref{sec:MNsection}, which is based on the geometry of \cite{Maldacena:2000mw}, to the general ansatz for half-maximal AdS$_5$ solutions of \cite{Lin:2004nb}. In particular, this would provide new consistent truncations containing the AdS$_5$ solutions of \cite{Gaiotto:2009gz}, describing M5-branes wrapped on Riemann surfaces with punctures. 
A similar construction is conceivable for the supergravity description of D3-branes wrapped on Riemann surfaces \cite{Maldacena:2000mw}, however in this case one would need to use the more complicated type IIB $\Ex{8}$ generalised geometry formalism, which is not fully developed yet (though see \cite{Hohm:2014fxa,Baguet:2016jph}).

M5-branes wrapped on Riemann surfaces also give rise to AdS$_5\times_{\rm w}M_6$ supergravity solutions preserving just one quarter of the supersymmetry, which are dual to $\mathcal{N}=1$ four-dimensional SCFTs~\cite{Maldacena:2000mw,Bah:2012dg}. Our general analysis can be used to  predict the form of the corresponding consistent truncations. For the quarter-supersymmetric solution of \cite{Maldacena:2000mw}, the structure is again $\U(1)$ but embedded in a different way in $\Ex{6}$. It is easy to see that in this case, there are only two singlet spinors, and so the truncation is to minimal five-dimensional supergravity. The scalar moduli space is 
\begin{equation}
\label{eq:MN-N=1}
   \Mscal = \frac{\Com{\U(1)}{\Ex{6}}}{\Com{U(1)}{\USp(8)}}
   = \bbR^+ \times \frac{\SO(3,1)}{\SO(3)} \times
      \frac{\SU(2,1)}{\SU(2)\times U(1)} \,,
\end{equation}
and there are five singlet vectors in the generalised tangent space. We see that the truncated theory is minimal five-dimensional supergravity coupled to four vector multiplets and a single hypermultiplet. The first factor in~\eqref{eq:MN-N=1} gives the homogeneous very special real geometry describing the four additional vector multiplets, while the second factor is the standard homogeneous quaternionic space for a single hypermultiplet. The singlet generalised vectors $K_\mathcal{A}$ are again constructed starting from the frames on $S^4$ but now the relevant twist connection is
\be
\Upsilon = \tfrac{1}{2}\, \upsilon \times_{\rm ad} (E_{14}-E_{23}) \,.
\ee
The details of this truncation will be discussed in future work. 
Although none of the truncations constructed in this way are expected to contain the domain wall solutions connecting the different AdS$_5\times_{\rm w}M_6$ supergravity solutions (which are dual to conjectured RG flows between the corresponding SCFTs), the generalised geometry approach may suggest how to make the twist ``dynamical'' so that it can evolve along the flow.

Half-maximal consistent truncations of massive type IIA supergravity can also be engineered by combining the formalism of the present paper with the one of \cite{Cassani:2016ncu,Ciceri:2016dmd}, where the maximally supersymmetric case was discussed.

Besides consistent truncations, a physically relevant motivation for developing half-maximal structures is to study the moduli space of half-maximal AdS solutions to supergravity theories, which is dual to the conformal manifold ({\it i.e.}\ the space of exactly marginal deformations) of SCFTs with eight Poincar\'e and eight conformal supercharges, in the large $N$ limit. In the quarter-supersymmetric case, a study of marginal deformations using generalised geometry was done in \cite{Ashmore:2016oug,Ashmore:2018npi}. The additional constraint of half-maximal supersymmetry may allow to go further in the analysis.  In particular, it may allow one to compare in great detail with field theory results, where the K\"ahler metric on the conformal manifold follows from the $S^4$ partition function \cite{Gerchkovitz:2014gta}, which is computable using supersymmetric localization.
It would also be interesting to compare with the results found in~\cite{Louis:2015dca} by means of a purely five-dimensional supergravity setup.

\section*{Acknowledgments}

We would like to thank Charlie Strickland-Constable and Oscar de Felice for collaboration in the initial stages of this work. We also thank Nikolay Bobev, Jerome Gauntlett, Nick Halmagyi, Emanuel Malek and Hagen Triendl for discussions. This work was supported in part by the EPSRC Programme Grant ``New Geometric Structures from String Theory'' EP/K034456/1, the EPSRC standard grant EP/N007158/1, and the STFC Consolidated Grant ST/L00044X/1. 
We acknowledge the Mainz Institute for Theoretical Physics (MITP) of the Cluster of Excellence PRISMA+ (Project ID 39083149) for hospitality and support during part of this work. DC would like to thank the LPTHE, the CNRS and the ``Research in Paris'' program of the Institut Henri Poincar\'e  for hospitality and support.

\appendix

\section{Type IIB $\mathbf{E_{6(6)}}$ generalised geometry}
\label{PreliminariesE66_IIB}

We briefly recall the exceptional geometry for type IIB compactifications on a five-dimensional manifold $M$, following the conventions of~\cite[App.~E]{Ashmore:2015joa}.
The type IIB generalised tangent bundle on $M$ has fibres transforming in the ${\bf 27}$ of $\Ex{6}$ and decomposes under the geometric $\GL(5)$  subgroup of  $\Ex{6}$ as 
\be
E \simeq TM\oplus T^*M \oplus \Lambda^5T^*M \oplus \Lambda^{\rm odd}T^*M\,,
\ee
where $\Lambda^{\rm odd}T^* = T^* \oplus \Lambda^3T^*\oplus \Lambda^5T^*$.
A generalised vector $V\in \Gamma(E)$ can be written as
\be
V = v + \lambda + \sigma  + \omega\,,
\ee
where $\omega = \omega_1 + \omega_3 + \omega_5$ is a poly-form of odd degree.
Alternatively, the generalised tangent bundle can be decomposed in a way that also makes the action of $\SL(2)$ manifest. The $\GL(5) \times \SL(2)$ covariant decomposition is
\begin{equation}
   E       \simeq TM \oplus ( S \otimes T^*M )  \oplus \Lambda^3 T^*M  \oplus (S \otimes \Lambda^5T^*M)\,,
\end{equation}
where  $S$ denotes an $\SL(2)$ doublet. In this picture a generalised vector can be expressed as 
\be
V = v + \lambda^\alpha + \rho + \sigma^\alpha\,, 
\ee
where the index $\alpha=\{+,-\}$ labels the states in the $\SL(2)$ doublet. In this paper we use the second description.

The dual generalised vector bundle decomposes accordingly as 
\begin{equation}
   E^* \simeq T^*M \oplus (S^* \otimes TM) \oplus \Lambda^3TM 
      \oplus (S^* \otimes\Lambda^5TM )\ ,
\end{equation}
and a generalised dual vectors $Z\in \Gamma(E^*)$ can be written as
\be
Z = \hat{v} + \hat{\lambda}_\alpha + \hat{\rho}  + \hat{\sigma}_\alpha \, . 
\ee

The adjoint bundle is defined as
\be 
{\rm ad} F = \mathbb{R} \oplus (T M \otimes T^*M) \oplus  (S \otimes S^*) \oplus (S \otimes \Lambda^2 TM)   \oplus (S \otimes \Lambda^2 T^*M)  \oplus  \Lambda^4 TM   \oplus  \Lambda^4 T^*M
\ee
with elements
\be
R = l + r + a^\alpha{}_\beta + \beta^\alpha + B^\alpha + \gamma  + C
\ee
where $l \in \Gamma(\mathbb{R})$, $r\in End (TM)$, $a^\alpha{}_\beta$  is an element of $SL(2)$, $\beta^\alpha$ and  $B^\alpha$ are  an $\SL(2)$ doublet of  bi-vectors and two-forms respectively, $\gamma$ is a four-vector and $C$ a four-form. 
The adjoint action of $R\in\Gamma({\rm ad})$ on $V\in \Gamma(E)$, denoted by $V^{\prime}=R\cdot V$, is defined as:
\begin{align}
v^{\prime} \,& =\, lv+r\cdot v+\gamma\lrcorner\rho+\epsilon_{\alpha\beta}\beta^{\alpha}\lrcorner\lambda^{\beta}\,,\nn\\[1mm]
\lambda^{\prime \alpha} \,& =\, l\lambda^{\alpha}+r\cdot\lambda^{\alpha}+a_{\phantom{\alpha}\beta}^{\alpha}\lambda^{\beta}-\gamma\lrcorner\sigma^{\alpha}+v\lrcorner B^{\alpha}+\beta^{\alpha}\lrcorner\rho\,,\nn\\[1mm]
\rho^{\prime} \,& =\, l\rho+r\cdot\rho+v\lrcorner C+\epsilon_{\alpha\beta}\beta^{\alpha}\lrcorner\sigma^{\beta}+\epsilon_{\alpha\beta}\lambda^{\alpha}\wedge B^{\beta}\,,\nn\\[1mm]
\sigma^{\prime \alpha} \,& =\, l\sigma^{\alpha}+r\cdot\sigma^{\alpha}+a_{\phantom{\alpha}\beta}^{\alpha}\sigma^{\beta}-C\wedge\lambda^{\alpha}+\rho\wedge B^{\alpha} \,,\label{eq:IIB_adjoint}
\end{align}
where $\epsilon_{\alpha\beta}$ is defined as $\epsilon_{+-}=-\epsilon_{-+}=1$, $\epsilon_{++}=\epsilon_{--}=0$, and for the definition of the $\mathfrak{gl}(5)$ action $r\cdot$ and of the contraction $\lrcorner$ we refer to \cite{Ashmore:2015joa}.

A generalised vector can be twisted by elements of the adjoint bundle. In particular,  the twisted generalised vector  $V = \rme^{B^\alpha + C} \check{V}$ is given by
\begin{align}
v &= \check v \,, \nn \\
\lambda^\alpha &= \check \lambda^\alpha +  \check{v} \lrcorner B^\alpha  \,, \nn \\
\rho &= \check \rho + \check{v} \lrcorner C + \epsilon_{\alpha \beta} \check \lambda^\alpha \wedge B^\beta +\frac{1}{2} \epsilon_{\alpha\beta}\,\check{v} \lrcorner B^\alpha \wedge B^\beta\,, \nn \\ 
\sigma^\alpha &= \check \sigma^\alpha  - C \wedge \check \lambda^\alpha + \check \rho \wedge B^\alpha - \frac{1}{2} \check{v} \lrcorner B^\alpha \wedge C + \frac{1}{2} \left(\check v\lrcorner C+ \epsilon_{\alpha\beta} \check\lambda^\beta \wedge B^\gamma\right) \wedge B^\alpha\, .
\label{twisted_vector}
\end{align}

Another bundle of interest is the bundle $N\simeq \det T^*M \otimes E^*$. This a sub-bundle of the symmetrised product of two copies of the generalised tangent bundle. Its fibres transform in the ${\bf 27^\prime}$ of $E_{6(6)}$  and its $\GL(5) \times \SL(2)$ decomposition reads
\be
\label{NbIIB} 
N \simeq (S^* \otimes \mathbb{R}) \oplus \Lambda^2 T^*M \oplus (S^* \otimes \Lambda^4 T^*M) \oplus (\det T^*M \otimes T^*M)\,,
\ee
with sections 
\be\label{sectionNIIB}
J = s_\alpha + \omega_2 + \omega_{4\,\alpha} + \varsigma \,.
\ee 

The $E_{6(6)}$ cubic invariant acting on  three generalised vectors is defined as 
\begin{equation}
\label{cubic_inv_differentV}
c(V,V',V'') = -\tfrac{1}{2} \left( \iota_v \rho' \wedge \rho'' + \epsilon_{\alpha\beta}\, \rho\wedge \lambda'{}^\alpha \wedge \lambda''{}^\beta - 2 \epsilon_{\alpha\beta}\,\iota_v \lambda'{}^\alpha  \sigma''{}^\beta \right) + \text{ symm.~perm.}\, . 
\end{equation}
The cubic invariant acting on dual vectors is
\begin{equation}
\label{scubic_inv_differentV}
c^\ast(Z,Z',Z'') = -\tfrac{1}{2} \left( \iota_{\hat v}  \hat{\rho}' \wedge  \hat{\rho}'' + \epsilon^{\alpha\beta}\, \hat{ \rho} \wedge \hat{ \lambda}'_\alpha \wedge \hat{\lambda}''_\beta - 2 \epsilon^{\alpha\beta}\,\iota_{\hat v}  \hat{\lambda}'_\alpha \hat{ \sigma}''_\beta \right) + \text{symm.~perm.}\, , 
\end{equation}
where $\epsilon^{\alpha\beta}$ is defined as a matrix with the same components as $\epsilon_{\alpha\beta}$.

The product between an element  $V \in {\bf 27}$ and  $Z \in {\bf \overline{27}}$  is defined as
\be
\label{pairing_vector_dualvector-app}
\GM{Z}{V} = \hat v_m v^m + \hat\lambda^{m}_\alpha\lambda_m^{\alpha} + \tfrac{1}{3!}\,\hat\rho^{mnp}\rho_{mnp} +  \tfrac{1}{5!}\,\hat\sigma^{mnpqr}_\alpha\sigma_{mnpqr}^\alpha \,.
\ee

The generalised Lie derivative between two twisted generalised vectors $V,V'$ is given by
\begin{align}\label{genLieE66second}
L_V V' \ =&\ \mathcal{L}_v v'  + (\mathcal{L}_v \lambda'^{\alpha} -\iota_{v'}\mathrm{d}\lambda^\alpha)  + (\mathcal{L}_v \rho' -\iota_{v^\prime}\mathrm{d}\rho + \epsilon_{\alpha\beta}\diff\lambda^\alpha\wedge \lambda'^\beta) \nn \\[1mm]
&\ + \mathcal{L}_v \sigma'^\alpha - \mathrm{d}\lambda^\alpha \wedge \rho' + \lambda'^\alpha\wedge \mathrm{d}\rho    \, .
\end{align}

The generalised metric is defined as
\be
\label{genmet5d}
G(V,V') =  v^m v'_m + h_{\alpha \beta} \lambda^{m\alpha}   \lambda'^\beta_m   + \frac{1}{3!} \rho^{mnp} \rho'_{mnp} + \frac{1}{5!} \sigma^{mnpqr} \sigma'_{mnpqr}\,,
\ee
where $V$ and $V^\prime$ are the twisted generalised vectors defined in \eqref{twisted_vector} and   the latin indices are lowered/raised using the ordinary metric $g_{mn}$ or its inverse
$g^{mn}$.
The matrix $h_{\alpha \beta}$ parameterises the coset $\SL(2)/\SO(2)$ and is given by
\be
h_{\alpha \beta} \,=\, \rme^\phi \begin{pmatrix}  C_0^2 + \rme^{-2 \phi} & - C_0 \\ - C_0 & 1 \end{pmatrix}  \,,
\ee
with inverse
 \be
 \label{habi}
h^{\alpha \beta} \,=\, \rme^\phi \begin{pmatrix}  1 &    C_0 \\ C_0 &   C_0^2 + \rme^{-2 \phi}   \end{pmatrix}    \,.
\ee
Note that $h^{\alpha\beta}=\epsilon^{\alpha\alpha'}\epsilon^{\beta\beta'}h_{\alpha'\beta'}$.

The inverse generalised metric can be obtained from a generalised local frame $ E_A\in \Gamma(E)$, $A=1,\ldots,27$, as
\begin{align}
G^{-1}  &=  \delta^{A B} {E}_A \otimes {E}_B  \nn \\
& = \delta^{a b} {E}_a \otimes {E}_b  +   \delta_{a b}  \delta^{\hat \alpha \hat \beta}  {E}^a_{\hat \alpha}  \otimes  {E}^b_{\hat \beta} +
  \tfrac{1}{3!} \delta_{a_1 b_1}    \delta_{a_2 b_2}   \delta_{a_3 b_3}   {E}^{a_1 a_2 a_3} \otimes  {E}^{b_1 b_2 b_3}  \nn \\
&  \quad   +   \tfrac{1}{5!}\delta_{a_1 b_1}   \cdots  \delta_{a_5 b_5}     \delta^{\hat \alpha \hat \beta}  {E}^{a_1 \ldots a_5}_{\hat \alpha}  \otimes  {E}^{b_1 \ldots b_5}_{\hat \beta}\,,
\end{align}
where $a,b,\ldots$ are flat $\GL(5)$ indices while $\hat\alpha,\hat\beta$ are flat $\SL(2)$ indices.
Starting from an untwisted frame $\check{E}_A$, the generalised frame $ E_A$ is defined as~\cite{Lee:2014mla}
\be
E_A = \rme^{\Delta}\,\rme^{\phi}\,\hat{f}_{\hat \alpha}{}^\alpha\,\rme^{B^\alpha + C}\cdot \check{E}_A \, , 
\ee
where $\hat{f}_{\hat\alpha}{}^\alpha= { \rme^{\phi/2}\ C_0\rme^{\phi/2}\choose 0\quad \rme^{-\phi/2}}$.
The action of the warp factor $\rme^{\Delta}$ is defined by exponentiating the adjoint element given by $l= \Delta$, while the dilaton action $\rme^{\phi}$ is defined by exponentiating the adjoint element given by $l + r = \frac{\phi}{4}(-1 + \mathbf{1})$ \cite{Ashmore:2015joa}.
Decomposing the flat index $A$ in $\GL(5)\times \SL(2)$ representations, the generalised frame may be written as
\begin{align}
&  E_a  = \rme^{\Delta} \left( \hat{e}_a + \iota_{ \hat{e}_a} B^\alpha +  \iota_{ \hat{e}_a} C + \tfrac{1}{2} \epsilon_{\alpha \beta }  \iota_{ \hat{e}_a} B^\alpha  \wedge B^\beta
+  \iota_{ \hat{e}_a} C  \wedge B^\alpha + 
\tfrac{1}{6} \epsilon_{\beta \gamma}\,  \iota_{ \hat{e}_a} B^\beta \wedge B^\gamma \wedge B^\alpha \right)  \nn \\[1mm]
&  E_{\hat \alpha}^a =  \rme^{\Delta}\,\rme^{-\phi/2} \left( \hat{f}_{\hat \alpha}{}^\alpha e^a +   \hat{f}_{\hat \alpha}{}^\alpha  \epsilon_{\alpha \beta}  e^a \wedge B^\beta
 -  \hat{f}_{\hat \alpha}{}^\alpha  C \wedge e^a + \tfrac{1}{2}  \hat{f}_{\hat \alpha}{}^\beta  \epsilon_{\beta \gamma}  e^a \wedge B^\gamma \wedge B^\alpha  \right) \nn \\[1mm]
& E^{a_1a_2a_3}  = \rme^{\Delta}\,\rme^{-\phi}\left(e^{a_1a_2a_3} +  e^{a_1a_2a_3}  \wedge B^\alpha \right) \nn \\[1mm]
& E_{\hat \alpha}^{a_1\ldots a_5} =   \rme^\Delta \rme^{-3\phi/2}  \hat{f}_{\hat \alpha}{}^\alpha \,e^{a_1\ldots a_5}\,,\label{SplitFrameE66}
\end{align}
where $e^{a_1\ldots a_n}= e^{a_1}\wedge \cdots \wedge e^{a_n}$.
Using this expression for the frame, we obtain for the different components of the inverse generalised metric:
\begin{align}\label{generalised_metric_IIB}
& (G^{-1})^{mn}  = \rme^{2\Delta} g^{mn}\nn \\[1mm] 
& (G^{-1})^{m}{}^\beta_{n}  
 = \rme^{2\Delta}g^{mp} B^\beta_{pn}  \nn \\[1mm]
& (G^{-1})^m{}_{npq}  = \rme^{2\Delta} g^{m r} \left(C_{r npq} + \tfrac{3}{2} \epsilon_{\alpha \beta} B^\alpha_{r [n} B^\beta_{ pq]} \right)  \nn \\[1mm]
& (G^{-1})^{m}{}^\beta_{npqrs}  =   \rme^{2\Delta} g^{m u} \left( 10 C_{u [npq} B^\beta_{r s]} + 5 \epsilon_{\gamma \delta} B^\gamma_{u [n} B^\delta_{pq} B^\beta_{rs]}\right)  \nn \\[1mm]
& (G^{-1})_{m \, n}^{\alpha\,\,\beta}  = \rme^{2\Delta}\left( \rme^{-\phi} h^{\alpha \beta} g_{mn} - B^\alpha _{m p} g^{pq} B^\beta_{qn} \right) \nn \\[1mm]
& (G^{-1})^\alpha_{m\, npq}  = \rme^{2\Delta}\left( 3 \rme^{-\phi} h^{\alpha \beta} \epsilon_{\beta \gamma} g_{m[n} B^\gamma_{pq]}  - B^\alpha_{mr}  g^{rs} C_{s npq}  -  \tfrac{3}{2} \epsilon_{\beta \gamma}   B^\alpha_{m r} g^{rs} B^\beta_{s [n} B^\gamma_{pq]} 
 \right)  \nn \\[1mm]
& (G^{-1})^{\alpha \, \beta}_{m  \,  n pq rs } = \rme^{2\Delta}\rme^{-\phi}\left( - 5 h^{\alpha \beta} g_{m[n} C_{pqrs]}  + 15 h^{\alpha \gamma}   \epsilon_{\gamma \delta} g_{m [n} B^\delta_{pq} 
B^\beta_{rs]}  \right) \nn \\[1mm]
& (G^{-1})_{mnp \, qrs} =  \rme^{2\Delta}\left[ g^{tu} \left(C_{t mnp} + \tfrac{3}{2} \epsilon_{\alpha \beta} B^\alpha_{t[m} B^\beta_{n p]}\right) \left(C_{u qrs} + \tfrac{3}{2} \epsilon_{\gamma \delta} B^\gamma_{u[q} B^\delta_{rs]}\right) \right. \nn \\[1mm]
& \hspace{35mm} \left. +\, 9\, \rme^{-\phi} h_{\alpha \beta} B^\alpha_{[mn } g_{p] [q} B^\beta_{rs]}  +  6\,\rme^{-2\phi} g_{mnp,qrs} \right] \,,
\end{align}
where in the last line we defined $g_{mnp,qrs} = g_{q[m}g_{n|r|}g_{p]s}$. 
 We will not need the expressions for the remaining components $(G^{-1})_{mnpqr \, stu}^\alpha$ and $(G^{-1})^{\alpha\quad\;\;\;\beta}_{mnpqr \, stuvw}$.

The warp factor $\rme^{\Delta}$ is in principle extracted by evaluating the determinant of the whole generalised metric. However, for type IIB we can follow the same shortcut given in~\cite{Cassani:2016ncu} for type IIA. We introduce
\begin{equation}
\label{eq:GB-metric}
	\mathcal{H}^{-1} \,=\, \begin{pmatrix} (G^{-1})^{mn} & (G^{-1})^m{}_n^+ \\[1mm]
		 (G^{-1})^{+}_m{}^n & (G^{-1})_{mn}^{++} \end{pmatrix}
		 \,=\, \rme^{2\Delta} \begin{pmatrix} g^{mn} & (g^{-1} B)^m{}_n \\[1mm]
		 -(B g^{-1})_m{}^n & (g - B g^{-1} B)_{mn} \end{pmatrix}\,,
\end{equation}
where $B^+$ is the NSNS two-form potential and observe that the matrix on the right hand side  has unit determinant. Therefore we obtain\footnote{The expressions above are given in
string frame.  In Einstein frame the term $g^{mn}$ in \eqref{eq:GB-metric} becomes $\rme^{\phi} g^{mn}$,  and \eqref{eq:GB-metric} becomes
\be
\label{detGB}
	\rme^{\Delta+ \frac{\phi}{4} }\, =\, (\det \mathcal{H})^{-\frac{1}{4d}}\,.
\ee}
\begin{equation}
\label{detGB}
	\rme^{\Delta}\, =\, (\det \mathcal{H})^{-\frac{1}{4d}}\,.
\end{equation}

\section{Generalised vectors in angular coordinates on $M_6$}\label{app:angular_coords_M6}

In this appendix we provide explicit expressions for the generalised vectors $K_{\mathcal{A}}$, $\mathcal{A}=0,1,\ldots,8$, defining the generalised $\U(1)$ structure on the six-manifold discussed in Section~\ref{sec:MNsection}. We start by relating the constrained coordinates $y_i$, $i=1,\ldots,5$, used in the main text to angular coordinates $(\theta,\psi,\chi,\phi)$ on a round $S^4$ of unit radius:
\begin{align}
y^1 &= \sin\theta\cos\psi\,, \nonumber\\
y^2 &= \sin\theta\sin\psi\,, \nonumber\\
y^3 &= \cos\theta\;\mu^3 = 
\cos\theta\,\cos\chi\,, \nonumber\\
y^4 &= \cos\theta\;\mu^4 = 
\cos\theta\,\sin\chi \cos\phi\,, \nonumber\\
y^5 &= \cos\theta\;\mu^5 =
\cos\theta\,\sin\chi \sin\phi \,.
\end{align}
Notice that $\psi$ parameterises $\U(1)$ rotations in the $1-2$ plane,
while $\chi,\phi$ parameterise $\SO(3)$ rotations in the $3-4-5$ space and thus describe a round $S^2$. The latter is equivalently described by the constrained coordinates $\mu^\alpha$, $\alpha=3,4,5$, satisfying $\delta_{\alpha\beta}\mu^\alpha\mu^\beta=1$; we  use either one or the other description according to convenience.
The round metric on $S^4$ and the associated volume form \eqref{metricvolS4_ycoords} read
\begin{align}
g_{4} 
&= R^2  \left(\dd \theta^2 +  \sin^2\theta\, \dd \psi^2 + \cos^2\theta \: g_{S^2} \right)\,,\nn\\[1mm]
{\rm vol}_{4}  &= R^4\cos^2\theta\sin\theta \,\dd\theta \wedge \dd\psi \wedge \vol_{S^2}\,,
\end{align}
where 
\begin{align}
g_{S^2} &= \delta_{\alpha\beta} \dd\mu^\alpha \dd \mu^\beta  =  \dd \chi^2 + \sin^2\chi \,\dd \phi^2  \,,\nn\\[1mm]
{\rm vol}_{S^2}  &=  \tfrac{1}{2}\epsilon_{\alpha\beta\gamma}\,\mu^\alpha\dd \mu^\beta\wedge \dd \mu^\gamma = \sin\chi \, \dd \chi \wedge\dd \phi
\end{align}
are the unit metric and volume form on the two-sphere identified above.
The $S^4$ Killing vectors generating the $\mathfrak{u}(1)\oplus \mathfrak{su}(2)$ algebra of interest are  expressed in terms of these angular coordinates as:
\begin{align}
v_{12} &= R^{-1}\,\tfrac{\partial}{\partial \psi}\,,\nonumber\\[1mm]
v_{45} &= R^{-1}\,\tfrac{\partial}{\partial \phi}\,,\nonumber\\[1mm]
v_{53} &=  R^{-1}\,\left(-\sin\phi\,\tfrac{\partial}{\partial \chi}- \cot\chi\cos\phi\,\tfrac{\partial}{\partial \phi}\right)\,,\nonumber\\[1mm]
v_{34} &= R^{-1}\left(\cos\phi\,\tfrac{\partial}{\partial \chi}- \cot\chi\sin\phi\,\tfrac{\partial}{\partial \phi}\right)\,.
\end{align}

For the M-theory three-form potential on $S^4$ satisfying \eqref{flux_MN_background}, we choose
\be
A =  - 3R^3 \, \psi\, \cos^2\theta \sin\theta \,\dd\theta \wedge\vol_{S^2}\,.
\ee

As for the Riemann surface ${\mathbb{\Sigma}}$, we do not need to introduce explicit coordinates; we rather use the one-forms $e_1,e_2$ satisfying \eqref{gSigmavolSigma}, \eqref{d_frame_Sigma}.

Evaluating the twisted $K$'s \eqref{gen_vec_S4xSigma} in these coordinates, we find the following expressions:
\begin{align}
K_0 &= \tfrac{1}{2}\, v_{12} + \tfrac{1}{2}\, R^2 \ct^3\,  \vol_{S^2}   \,, \nn \\[3mm]
K_1 \!+\! \ii K_2&= R\,\rme^{\ii \,\psi} \left(e_1+\ii \, e_2 \right)\wedge \left(\ct \,\dd \theta + \ii \,\st \,(\dd \psi + \upsilon)\right)     + R^4\, \rme^{\ii\, \psi} \, \ct^2\left(e_1+\ii\, e_2\right)\wedge
  \nn \\[1mm]
&\quad\wedge \big[ \left( \st^2 \left(1-3\,\ii\,\psi\right)\left(\dd \psi + \upsilon\right) +\ct^2\, \upsilon \right) \wedge \dd\theta +\ii\, \ct\, \st \,\upsilon \wedge \dd\psi \big] \wedge\vol_{S^2}\,,   \nn \\[3mm]
K_3 &= v_{45} - R^2\, \dd\left(\ct\, \cc\right)\wedge (\dd\psi+\upsilon) + R^2\, \dd\left[ \cc\, \dd\left(\psi\, \ct^3\right) \right]+ \ct\, \cc \vol_{\mathbb{\Sigma}}     \nn \\[1mm]
& - R^3\ct \st \vol_{\mathbb{\Sigma}} \wedge \left(  \ct \st \cc\, \dd\psi\wedge \vol_{S^2} + \sc^2\, \dd\theta\wedge\dd\psi\wedge\dd\phi  + 3\,\psi\, \ct^2\, \cc \, \dd\theta\wedge \vol_{S^2} \right) \nn\\[1mm]
& - R^5\,\ct^3 \,\st\,\cc \,\upsilon \wedge \dd\theta\wedge\dd\psi\wedge \vol_{S^2}   \,, \nn \\[3mm]
K_4 &= v_{53} - R^2 \dd(\ct\, \sc \, \cp)\wedge (\dd\psi+ \upsilon) + R^2\, \dd \!\left[  \sc\, \cp \,\dd\left(\psi\, \ct^3\right) \right] + \ct\, \sc\, \cp\vol_{\mathbb{\Sigma}}    \nn \\[1mm]
& + R^3\,\ct\st \vol_{\mathbb{\Sigma}} \wedge \big[ \cp \left( \cc \sc\dd\theta\wedge\dd\psi\wedge\dd\phi -\ct\,\st\,\sc \dd\psi\wedge\vol_{S^2} -3\,\psi\,\ct^2\,\sc\,\dd\theta\wedge \vol_{S^2} \right)  \nn\\[1mm]
&  + \sp \,\dd\theta\wedge\dd\psi\wedge\dd\chi \big] - R^5\,\ct^3 \,\st\,\sc \,\cp\,\upsilon\wedge\dd\theta\wedge\dd\psi \wedge\vol_{S^2}  \,,   \nn \\[3mm]
K_5 &= v_{34}  - R^2\, \dd(\ct\, \sc \sp)\wedge (\dd\psi+ \upsilon) + R^2\, \dd \!\left[  \sc \sp \,\dd\left(\psi\, \ct^3\right) \right] + \ct\, \sc \sp\vol_{\mathbb{\Sigma}}    \nn \\[1mm]
& + R^3\,\ct\st \vol_{\mathbb{\Sigma}} \wedge \big[ \sp \left( \cc \sc\dd\theta\wedge\dd\psi\wedge\dd\phi -\ct\,\st\,\sc\, \dd\psi\wedge\vol_{S^2} -3\,\psi\,\ct^2\,\sc\,\dd\theta\wedge\vol_{S^2} \right)  \nn\\[1mm]
&  - \cp \,\dd\theta\wedge\dd\psi\wedge\dd\chi \big] - R^5\,\ct^3 \,\st\,\sc \,\sp\,\upsilon\wedge\dd\theta\wedge\dd\psi\wedge\vol_{S^2}   \,,  \label{explicit_K_MN}
\end{align}
with $K_6,K_7,K_8$ being obtained from $K_3,K_4,K_5$, respectively, by sending $\vol_{\mathbb{\Sigma}} \to -\vol_{\mathbb{\Sigma}}\,$. In these formulae, we are using the shorthand notation $\ct =\cos\theta$, $\st=\sin\theta$, and similarly for the angles $\chi,\phi$. The terms in \eqref{explicit_K_MN} proportional to $\psi$ are those coming from the action of the three-form $A$ in the $S^4$ frames \eqref{frames_S4}.
The terms proportional to $\upsilon$ are those generated by the twist \eqref{twistedKs} (that is, setting $\upsilon=0$ we recover the generalised vectors on the direct product ${\mathbb{\Sigma}} \times S^4$).
We see that the latter transformation shifts $\dd \psi$ by the connection one-form on ${\mathbb{\Sigma}}$, such that $\dd\psi\to(\dd\psi+\upsilon)$, and also introduces some additional five-form parts in the generalised vectors.

The weighted dual vectors $J^{\mathcal A}\in \Gamma(N)$ that give the ansatz for the supergravity two-forms can be computed from the $K_{\mathcal{A}}$ using \eqref{K*fromK}. We find:
\begin{align}
J^0&=2R^4 \, c_\theta s_\theta \vol_{\mathbb{\Sigma}}\wedge\,\dd\theta\wedge\dd\psi +2 R(\upsilon+ 3\psi\, c_\theta s_\theta\, \dd\theta +s_\theta^2\,\dd\psi)\otimes \text{vol}_6 \,,\nn\\[1mm]
J^1+\ii J^2 &=  \frac{R}{2}s_\theta \,\ii \,\rme^{\ii\, \psi} (e_1+\ii\, e_2)  - \frac{R^4}{2} \,\rme^{\ii\,\psi} (e_1 + \ii\, e_2)\wedge\left[  c^2_\theta\, \dd\theta + \ii\, s_\theta \,  \dd\left(\psi\, c^3_\theta\right)  \right]\wedge\vol_{S^2} \,,\nn\\[1mm]
J^3&=\frac{R}{2}\,\dd( c_\theta c_\chi)+\frac{R^4}{2}c_\theta  \big[ s_\theta s_\chi^2 \vol_{\mathbb{\Sigma}} \wedge\,\dd\theta\wedge\dd\phi - c_\theta c_\chi(\vol_{\mathbb{\Sigma}} +c_\theta s_\theta \,\dd\theta\wedge\dd\psi)\wedge\vol_{S^2}\big]\nn\\[1mm]
&\quad -\frac{R}{2} c^2_\theta \,s^2_\chi\, \dd\phi \otimes  \text{vol}_6\,,\nn\\
J^4&=\frac{R}{2}\dd(c_\theta c_\phi s_\chi )- \frac{R^4}{2}c_\theta \big [s_\theta c_\phi c_\chi s_\chi \vol_{\mathbb{\Sigma}}\wedge\,\dd\theta\wedge\dd\phi + s_\theta s_\phi \vol_{\mathbb{\Sigma}}\wedge\,\dd\theta\wedge\dd\chi\nn\\
&\quad + c_\theta c_\phi s_\chi (\vol_{\mathbb{\Sigma}} + c_\theta s_\theta\,\dd\theta \wedge\dd\psi )\wedge \vol_{S^2} \big] +\frac{R}{2} c^2_\theta (s_\phi\,\dd\chi +c_\phi c_\chi s_\chi\, \dd\phi )\otimes  \text{vol}_6\,,\nn\\[1mm]
J^5&=\frac{R}{2}\dd(c_\theta s_\phi s_\chi )+\frac{R^4}{2}c_\theta \big[- s_\theta  c_\chi s_\chi s_\phi \vol_{\mathbb{\Sigma}} \wedge\,\dd\theta\wedge\dd\phi+s_\theta c_\phi \vol_{\mathbb{\Sigma}} \wedge\,\dd\theta\wedge\dd\chi\nn\\
&\quad -c_\theta s_\phi s_\chi(\vol_{\mathbb{\Sigma}} +c_\theta s_\theta\,\dd\theta\wedge\dd\psi)\wedge  \vol_{S^2}\big] +\frac{R}{2} c^2_\theta (-c_\phi \,\dd\chi +c_\chi s_\phi s_\chi\, \dd\phi )\otimes  \text{vol}_6 \,,
\end{align}
and again $J^6,J^7,J^8$ are obtained from $J^3,J^4,J^5$, respectively, by sending $\vol_{\mathbb{\Sigma}} \to - \vol_{\mathbb{\Sigma}}$ and $\vol_6\to-\vol_6$.

\bibliographystyle{JHEP}
\bibliography{Bibliography}
\end{document}